\documentclass[iop]{emulateapj}
\usepackage{hyperref}
\usepackage{graphicx}
\pdfoutput=1
\usepackage{dcolumn}
\usepackage{bm}
\usepackage{float}
\usepackage{listings}
\usepackage{color}
\usepackage{textcomp}
\usepackage{times}
\usepackage{graphics}
\usepackage{amssymb}
\usepackage{latexsym}  
\usepackage{amsmath}
\usepackage{fancyhdr}
\usepackage[figuresright]{rotating}
\usepackage{array}
\usepackage{multirow}
\usepackage{afterpage}

\usepackage{enumitem}
\newlist{inparaenum}{enumerate}{2}
\setlist[inparaenum]{nosep}
\setlist[inparaenum,1]{label=\bfseries\arabic*.}
\setlist[inparaenum,2]{label=\arabic{inparaenumi}\emph{\alph*})}

\newcommand{\degree}{\ensuremath{^\circ}}
\newcommand{\ps}{peaked-spectrum }

\newcommand{\al}{$\alpha_{\mathrm{low}}$}
\newcommand{\ah}{$\alpha_{\mathrm{high}}$}
\newcommand{\athin}{$\alpha_{\mathrm{thin}}$}
\newcommand{\athick}{$\alpha_{\mathrm{thick}}$}
\shorttitle{Extragalactic Peaked-Spectrum Radio Sources at Low frequencies}
\shortauthors{\textsc{Callingham et al.}}
\submitted{Accepted to ApJ 2017 Jan. 04}

\begin{document}

\title{Extragalactic Peaked-Spectrum Radio Sources at Low Frequencies}
	
\author{J.~R.~Callingham\altaffilmark{1,2,3}, R.~D.~Ekers\altaffilmark{2}, B.~M.~Gaensler\altaffilmark{4,1,3}, J.~L.~B.~Line\altaffilmark{6,3}, N.~Hurley-Walker\altaffilmark{5}, E.~M.~Sadler\altaffilmark{1,3}, S.~J.~Tingay\altaffilmark{7,5}, P.~J.~Hancock\altaffilmark{5,3}, M.~E.~Bell\altaffilmark{2,3}, K.~S.~Dwarakanath\altaffilmark{8}, B.-Q.~For\altaffilmark{9}, T.~M.~O.~Franzen\altaffilmark{5}, L.~Hindson\altaffilmark{10}, M.~Johnston-Hollitt\altaffilmark{10}, A.~D.~Kapi\'nska\altaffilmark{9,3}, E.~Lenc\altaffilmark{1,3}, B.~McKinley\altaffilmark{6,3}, J.~Morgan\altaffilmark{5}, A.~R.~Offringa\altaffilmark{11}, P.~Procopio\altaffilmark{6,3}, L.~Staveley-Smith\altaffilmark{9,3}, R.~B.~Wayth\altaffilmark{5,3}, C.~Wu\altaffilmark{9}, Q.~Zheng\altaffilmark{10}}

\affil{$^1$Sydney Institute for Astronomy (SIfA), School of Physics, The University of Sydney, NSW 2006, Australia}
\affil{$^2$CSIRO Astronomy and Space Science (CASS), Marsfield, NSW 2122, Australia}
\affil{$^3$ARC Centre of Excellence for All-Sky Astrophysics (CAASTRO)}
\affil{$^4$Dunlap Institute for Astronomy \& Astrophysics, University of Toronto, Toronto, ON, M5S 3H4, Canada}
\affil{$^5$International Centre for Radio Astronomy Research (ICRAR), Curtin University, Bentley, WA 6102, Australia}
\affil{$^6$School of Physics, The University of Melbourne, Parkville, VIC 3010, Australia}
\affil{$^7$Istituto Nazionale di Astrofisica (INAF), Istituto di Radioastronomia, Via Piero Gobetti, Bologna, 40129, Italy}
\affil{$^{8}$Raman Research Institute (RRI), Bangalore 560080, India}
\affil{$^{9}$International Centre for Radio Astronomy Research (ICRAR), The University of Western Australia, Crawley, WA 6009, Australia}
\affil{$^{10}$School of Chemical \& Physical Sciences, Victoria University of Wellington, Wellington 6140, New Zealand}
\affil{$^{11}$Netherlands Institute for Radio Astronomy (ASTRON), Dwingeloo, The Netherlands}

\email{j.callingham@physics.usyd.edu.au}
	
\begin{abstract}

We present a sample of 1,483 sources that display spectral peaks between 72\,MHz and 1.4\,GHz, selected from the GaLactic and Extragalactic All-sky Murchison Widefield Array (GLEAM) survey. The GLEAM survey is the widest fractional bandwidth all-sky survey to date, ideal for identifying peaked-spectrum sources at low radio frequencies. Our peaked-spectrum sources are the low frequency analogues of gigahertz-peaked spectrum (GPS) and compact-steep spectrum (CSS) sources, which have been hypothesized to be the precursors to massive radio galaxies. Our sample more than doubles the number of known peaked-spectrum candidates, and 95\% of our sample have a newly characterized spectral peak. We highlight that some GPS sources peaking above 5\,GHz have had multiple epochs of nuclear activity, and demonstrate the possibility of identifying high redshift ($z > 2$) galaxies via steep optically thin spectral indices and low observed peak frequencies. The distribution of the optically thick spectral indices of our sample is consistent with past GPS/CSS samples but with a large dispersion, suggesting that the spectral peak is a product of an inhomogeneous environment that is individualistic. We find no dependence of observed peak frequency with redshift, consistent with the peaked-spectrum sample comprising both local CSS sources and high-redshift GPS sources. The 5\,GHz luminosity distribution lacks the brightest GPS and CSS sources of previous samples, implying that a convolution of source evolution and redshift influences the type of peaked-spectrum sources identified below 1\,GHz. Finally, we discuss sources with optically thick spectral indices that exceed the synchrotron self-absorption limit.

\end{abstract}
	
	
\section{Introduction}

Gigahertz-peaked spectrum (GPS), compact steep spectrum (CSS), and high frequency peaked (HFP) sources are a class of radio-loud active galactic nuclei (AGN) that have been argued to be the young precursors to massive radio-loud AGN, such as Centaurus\,A and Cygnus\,A \citep{1990A&A...231..333F,Odea1991,2000A&A...363..887D,2005A&A...432...31T,2010MNRAS.408.2261K}. GPS and HFP sources are defined as having a peak in their radio spectra and steep spectral slopes either side of the peak. They are also often found with small linear sizes and low radio polarization fractions \citep{Odea1991}. CSS sources are thought to be a related class that has similar properties to GPS and HFP sources but peak frequencies below the traditional gigahertz selection frequencies \citep{1990A&A...231..333F}. Hence, the main differentiation between GPS, CSS and HFP sources is the frequency of the spectral peak and the largest linear size. GPS and HFP sources have linear sizes $\lesssim$\,1\,kpc and peak frequencies of $\sim$\,1 $-$ 5\,GHz, and $\gtrsim$\,5\,GHz, respectively \citep{Odea1998,2000A&A...363..887D}. In comparison, CSS sources have linear sizes of $\sim$\,$1 - 20$\,kpc and are thought to have the lowest peak frequencies $\lesssim$\,500\,MHz, but until recently low radio frequency observations have been lacking to confirm such a situation \citep{1990A&A...231..333F}.

The argument that GPS, CSS, and HFP sources represent the first stages of radio-loud AGN evolution was inferred by very long baseline interferometry (VLBI) observations of these sources, revealing small scale morphologies reminiscent of large scale radio lobes of powerful radio galaxies, with two steep-spectra lobes surrounding a flat spectrum core \citep{1980ApJ...236...89P,1994ApJ...432L..87W,Stanghellini1997,2006A&A...450..959O,2012ApJS..198....5A}. Additional multi-epoch VLBI observations measured the motion of the hotspots, providing indirect evidence for ages $\lesssim 10^{5}$\,yrs \citep{Owsianik1998,Polatidis2003,Gugliucci2005}. The `youth' scenario for these sources is further supported by high frequency spectral break modeling \citep{Murgia1999,Orienti2010} and the discovery of the empirical relation between rest frame turnover frequencies and linear size \citep{Odea1997,Odea1998}. This suggests HFP sources evolve into GPS sources, which in turn evolve into CSS sources, and then finally grow to reach the size of FR\,I and FR\,II radio galaxies \citep{1974MNRAS.167P..31F,2010MNRAS.408.2261K}. 

However, the `frustration' hypothesis, which implies that these sources are not young but are confined to small spatial scales due to unusually high nuclear plasma density \citep{vanBreugel1984,Bicknell1997}, has seen a resurgence in explaining the properties of the GPS, CSS, and HFP population \citep[e.g.][]{Peck1999,Kameno2000,Marr2001,Tingay2003,Orienti2008,Marr2014,Tingay2015,Callingham2015}. The primary reasons that a debate remains about the nature of GPS, CSS, and HFP sources is because there appears to be an overabundance of these sources relative to the number of large radio AGN \citep{Odea1997,Readhead1996,2000MNRAS.319..445S,An2012}, and detailed spectral and morphological studies of individual sources have demonstrated that several of these sources are confined to a small spatial scale due to a dense ambient medium and a cessation of AGN activity \citep[e.g.][]{Peck1999,Orienti2010,Callingham2015}. It is also possible that both the `youth' and `frustration' scenarios may apply to the GPS, CSS, and HFP population, since sources with intermittent AGN activity may never break through a dense nuclear medium but young sources with constant AGN activity could evolve past the inner region of the host galaxy \citep{An2012}. 

One method that can deduce whether a GPS, CSS, or HFP source is frustrated or young is by identifying whether synchrotron self-absorption (SSA) or free-free absorption (FFA) is responsible for the turnover in the radio spectrum \citep[e.g.][]{Tingay2003,Marr2014,Callingham2015}. This is because the turnover in a source's spectrum will likely be dominated by FFA when confined to a small spatial scale by a dense medium \citep{Bicknell1997,Kuncic1998}. To successfully discriminate between SSA and FFA requires comprehensively sampling the spectrum of the source below the turnover, ideally with the observations below the spectral peak occurring simultaneously \citep{Odea1991,Odea1997}. Previous studies of samples of GPS, CSS, and HFP sources have often had only a single flux density measurement below the spectral peak, and composed of multi-epoch data with sparse frequency sampling, such that differentiation between FFA and SSA have been ambiguous for sources in large samples \citep[e.g.][]{Odea1998,2000MNRAS.319..445S,2000A&A...363..887D,2002MNRAS.337..981S,2004A&A...424...91E,Randall2011}. Other methods of differentiating between SSA and FFA, such as spectral variability \citep[][]{Tingay2015} or the change in circular polarization over the spectral peak \citep{1971Ap&SS..12..172M}, have also suffered from having incomplete, multi-epoch data below the turnover.

In addition to GPS, CSS, and HFP sources, there have been recent studies of a related class of radio-loud AGN referred to as megahertz-peaked spectrum (MPS) sources \citep{2004NewAR..48.1157F,2015MNRAS.450.1477C,2016MNRAS.459.2455C}. These sources have the same spectral shape as GPS, CSS, and HFP sources but have an observed turnover frequency below 1\,GHz. MPS sources are believed to be a combination of nearby CSS sources, and GPS and HFP sources at high redshift such that the turnover frequency has shifted below a gigahertz due to cosmological evolution \citep{2015MNRAS.450.1477C}. In particular, \citet{2015MNRAS.450.1477C} and \citet{2016MNRAS.459.2455C} have demonstrated that low-radio frequency selection criteria can identify non-beamed sources located at $z > 2$ and which appear young due to their small linear size. So far, investigations of MPS sources have been limited to small sections of the sky and moderate sample sizes.

We have entered a new era in radio astronomy where the limitations of small fractional bandwidth at low radio frequencies have been lifted, with the Murchison Widefield Array \citep[MWA;][]{Tingay2013}, the Giant Metrewave Radio Telescope \citep[GMRT;][]{1991ASPC...19..376S}, and the LOw-Frequency ARray \citep[LOFAR;][]{vanHaarlem2013} now operational. With the all-sky surveys at these facilitates nearing completion, such as the GaLactic and Extragalactic All-sky Murchison Widefield Array \citep[GLEAM;][]{2015PASA...32...25W} survey, the TIFR GMRT Sky Survey \citep[TGSS;][]{2016arXiv160304368I}, and the LOFAR Multifrequency Snapshot Sky Survey \citep[MSSS;][]{2015A&A...582A.123H}, astronomers now have unprecedented access to the radio sky below 300\,MHz. While the first surveys in radio astronomy were conducted at low frequencies \citep[e.g.][]{1958AuJPh..11..360M,1959MmRAS..68...37E}, MSSS and the GLEAM survey represent a significant step forward in the field because they have surveyed the sky with wide fractional bandwidths and much higher sensitivity. In particular, the GLEAM survey represents the widest continuous fractional bandwidth all-sky survey ever produced, with twenty contemporaneous flux density measurements between 72 and 231\,MHz for approximately 300,000 sources in the extragalactic catalog produced from the GLEAM survey \citep{2017MNRAS.464.1146H}. Hence, the GLEAM extragalactic catalog is a rich dataset to study sources that have a spectral peak at low frequencies.

The purpose of this paper is to use the GLEAM extragalactic catalog to construct the largest sample of peaked-spectrum sources to date, with contemporaneous observations at and below the spectral peak. Such a statistically significant sample has unparalleled frequency coverage below the turnover of GPS, CSS and HFP sources, providing a database for a comprehensive spectral comparison of the different absorption models, a test of whether sources with low frequency spectral peaks are preferentially found at high redshift, and analysis of how many peaked-spectrum sources constitute the wider radio-loud AGN at low radio frequencies. Note that in this paper we use the term `peaked-spectrum' to collectively refer to GPS, CSS, HFP, and MPS sources, and the terms `spectral peak' and `spectral turnover' interchangeably. 

The relevant surveys used in the source selection, the cross-matching routine, and spectral modeling procedure performed are outlined in \S\,\ref{sec:surveys}, \S\,\ref{sec:crossmatch}, and \S\,\ref{sec:spec_mod}, respectively. In \S\,\ref{sec:selection}, we discuss the selection criteria implemented to identify \ps sources. Comparisons of the identified \ps sources to known GPS, CSS, and HFP sources, and USS sources, are presented in \S\,\ref{sec:gpscsscomp} and \S\,\ref{sec:uss}, respectively. Relevant observed and intrinsic spectral features of the \ps samples are outlined in \S\,\ref{sec:spec_prop}. Finally, we introduce and debate the nature of sources with radio spectra near the limit of SSA in \S\,\ref{sec:extreme_spec}. In this paper we adopt the standard Lambda Cold Dark Matter ($\Lambda$CDM) cosmological model, with parameters $\Omega_{\mathrm{M}} = 0.27$, $\Omega_{\mathrm{\Lambda}} = 0.73$, and Hubble constant $H_{0} = 70$\,km\,s$^{-1}$\,Mpc$^{-1}$ \citep{2013ApJS..208...19H}.

\section{Descriptions of the surveys used in the selection of peaked-spectrum sources}
\label{sec:surveys}

The sensitivity and frequency coverage of the surveys used for selecting peaked-spectrum sources impact the type of sources identified. Most previous studies \citep[e.g.][]{1990A&A...231..333F,Odea1991,1998A&AS..131..303S,1998A&AS..131..435S,2000A&A...363..887D,2010MNRAS.408.2261K} used surveys that observed the sky at a single frequency around or above 1\,GHz. Since the inverted spectrum below the spectral turnover distinguishes a peaked-spectrum source, the low frequency survey used for selection dictates the type of sources identified, while higher frequency surveys are used to confirm the turnover and measure the spectral slope in the optically thin regime. Additionally, it is ideal if the higher frequency surveys have better sensitivities compared to the low frequency survey to ensure that the selected peaked-spectrum sample has a completeness set only by the low frequency data.

\begin{figure}
\begin{center}
\includegraphics[scale=0.4]{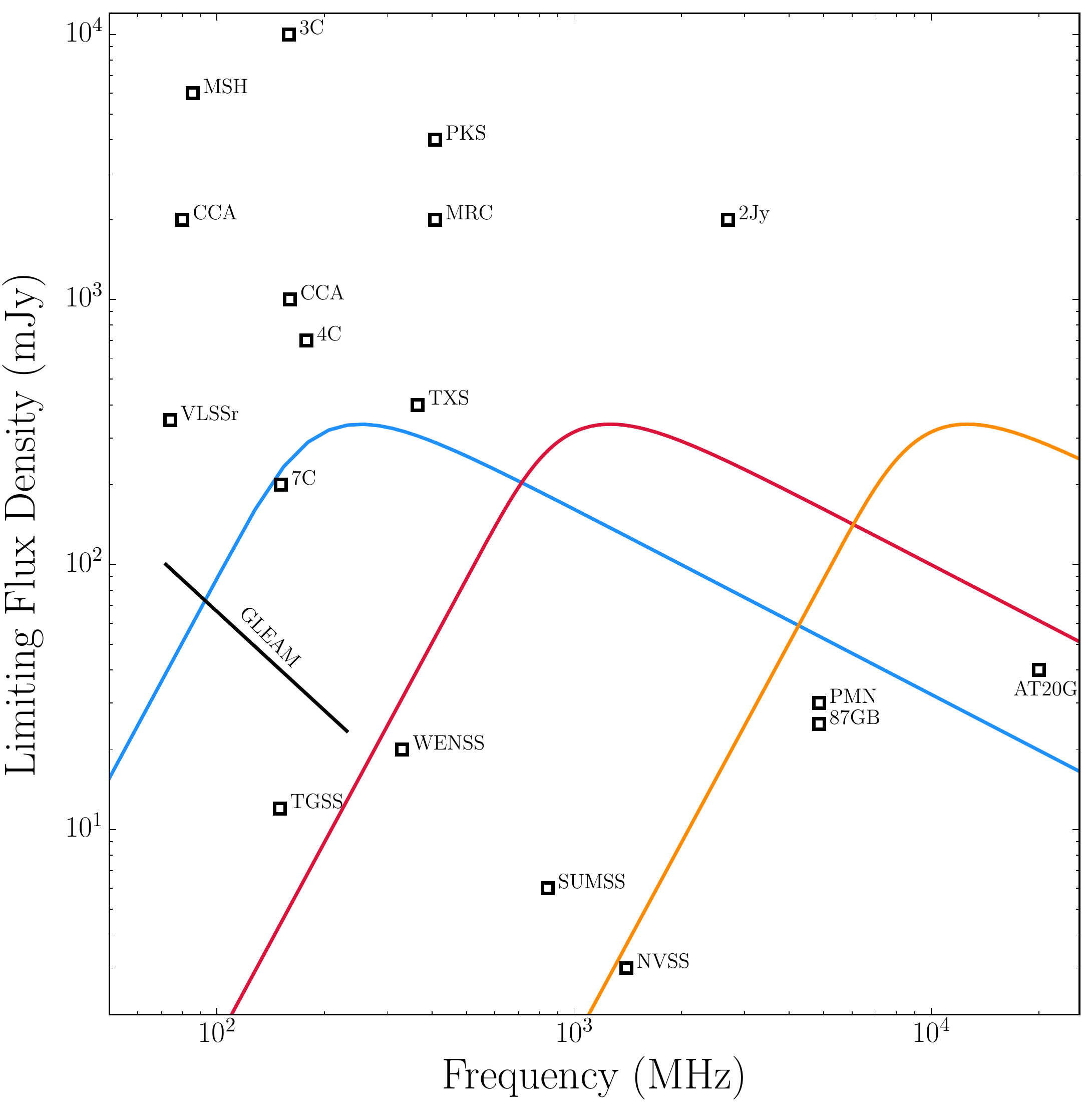}
\caption{The different frequencies and limiting sensitivities for the major radio surveys. The GLEAM survey is shown as a black line due to its variable limiting sensitivities between 72 and 231\,MHz. The blue, red, and orange curves represent the SSA spectra of peaked-spectrum sources that peak at 200, 1000, and 10000\,MHz, respectively. Therefore, this study is sensitive to such \ps sources portrayed by the blue spectrum but not to previously identified \ps sources portrayed by the red or orange spectra. The other plotted surveys not previously introduced are as follows: Mills, Slee, and Hill \citep[MSH;][]{1958AuJPh..11..360M,1960AuJPh..13..676M,1961AuJPh..14..497M} survey, Cambridge 3C \citep{1959MmRAS..68...37E}, 4C \citep{1965MmRAS..69..183P}, and 7C \citep{2007MNRAS.382.1639H} surveys, Culgoora circular array \citep[CCA;][]{1995AuJPh..48..143S} survey, VLA Low-frequency Sky Survey redux \citep[VLSSr;][]{2014MNRAS.440..327L} survey, Texas survey \citep[TXS;][]{1996AJ....111.1945D}, Parkes \citep[PKS;][]{1969AuJPA...7....3E} survey, Molonglo Reference Catalogue \citep[MRC;][]{Large1981,1991Obs...111...72L}, Westerbork Northern Sky Survey \citep[WENSS;][]{1997A&AS..124..259R}, 2\,Jy survey \citep{1985MNRAS.216..173W}, Parkes-MIT-NRAO \citep[PMN;][]{1994ApJS...90..173G,Wright1994} survey, MIT-Green Bank 5\,GHz \citep[87GB;][]{1991ApJS...75.1011G} survey, and Australia Telescope 20GHz \citep[AT20G;][]{Murphy2010} survey.}
\label{fig:survey_sens}
\end{center}
\end{figure}

The GLEAM extragalactic catalog represents a significant advance in selecting peaked-spectrum sources, since it is constituted of sources that were contemporaneously surveyed with the widest fractional radio bandwidth to date, with twenty flux density measurements between 72 and 231\,MHz. With such frequency coverage, selection below and at the spectral peak can be performed solely using one survey. To increase the validity of the detection of the peak in the spectra of these sources, to remove flat-spectrum sources, and to measure the slope above the turnover, we also use the NRAO VLA Sky Survey \citep[NVSS;][]{Condon1998} and the Sydney University Molonglo Sky Survey \citep[SUMSS;][]{Bock1999,Mauch2003}. Since the combination of NVSS and SUMSS cover the entire GLEAM survey, and are an order of magnitude more sensitive, this study is sensitive to \ps sources that peak anywhere between 72\,MHz and 843\,MHz\,/\,1.4\,GHz. Examples of the types of \ps sources that this study and previous studies identify are highlighted in Figure \ref{fig:survey_sens}. Details of the surveys used to select \ps sources in this study are provided below.

\subsection{GaLactic and Extragalactic All-sky Murchison Widefield Array (GLEAM) Survey}

The GLEAM survey was formed from observations conducted by the MWA, which surveyed the sky between 72 and 231\,MHz from August 2013 to July 2014 \citep{2015PASA...32...25W}. The MWA is a low radio frequency aperture array that is composed of 128 32-dipole antenna ``tiles'' spread over a $\approx$\,10\,km area in Western Australia \citep{Tingay2013}. The extragalactic catalog formed from the GLEAM survey consists of 307,455 sources south of declination $+30\degree$, excluding Galactic latitudes $|b| < 10\degree$, the Magellanic clouds, within 9$\degree$ of Centaurus A, and a 859 square degree section of the sky centered at RA 23\,h and declination $+15\degree$. The positions of the sources reported are accurate to within $\approx$\,30\arcsec and the catalog is $\approx$\,90\% complete at 0.16\,Jy \citep{2017MNRAS.464.1146H}. The sources in the catalog have twenty flux density measurements between 72 and 231\,MHz, mostly separated by 7.68\,MHz.

While the data reduction process that was performed to produce the GLEAM extragalactic catalog is discussed in detail in \citet{2017MNRAS.464.1146H}, we summarize the details here considering the importance of the GLEAM extragalactic catalog to this study. At seven independent declination settings, the GLEAM survey employed a two-minute ``snapshot'' observing mode. \textsc{Cotter} \citep{2015PASA...32....8O} was used to process the visibility data, and any radio frequency interference was excised by the \textsc{AOFlagger} algorithm \citep{2012A&A...539A..95O}. For the five instantaneous observing bandwidths of 30.72\,MHz, which are observed approximately two minutes apart, an initial sky model was produced by observing bright calibrator sources. \textsc{WSClean} \citep{2014MNRAS.444..606O} was used to perform the imaging of the observations, implementing a ``robust'' parameter of $-1.0$ \citep{Briggs1995}, which is close to uniform weighting. Each snapshot observation had multi-frequency synthesis applied across the instantaneous bandwidth, and then \textsc{clean}ed to the first negative \textsc{clean} component. A self-calibration loop was then applied to each of the images. The shallowly \textsc{clean}ed 30.72\,MHz bandwidth observations were divided into four 7.68-MHz sub-bands and jointly \textsc{clean}ed, resulting in a RMS of $\approx$\,100 to $\approx$\,20\,mJy\,beam$^{-1}$ for 72 to 231\,MHz, respectively. 

The 408\,MHz Molonglo Reference Catalogue \citep[MRC;][]{Large1981,1991Obs...111...72L}, scaled to the respective frequency, was used to set an initial flux density scale for the images and to apply an astrometric correction. The snapshots for an observed declination strip were mosaicked, with each snapshot weighted by the square of the primary beam response. Due to inaccuracies in the primary beam model, the remaining declination dependence in the flux density scale was corrected using the Very Large Array Low-Frequency Sky Survey Redux \citep[VLSSr;][]{2014MNRAS.440..327L}, MRC, and NVSS, and to place the survey on the \citet{Baars1977} flux density scale. It is estimated that the flux density calibration is internally accurate to within 2--3\% and accurate to 8-13\% when comparing the GLEAM flux densities to other surveys \citep{2017MNRAS.464.1146H}.

For each mosaic, a deep wideband image covering 170-231\,MHz was formed, with a resolution of $\approx$\,2\arcmin, to provide a higher signal-to-noise ratio and more accurate source positions than what can be attained from a single 7.68-MHz sub-band image. The sources in this wideband image were then convolved with the appropriate synthesized beam and used as priors for characterizing the flux density of the sources at each of the sub-band frequencies, using the source finding and characterization program \textsc{Aegean}\footnote{\url{https://github.com/PaulHancock/Aegean}} v1.9.6 \citep{2012MNRAS.422.1812H}.

\subsection{NRAO VLA Sky Survey (NVSS)}

NVSS is a 1.4\,GHz continuum survey that was conducted by the Very Large Array (VLA) between 1993 and 1996 \citep{Condon1998}. It covers the entire sky north of a declination of $-$40\degree\,and at a resolution of $\approx$\,45\arcsec. The catalog produced from the survey has a total of 1,810,672 sources, which is 100\% complete above 4\,mJy. The positions of the sources are accurate to within 1\arcsec.

\subsection{Sydney University Molonglo Sky Survey (SUMSS)}

SUMSS is a continuum survey designed to have similar frequency, resolution, and sensitivity to NVSS but to cover the sky below the declination limit of NVSS. SUMSS was conducted by the Molonglo Observatory Synthesis Telescope \citep[MOST;][]{Mills1981,Robertson1991} at 843\,MHz between 1997 and 2003, covering the sky south of a declination of $-$30\degree, excluding Galactic latitudes below $10\degree$ \citep{Bock1999,Mauch2003}. The resolution of the survey varied with declination $\delta$ as $45\arcsec\,\times\,45\arcsec\,\mathrm{cosec}|\delta|$. The catalog consists of 211,063 sources, with a limiting peak brightness of 6\,mJy beam$^{-1}$ for sources with declinations below $-50\degree$, and 10\,mJy\,beam$^{-1}$ for sources with declinations above $-50\degree$. Positions in the catalog are accurate to within 1\arcsec $-$ 2\arcsec~for sources with flux densities greater than 20\,mJy\,beam$^{-1}$, and are always better than 10\arcsec. The survey is believed to be 100\% complete above $\approx$\,8\,mJy south of a declination of $-50\degree$, and above $\approx$\,18\,mJy for sources between declinations of $-50\degree$ and $-30\degree$. 

\subsection{Additional radio surveys}

While GLEAM, NVSS, and SUMSS were the only surveys used for selecting \ps sources, once a \ps source was identified it was cross-matched to other all-sky radio surveys that covered any part of the GLEAM survey region. This included the 74\,MHz VLSSr, 408\,MHz MRC, and the Australia Telescope 20GHz \citep[AT20G;][]{Murphy2010} survey. These additional surveys where not used in any of the following spectral modeling, unless otherwise explicitly stated, but will be shown in any spectral energy distributions to help identify if a spectral fit to the GLEAM and NVSS/SUMSS data is accurate. 

Note that at the time of writing the 150\,MHz TGSS-Alternative data release 1 \citep[TGSS-ADR1;][]{2016arXiv160304368I} was released and undergoing review, including refining the uniformity of its flux density scale. The identified \ps sources were also cross-matched to TGSS-ADR1 but TGSS-ADR1 was not used for the selection of \ps sources despite a significant improvement in sensitivity and resolution compared to the GLEAM survey. This is largely because TGSS-ADR1 only surveyed the sky at a single frequency with a comparatively small bandwidth of 17\,MHz, thus increasing the potential for a sample to be biased by variable sources.  

\section{Cross-matching routine}
\label{sec:crossmatch}

The Positional Update and Matching Algorithm \citep[PUMA;][]{Line2016} was used to assess the probability of a cross-match between the GLEAM extragalactic catalog and NVSS/SUMSS. PUMA is an open source cross-matching software\footnote{\url{https://github.com/JLBLine/PUMA}}, specifically designed for matching low-frequency radio ($\lesssim 1\,$GHz) catalogs that have varying resolutions. It implements a Bayesian positional matching approach that uses catalog source density, sky coverage, and positional errors as a prior, to calculate the probability of a true match for any cross-match \citep{Budavari2008}. 

As the surveying telescopes used to create the all-sky catalogs have differing resolutions, multiple matches are common-place when cross-matching the different catalogs \citep[see e.g.][]{Carroll2016}. Confused matches can manifest in two different ways: multiple sources from a higher resolution catalog appear to match a single source from a lower resolution catalog, when really only one source truly matches; a lower resolution catalog is blending multiple components together and so multiple sources from a higher resolution catalog do truly match a single lower resolution source. 

In this work, the GLEAM extragalactic catalog was used as the base catalog, and was individually cross-matched to SUMSS and NVSS, with an angular cut-off of $2\arcmin20\arcsec$, which is approximately the full-width half-maximum of the MWA beam in the wide-band image. All possible matches were retained, and the results combined to create groups of possible cross-matches to each GLEAM source. Within each group, the positional probability of a true match was calculated for each cross-match (a combination of sources that only includes one source from each catalog). Using these positional results, we then selected the sources that PUMA assigned as $\tt{isolated}$, implying only one source from each catalog lay within $2\arcmin20\arcsec$. These cases were accepted if all matched sources lay within $1\arcmin10\arcsec$ of the GLEAM source position or if the positional probability of the cross-match was $>0.99$. We excluded cases where multiple sources were matched to a GLEAM source, since we are interested in \ps sources, which are defined to be unresolved at the resolution of the surveys used in this study. Additional details about the cross-matching of the GLEAM sample are presented in \S\,\ref{sec:selection}.

\section{Spectral modeling procedure}
\label{sec:spec_mod}

Selecting and assessing the spectral properties of \ps sources requires fitting their spectra. The parameter values of various models fit in this study were assessed using the Bayesian model inference routine outlined in \citet{Callingham2015}. In summary, a Markov chain Monte Carlo (MCMC) algorithm was used to sample the posterior probability density functions of the various model parameters. The parameter values were accepted when the applied Gaussian likelihood function was maximized under physically sensible uniform priors. The affine-invariant ensemble sampler of \citet{Goodman2010}, via the Python package $\tt{emcee}$ \citep{ForemanMackey2013}, was implemented. We utilized the simplex algorithm to direct the walkers to the maximum of the likelihood function \citep{Nelder1965}. 

When fitting within the GLEAM band, we assumed that the flux density measurements were independent and the uncertainties were Gaussian. However, the known correlation between the sub-band flux densities within the GLEAM band \citep[see \S\,5.4 of][]{2017MNRAS.464.1146H} had to be modeled when GLEAM data were fit simultaneously with other surveys, to ensure that any spurious trends present in the GLEAM flux density measurements did not influence any physical relations. It is not possible to calculate the exact form of the covariance matrix that would describe the correlation between the GLEAM points, but it can be approximated using Gaussian processes with a Mat\'{e}rn covariance function \citep{Rasmussen2006}. The Mat\'{e}rn covariance function produces a stronger correlation between flux density measurements closer in frequency space than further away, as is physically expected for the GLEAM correlation since it largely arises from a complex interaction of multi-frequency \textsc{clean}, self-calibration, and side-lobe confusion \citep{2016MNRAS.459.3314F,2017MNRAS.464.1146H}.

\subsection{Spectral models}

In this study, the spectra of sources are fit with four different spectral models to help select and characterize \ps sources. Firstly, we use the standard non-thermal power-law model of the form:

\begin{equation}\label{eqn:powlaw}
 S_{\nu} = a\nu^{\alpha},
\end{equation} 

\noindent where $a$, in Jy, characterizes the amplitude of the synchrotron spectrum, $\alpha$ is the synchrotron spectral index, and $S_{\nu}$ is the flux density at frequency $\nu$, in MHz. Since the GLEAM survey has a large fractional bandwidth, and radio sources are known to show curvature in their observed spectra when closely sampled \citep{1999AJ....117..677B}, we also fit the curved power-law model characterized as 

\begin{equation}\label{eqn:curve}
 S_{\nu} = a\nu^{\alpha}e^{q(\ln\nu)^{2}},
\end{equation} 

\noindent where $q$ parametrizes the spectral curvature and $\nu_{\mathrm{p}} = e^{-\alpha/2q}$ is the frequency at which the spectrum peaks. Significant curvature is represented by values of $|q| > 0.2$, and the spectral curvature flattens towards a standard power-law as $q$ approaches zero. While such a parameterization of curvature might not seem physically motivated, \citet{2012MNRAS.421..108D} have shown that $q$ can be directly related to the magnetic field strength, energy density, and electron density of lobe-dominated sources. 

Additionally, the following generic curved model was used to characterize the entire spectrum of a \ps source:

\begin{equation}\label{eqn:gen}
 S_{\nu} = \dfrac{S_{\mathrm{p}}}{(1-e^{-1})}\left(1 - e^{-\left(\nu / \nu_{\mathrm{p}}\right)^{\alpha_{\mathrm{thin}} - \alpha_{\mathrm{thick}}}}\right)\left(\dfrac{\nu}{\nu_{\mathrm{p}}}\right)^{\alpha_{\mathrm{thick}}},
\end{equation} 

\noindent where $\alpha_{\mathrm{thick}}$ and $\alpha_{\mathrm{thin}}$ are the spectral indices in the optically thick and optically thin regimes of the spectrum, respectively. $S_{\mathrm{p}}$ is the flux density at the frequency $\nu_{\mathrm{p}}$ \citep{1998A&AS..131..435S}. When $\alpha_{\mathrm{thick}} = 2.5$, this model reduces to a homogeneous SSA source. Equation \ref{eqn:gen} does not assume the underlying absorption mechanism is SSA or FFA, but does require that the slope of the spectrum above and below the spectral peak be modeled by a power-law \citep[similar to e.g.][]{Bicknell1997}. Note that this model is only used to describe the spectrum of a source, not to assess whether SSA or FFA is responsible for the turnover. 

In rare cases, the spectrum of a source is not well modeled by a turnover with power-law slopes on either side. To describe the spectra of these complex sources, it is assumed that the particle population producing the non-thermal power-law spectrum is surrounded by a homogeneous ionized screen of plasma such that

\begin{equation}\label{eqn:ffa}
	S_{\nu} = a \nu^{\alpha} e^{(\nu/\nu_{\mathrm{p}})^{-2.1}}.
\end{equation}

This homogeneous FFA model is used to model the spectra of such sources because it produces an exponential attenuation below the spectral peak, as opposed to the power-law relations described by Equation \ref{eqn:gen}. The FFA model, and the spectra of the sources that the model is used to describe, are discussed in more detail in \S\,\ref{sec:extreme_spec}.

\section{Peaked-spectrum source selection criteria}
\label{sec:selection}

A selection criterion that is effective in selecting a particular type of source, and which is well-defined such that any introduced bias is easily quantified, is required in order to produce a complete and reliable sample. For identifying \ps sources, the selection criterion involves characterizing whether the distinguishing feature of a spectral peak occurs in a source's spectrum. 

Previous studies have made the assessment of a spectral peak in radio color-color phase space \citep[e.g.][]{2006MNRAS.371..898S,2011MNRAS.412..318M}, where radio color-color phase space is defined by the spectral index derived between two high frequencies \ah~and the spectral index derived between two lower frequencies \al. If \al~had an opposite sign to \ah, it was assumed that a peak occurs in the frequency range somewhere between the frequencies in which \al~and \ah~were derived. We also utilize radio color-color space for identifying sources that peak somewhere between the end of the GLEAM frequency coverage and the start of the frequency coverage of SUMSS/NVSS. However, due to the large fractional bandwidth of the GLEAM survey, sources that have a turnover between 72 and 231\,MHz could be missed in radio color-color phase space since a power-law does not accurately describe their spectra. Therefore, we will also identify \ps sources through a direct measurement of their curvature in the GLEAM band. This is outlined below in \S\,\ref{subsec:sky_res_flux} and \S\,\ref{subsec:spec_class}.

\subsection{Sky area, resolution, and flux density limits}
\label{subsec:sky_res_flux}

Before making a distinction based on the spectral properties of a source, we must first make resolution, cross-matching and flux density cuts to the GLEAM extragalactic catalog to ensure that a reliable and complete \ps sample with well understood biases is derived. The selection criteria employed are summarized in Table \ref{table:sel_criteria} and detailed below:

\begin{table*}
	\small
	\caption{\label{table:sel_criteria} A summary of the applied selection criteria and the number of sources that remained after each stage of selection. Italicized numbers indicate the subset of sources selected from the previous non-italicized number. The details of the selection process are discussed in \S\,\ref{sec:selection}. With regard to the \ps samples, ``high frequency'' refers to sources with a spectral peak above a frequency of $\approx 180$\,MHz, while ``low frequency'' refers to sources with a spectral peak below a frequency of $\approx 180$\,MHz. $b$ and $\delta$ represent Galactic latitude and declination, respectively.}
	\begin{center}
		\begin{tabular}{ccc}
		\hline
		\hline
Selection step & Selection criterion & Number of sources \\
		\hline			
0 & Total GLEAM extragalactic catalog ($|b| \geqslant 10\degree$, $\delta \leqslant +30\degree$) & 307,456 \\
1 & Unresolved in the GLEAM wideband image $ab / (a_{\mathrm{psf}}b_{\mathrm{psf}}) \leqslant 1.1$  & 210,365 \\
2 & $\delta \geqslant -80\degree$ & 208,595 \\
3 & $S_{200\mathrm{MHz,wide}} \geqslant 0.16$\,Jy & 98,329 \\
4 & Sources with 8 or more GLEAM flux density \\ & measurements with a SNR $\geqslant3$ & 96,698 \\
5 & NVSS and/or SUMSS counterpart & 96,628 \\
6 & Peaked-spectrum selection & 1,483 \\
6a & High frequency soft sample \\ & \al~$\geqslant 0.1$ and \ah~$\leqslant -0.5$ & \emph{207} \\ 
6b & High frequency hard sample \\ & $\alpha_{\mathrm{low}} \geqslant 0.1$ and $-0.5 < \alpha_{\mathrm{high}} \leqslant 0$ & \emph{506} \\
6c & GPS sample \\ & \al~$\geqslant 0.1$ and \ah~$> 0$ & \emph{261} \\  
6d & Low frequency soft sample \\ & $\alpha_{\mathrm{low}} < 0.1$, \ah~$\leqslant -0.5$, 72\,MHz\,$\leqslant$\,$\nu_{\mathrm{p}}$\,$\leqslant$\,231\,MHz, $q \leqslant -0.2$, and $\Delta q \leqslant 0.2$ & \emph{394} \\  
6e & Low frequency hard sample \\ & $\alpha_{\mathrm{low}} < 0.1$, $-0.5 < \alpha_{\mathrm{high}} \leqslant 0$, 72\,MHz\,$\leqslant$\,$\nu_{\mathrm{p}}$\,$\leqslant$\,231\,MHz, $q \leqslant -0.2$, and $\Delta q \leqslant 0.2$ & \emph{115} \\  

		\hline\end{tabular}                           
\end{center}                                                                               
\end{table*}

\begin{enumerate}

\item Any source that is resolved in the GLEAM wideband image, centered on 200\,MHz, was eliminated since \ps sources are found to have small spatial scales. The wideband image was used to perform this cut because it achieves the highest resolution of the GLEAM survey of $\approx$\,2\arcmin, and all sources in the GLEAM survey are found within the wideband image. We determined whether a source was resolved in the wideband image by the criterion $ab / (a_{\mathrm{psf}}b_{\mathrm{psf}}) \leqslant 1.1$, where $a$, $b$, $a_{\mathrm{psf}}$, and $b_{\mathrm{psf}}$ are the semi-major and semi-minor axes of a source and the point spread function, respectively. While this resolution limit is signal-to-noise dependent, the flux density cut outlined in step 3 below ensures that this step has not removed potential low signal-to-noise \ps sources. This resolution cut reduces the total GLEAM extragalactic catalog by approximately one-third to 208,595 sources. 

\item Sources that are located within 10\degree\,of the south celestial pole were removed because of greater than 80\% uncertainties in the GLEAM flux density scale, and greater than 1\arcmin~uncertainties in the GLEAM positions. Such large uncertainties are mostly due to blurring of the source resulting from mosaicking many hours of data with different ionospheric conditions \citep{2017MNRAS.464.1146H}. This decreased the sample by an additional 1,770 sources, or 0.8\%. Note that sources between declinations of $-72\degree$ and $-80\degree$ have larger GLEAM systematic uncertainties then the rest of the survey to reflect the larger uncertainty in the flux density scale for this part of the sky.

\item A flux density cut was made to provide a reliable \ps sample and to evaluate its completeness. The GLEAM extragalactic catalog is estimated to be $\approx$\,90\% complete and 99.98\% reliable at 0.16\,Jy based on the wideband image. Therefore, we only investigated sources which had $S_{200\mathrm{MHz,wide}} \geqslant 0.16$\,Jy, where $S_{200\mathrm{MHz,wide}}$ is the flux density in the wideband image. Imposing this flux density limit also guarantees that the sample is only formed from sources with signal-to-noise ratios (SNRs) greater than 20, limiting the impact of classical confusion on the spectra of the sources \citep{1973ApJ...183....1M}. The sample was approximately halved to 98,329 sources after this flux density cut.

\item The previous flux density cut selects a reliable sample from the wideband image but does not account for local variations in the noise within the sub-band images. Large variations in the local RMS noise with frequency in the GLEAM survey are often due to calibration issues near bright sources or the edge of the survey. Additionally, sources with low SNR will suffer from non-Gaussianity in their uncertainties \citep{2017MNRAS.464.1146H}. Since accurate spectra across the entirety of the GLEAM band are needed to reliably select \ps sources, we required that a sub-band flux density have a SNR~$\geqslant3$ to be used in the spectral fitting. If a source was left with seven or less GLEAM flux density measurements, which represents less than a third of the total GLEAM bandwidth, it was excluded from the sample. This step removed 1,739 sources, $\approx$\,1.8\% of the sources from the previous step. 

\item The remaining sources were cross-matched to NVSS and SUMSS using the cross-matching routine outlined in \S\ref{sec:crossmatch}. Since NVSS and SUMSS are over two orders of magnitude more sensitive than the GLEAM survey, 99.93\% of the sample have a NVSS/SUMSS counterpart. The 70 sources that do not have a counterpart are retained for follow-up investigations. Note that GLEAM sources that are located between declinations of $-$30\degree\,and $-$40\degree\,have two counterparts due to a 10\degree\,declination overlap between NVSS and SUMSS.  
\end{enumerate}

\subsection{Spectral classification}
\label{subsec:spec_class}

At this step in the selection process, it is possible to define the position of the sources in a radio color-color phase space that is characterized from 72\,MHz to 843\,MHz\,/\,1.4\,GHz. Using the modelling procedure detailed in \S\,\ref{sec:spec_mod}, we fit Equation \ref{eqn:powlaw} to the twenty GLEAM flux density points to derive the low frequency spectral index \al. Similarly, the high frequency spectral index \ah~was derived by fitting a power-law to the SUMSS and/or NVSS flux density point(s) and to the two GLEAM sub-band flux densities centered on 189 and 212\,MHz. These two GLEAM sub-band frequencies were chosen because they are near the top of the overall GLEAM band and are from two different GLEAM observing bands. Since systematic fluctuations introduced from the data reduction procedure are largely observing band based (see e.g. Figure 18 of Hurley-Walker et al. 2016), the selection of two sub-band frequencies from different observing bands minimizes the impact of any systematic and statistical variations in calculating \ah. 

\begin{figure*}
\begin{center}
\includegraphics[scale=0.57]{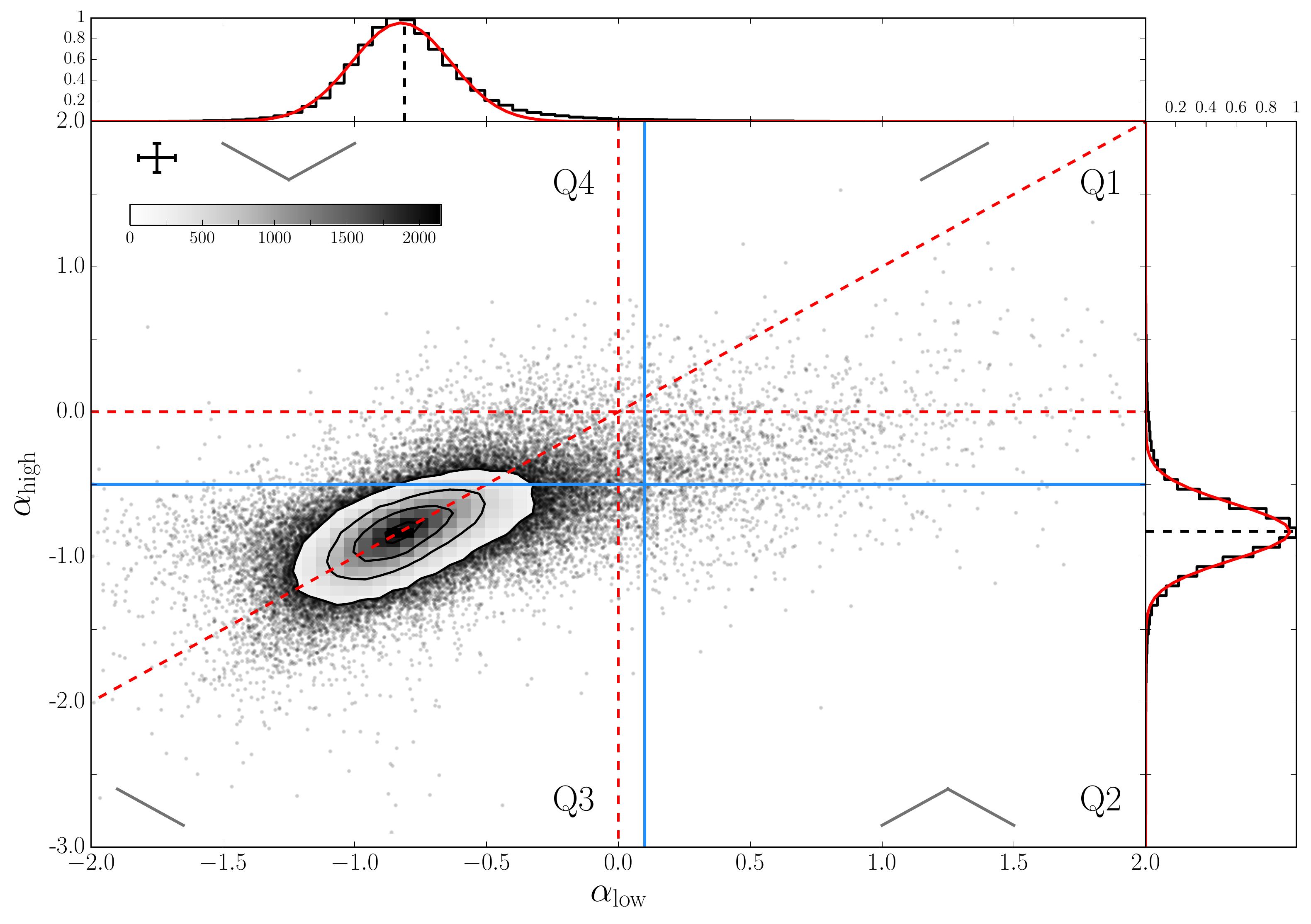}
\caption{Radio color-color diagram for the 96,628 GLEAM sources that remain after step five of the selection process that is detailed in Table \ref{table:sel_criteria}. \al~is derived from the twenty GLEAM data points between 72 and 231\,MHz. \ah~was calculated between the GLEAM flux density measurements centered on 189 and 212\,MHz and NVSS and/or SUMSS. Contours and a density map are plotted for the region surrounding (\al,\,\ah)\,$\approx-0.8$ due to the large number of points. The color in the density map conveys the number of sources each colored pixel, corresponding to the values of the color bar at the top left of the plot. The contour levels represent 100, 500, 1000, and 2000 sources, respectively. At \ah~= $-0.5$, the horizontal solid blue line delineates the limit below which sources are identified as \ps in the literature. The vertical solid blue line at \al~= $0.1$ marks the separation between the high and low frequency \ps samples. The dashed red lines represent spectral indices of zero and the one-to-one relation of \al~and \ah. Shown in gray at the corners of the plot are mock spectra of the sources for that quadrant. Individual error bars are not plotted to avoid confusion, but the median error bar size for the sample is illustrated at the top left of the figure. The first, second, third, fourth quadrants discussed in the text are labeled by Q1, Q2, Q3, and Q4, respectively. The two histograms to the top and to the right of the diagram represent the one-dimensional distributions of \al~and \ah, normalized by the maximum value of the distribution. The median value, and standard deviation, of \al~and \ah~are $-0.81 \pm 0.19$ and $-0.82 \pm 0.21$, respectively. The median values of \al~and \ah~are shown by dashed black lines. Over-plotted on the histograms, in red, are Gaussian fits to the distributions.}
\label{fig:color_color}
\end{center}
\end{figure*}

The radio color-color phase space from 72\,MHz to 843\,MHz\,/\,1.4\,GHz for 96,628 sources is presented in Figure \ref{fig:color_color}, and represents the most populated radio color-color plot produced to date. As expected from previous spectral index studies at these frequencies \citep{2014MNRAS.440..327L,2016MNRAS.463.2997M,2017MNRAS.464.1146H}, $\approx$\,70\% of sources cluster around (\al,\,\ah)\,$\approx-0.8 \pm 0.2$, which is located in the third quadrant of the diagram, as represented by the symbol $\mathrm{Q}3$ in Figure \ref{fig:color_color}, corresponding to spectra described by an optically thin synchrotron power-law. The sources in the first quadrant have a positive spectral index that extends from 72\,MHz to 843\,MHz\,/\,1.4\,GHz, and are likely dominated by GPS or HFP sources that are peaking near or above 1\,GHz. The fourth quadrant is occupied by sources that exhibit convex spectra, which are likely composite sources with a steep-spectrum power-law component at low frequencies and an inverted component at high frequency, which could indicate multiple epochs of AGN activity. 

Sources that display concave spectra near the top, or above, the GLEAM band are located in the second quadrant of Figure \ref{fig:color_color}, and thus it is possible to select \ps sources from this region. Note that in the literature, GPS, CSS, and HFP have been defined as having \ah~$\leqslant -0.5$ \citep{Odea1998}. Such a definition will also be applied to isolate \ps sources in this study for ease of comparison to literature samples, but the continuous distribution in \ah~ across \ah~$= -0.5$ in Figure \ref{fig:color_color} suggests such a definition is arbitrary. We have included Figure \ref{fig:color_color_guide} to help guide the reader in identifying the different areas of Figure \ref{fig:color_color} used to isolate \ps sources, as outlined in the next step of the selection process:

\begin{figure}
\begin{center}
\includegraphics[scale=0.275]{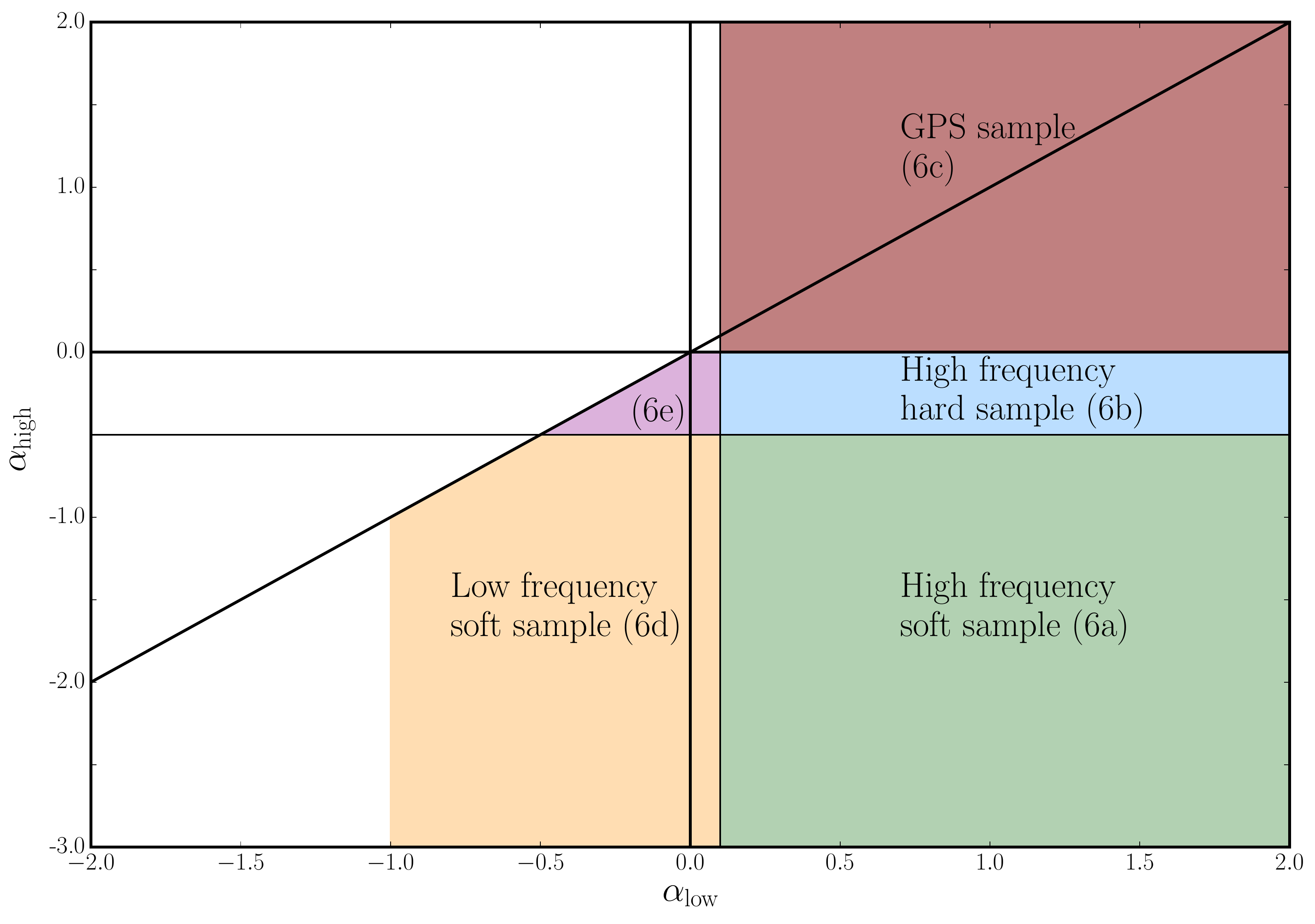}
 \caption{A schematic of the radio color-color plot of Figure \ref{fig:color_color}. The colored areas represent the regions in which \ps sources were selected for the different samples outlined in step 6 of the selection process. The green, blue, yellow, purple, and maroon sections are the areas in which the high frequency soft (step 6a), high frequency hard (step 6b), GPS (step 6c), low frequency soft (step 6d), and low frequency hard (step 6e) samples were selected, respectively. Note that high frequency here implies the peak of the source generally occurs above frequencies of $\approx$180\,MHz. The areas highlighted for the low frequency samples are only indicative and $\approx$\,5\% of sources outside these regions are located above the one-to-one line.}
\label{fig:color_color_guide}
\end{center}
\end{figure}

\begin{enumerate}[resume]

\item[6a.] Sources in the second quadrant of Figure \ref{fig:color_color} have a turnover in their spectra somewhere between $\approx$\,200\,MHz and 843\,MHz\,/\,1.4\,GHz. Since the original definition of GPS, CSS, HFP sources requires \ah~$\leqslant -0.5$ \citep{Odea1998}, which is shown by the solid blue line in Figure \ref{fig:color_color}, we use this limit to select \ps sources with \ah~$\leqslant -0.5$. We also require \al~$\geqslant 0.1$, instead of \al~$\geqslant 0$, because it is significantly more reliable in selecting \ps sources as the cut at \al~$\geqslant 0.1$ minimizes the contamination of flat spectrum sources, and because the median uncertainty in \al~is $\approx$\,0.1. Sources with \al~$< 0.1$ are discussed below. This sample is referred to as the high frequency soft sample and contains 207 sources.

\item[6b.] It is possible for \ps sources to exist in the second quadrant above the limit of \ah~$= -0.5$. Such sources have wider spectral widths, or higher frequency peaks, than those \ps sources identified in step 6a. Therefore, we also select \ps sources with \al~$\geqslant 0.1$ and $-0.5 < \alpha_{\mathrm{high}} \leqslant 0$, and refer to this collection of sources as the high frequency hard sample. Such a sample may be more contaminated by variable flat-spectrum sources than the soft sample due to the shallower dependence on \ah. There are a total of 506 sources in this high frequency hard sample.

\item[6c.] Sources located in quadrant 1 of Figure \ref{fig:color_color} could be GPS, HFP, and CSS sources that peak above 843\,MHz\,/1.4\,GHz. Therefore, sources with \al~$\geqslant 0.1$ and \ah~$> 0$ are also isolated. This sample is referred generally to as the GPS sample, and it contains 261 sources. As a spectral peak in these sources is not directly detected, they are isolated to largely provide spectral coverage below the turnover for known GPS sources.

\end{enumerate}

\begin{figure}
\begin{center}
\includegraphics[scale=0.3]{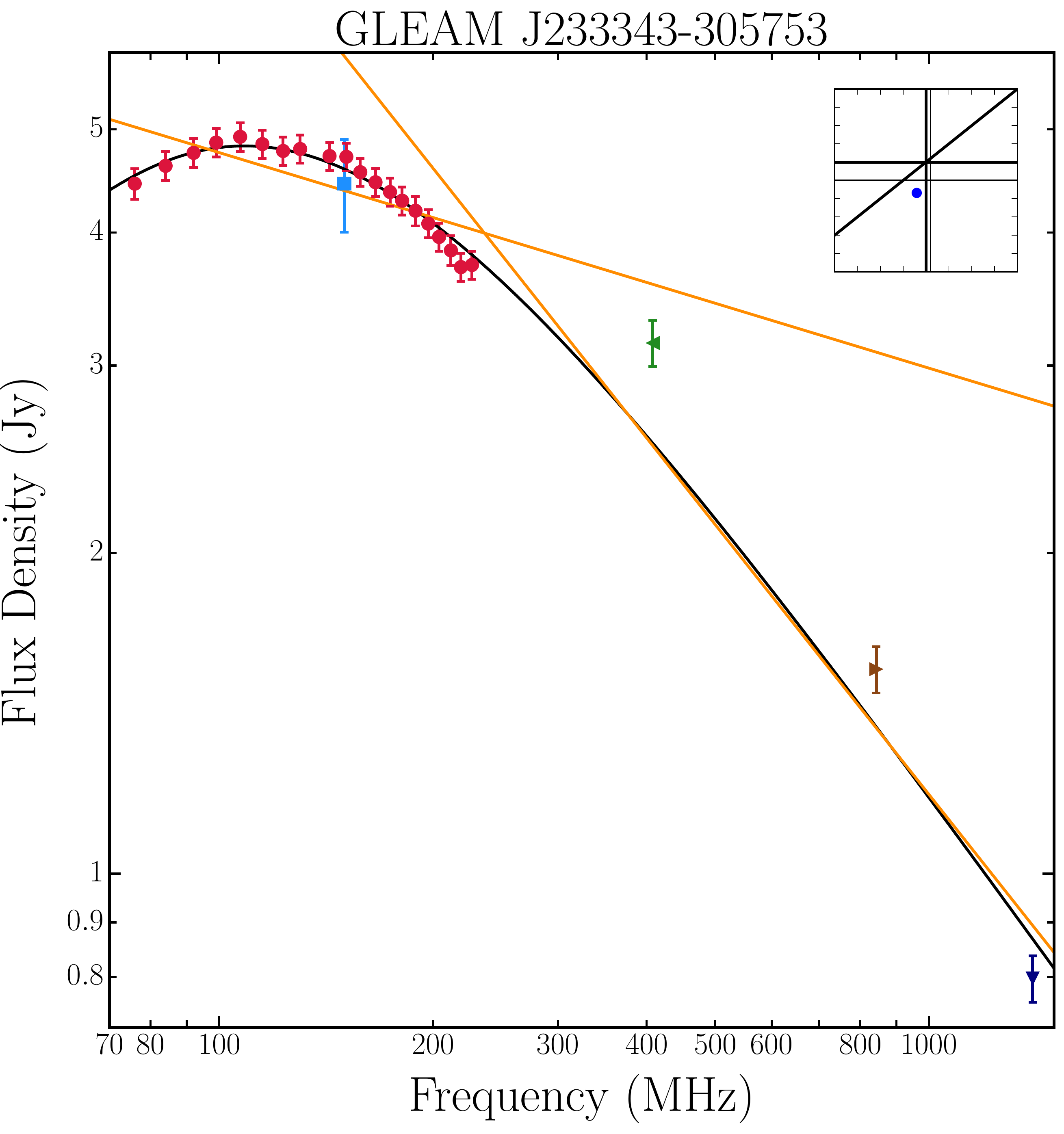}
 \caption{Spectral energy distribution of GLEAM~J233343-305753 from 72\,MHz to 1.4\,GHz. The red circles, blue square, green leftward-pointing triangle, brown rightward-pointing triangle, and navy downward-pointing triangle represent data points from GLEAM, TGSS-ADR1, MRC, SUMSS, and NVSS, respectively. The plot inset in the top-right corner is the color-color diagram of Figure \ref{fig:color_color}, with the position of GLEAM~J233343-305753 marked by a blue circle. Evidently, \al~was calculated to be negative for this source due to the curvature in the GLEAM data. The fit of the generic curved spectral model of Equation \ref{eqn:gen} to only the GLEAM, SUMSS, and NVSS flux density points is shown by the black curve. The orange lines represent the power-law fits from which \ah~and \al~have been derived.} 
\label{fig:curved_sed}
\end{center}
\end{figure}

Since the spectra of the sources are being fit across a large fractional bandwidth to calculate \al, any spectral index derived from a power-law model fit can be artificially flattened if a source displays spectral curvature within the GLEAM band. This means that sources that display a peak between 72 and 231\,MHz are shifted towards the third quadrant of Figure \ref{fig:color_color}, and can be calculated to have a negative \al~if the curvature is significant. An example of a negative \al~being derived for a source with significant curvature in the GLEAM band, such that the source is not located in the second quadrant of Figure \ref{fig:color_color}, is provided in Figure \ref{fig:curved_sed}. 

\begin{figure}
\begin{center}
\includegraphics[scale=0.4]{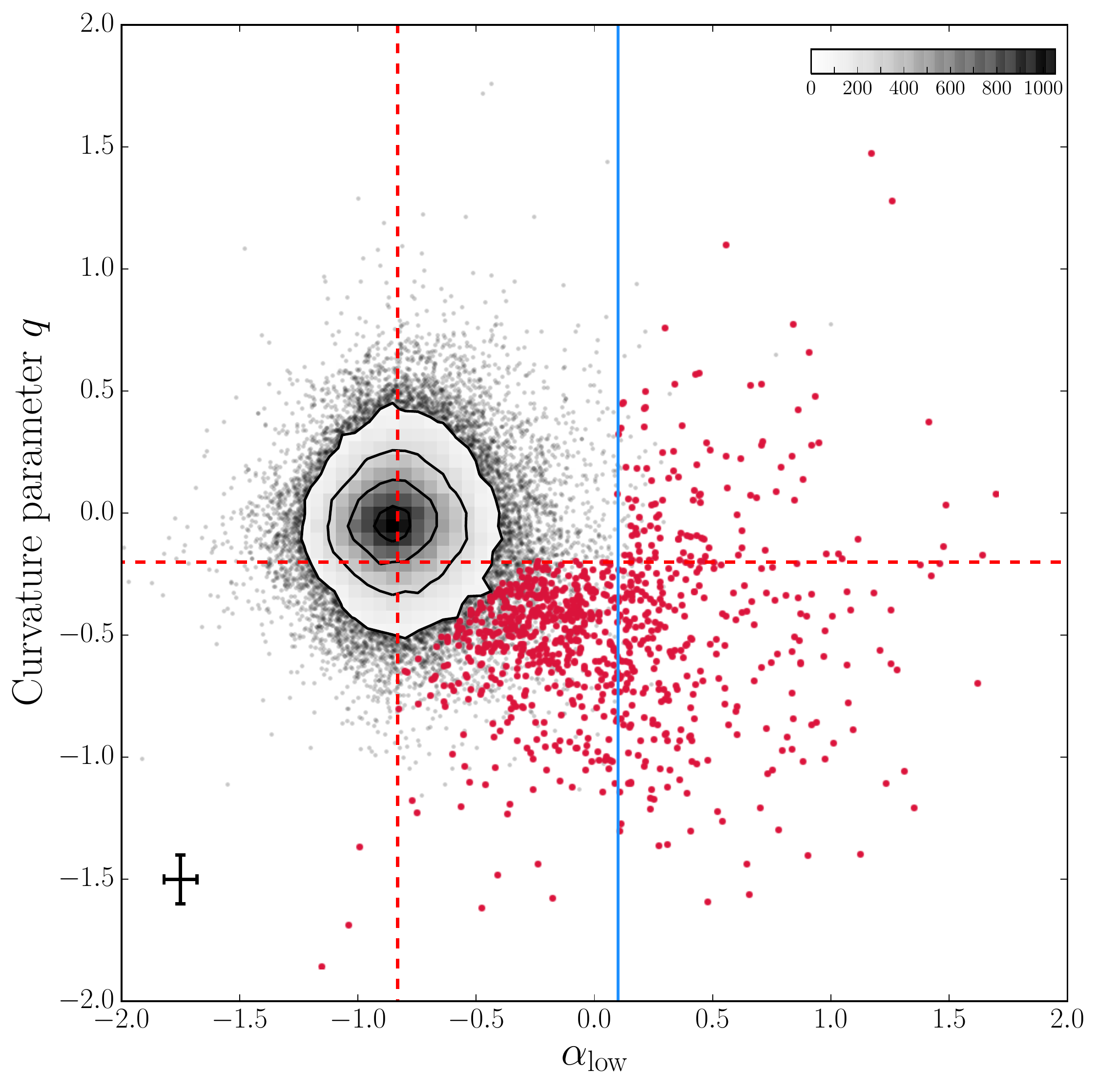}
 \caption{The distribution of the curvature parameter $q$ against \al~for the 96,628 sources remaining after step 5 of the selection process. Only sources with $\Delta q \leqslant 0.3$ are plotted. The vertical and horizontal red dashed lines represent the median of \al~and the cut-off used to indicate significant concave curvature $q = -0.2$, respectively. The color of the density map corresponds to the number of sources in each pixel, as set by the color bar in the top right-hand corner. The contour levels are at 50, 150, 450, and 850 sources. Peak-spectrum sources that are selected in step 6 are plotted in red, with sources above the blue line \al~$=0.1$ mostly identified based on their power-law spectral properties. Sources below this line were selected based on a spectral peak in the GLEAM band. The rough diagonal edge of the \ps source distribution around \al~$\approx -0.5$ and $q \approx-0.5$ is due to the requirement of more significant curvature in the GLEAM band to select lower signal-to-noise sources. The median uncertainties for $q$ and \al~are shown at the bottom-left corner of plot.}
\label{fig:curvature}
\end{center}
\end{figure}

The curvature model described by Equation \ref{eqn:curve} was also fit to the GLEAM data to test for any evidence of spectral curvature between 72 and 231\,MHz. Approximately 80\% of sources remaining at step 5 of the selection process show zero or negligible curvature in their spectra covered by the GLEAM band, with only 20,322 sources having a reliable value of the curvature parameter $|q| \geqslant 0.2$, which \citet{2012MNRAS.421..108D} class as significant curvature. The distribution of $q$ against \al~is presented in Figure \ref{fig:curvature} for sources with $\Delta q \leqslant 0.2$, where $\Delta q$ is the uncertainty in $q$. The requirement of $\Delta q \leqslant 0.2$ ensures the measurement of $q$ is reliable, and is equivalent to the signal-to-noise cut made in step 3.

Due to the impact that curvature has on calculating \al, selecting \ps sources in radio color-color phase space is unreliable for sources that display a peak in the GLEAM band. Hence, a source that had \al~$< 0.1$, which was the limit in \al~used in radio color-color phase space to select \ps sources, was also classified as a \ps source if a turnover was identified in the GLEAM band.

The information used to assess a spectral peak in a radio band is set by the signal-to-noise of the bandwidth available, with the maximum amount of information to detect a turnover transpiring when the peak occurs in the middle of the band. As a peak shifts to the edge of the GLEAM band, the information to reliably detect it declines and is dictated by the lever arm closest to the edge of the band. For example, a source that peaks at the central GLEAM frequency of 151\,MHz has ten spectral data points to assess whether the peak is real, while a source that peaks at 85\,MHz or 220\,MHz only has two data points to make the same assessment. In particular, the reliability in identifying a peak at the edge of the band is significantly impacted by statistical fluctuations.

Therefore, if the spectral peak of a source is measured at data point $N_{\mathrm{p}}$, which is above the central frequency data point $N_{\mathrm{c}}$, then a detection of a spectral turnover is considered reliable if

\begin{subequations}
\begin{equation}\label{eqn:curved_sources_lim1}
	\Delta q \leqslant 0.2 -\dfrac{\left(\sum\limits_{i=N_{\mathrm{p}}}^{N_{\mathrm{H}}}\sigma_{i}^{2}\right)^{1/2}} {\sum\limits_{i=N_{\mathrm{p}}}^{N_{\mathrm{H}}}S_{i}},
\end{equation}

\noindent where $\sigma_{i}$ and $S_{i}$ are the local rms noise and flux density in each subband $i$, respectively. $N_{\mathrm{H}}$ represents the highest frequency data point in the band. If $N_{\mathrm{p}} \leqslant N_{\mathrm{c}}$ then the inverse of the sum of the previous equation occurs

\begin{equation}\label{eqn:curved_sources_lim2}
	\Delta q \leqslant 0.2 -\dfrac{\left(\sum\limits_{i=N_{\mathrm{L}}}^{N_{\mathrm{p}}}\sigma_{i}^{2}\right)^{1/2}} {\sum\limits_{i=N_{\mathrm{L}}}^{N_{\mathrm{p}}}S_{i}},
\end{equation}
\end{subequations}

\noindent where $N_{\mathrm{L}}$ is the lowest frequency data point in the band. The second terms of Equations \ref{eqn:curved_sources_lim1} and \ref{eqn:curved_sources_lim2} represent the combination of the signal-to-noise ratios in each of the sub-bands above or below the spectral peak. The magnitude of these terms decrease as the number of points available to assess the reliability of a turnover increases. We chose to subtract these terms from 0.2 as the number of false detections of spectral peaks at 151\,MHz is below 1\% when $\Delta q = 0.2$. This is equivalent to a signal-to-noise cut of 30 in the wideband image. Note that the functional form of Equations \ref{eqn:curved_sources_lim1} and \ref{eqn:curved_sources_lim2} assumes that the noise between the sub-bands is independent. While this is not the case for the GLEAM data at low frequencies due to confusion, since we only test high signal-to-noise sources for a spectral peak, the impact of confusion is minimized on the spectrum and the noise can be approximated as Gaussian and independent \citep{2016MNRAS.459.3314F,2017MNRAS.464.1146H}. 

\begin{figure}
\begin{center}
\includegraphics[scale=0.4]{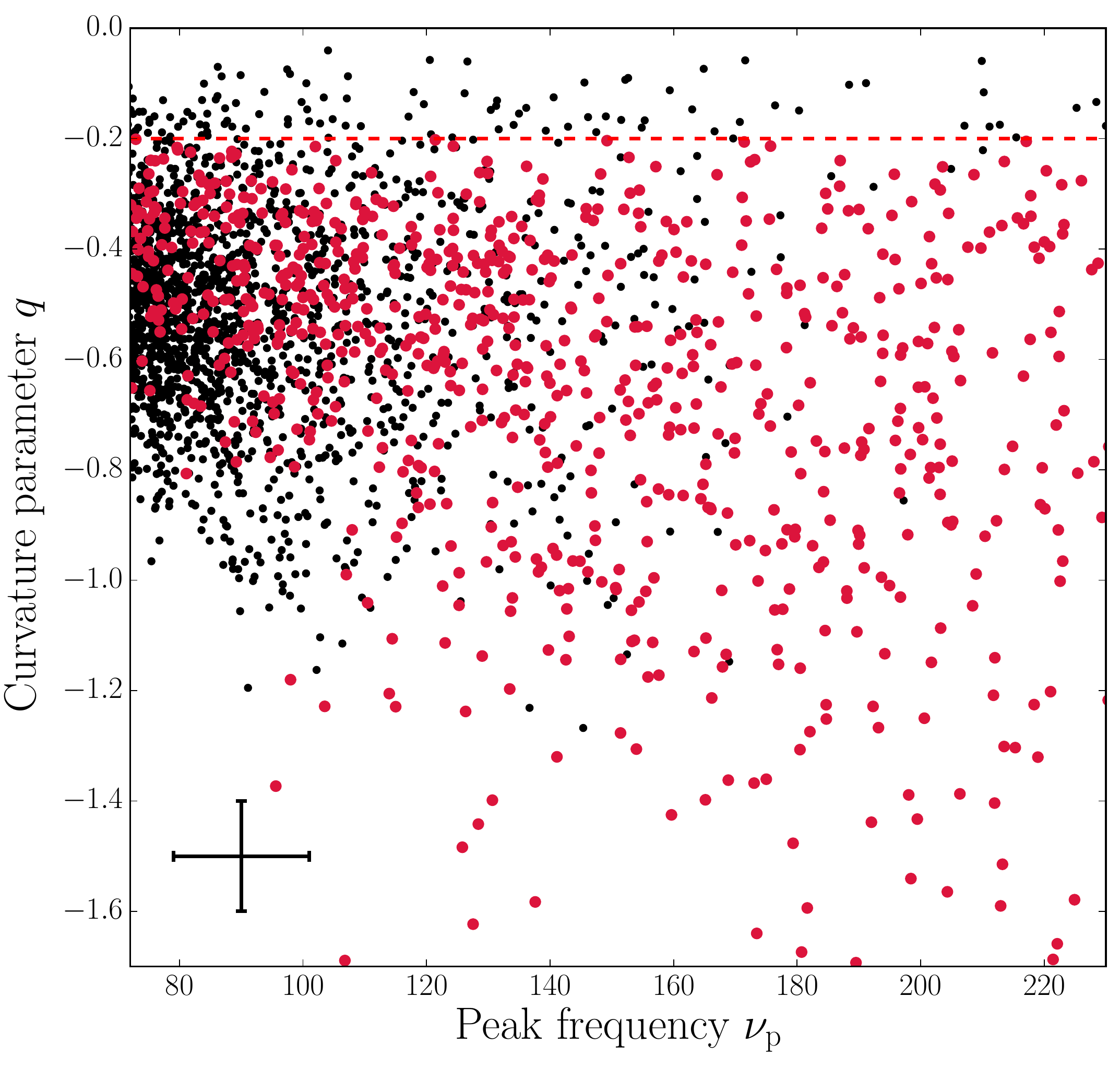}
 \caption{The distribution of the curvature parameter $q$ against the frequency of the peak in the GLEAM band $\nu_{\mathrm{p}}$ for the 96,628 sources remaining after step 5 of the selection process. All sources with $\Delta q \leqslant 0.3$ are plotted in black to provide an indication the impact the noise has on selecting \ps sources. Peaked-spectrum sources selected in step 6 are over-plotted in red, with all the sources with $\nu_{\mathrm{p}}$ below $\approx$180\,MHz selected on the basis of a peak in the GLEAM band. The concentration of black points toward low $\nu_{\mathrm{p}}$ is due to noise within the GLEAM band. Sources above $\approx$180\,MHz are largely identified as \ps from their position in radio color-color phase space. The horizontal red dashed line corresponds to limit of $q$ below which curvature in the GLEAM band was considered significant. The median uncertainties for $q$ and $\nu_{\mathrm{p}}$ are plotted at the bottom-left corner of diagram.}
\label{fig:mwa_turnover}
\end{center}
\end{figure}

The distribution of the curvature parameter, $q$, against the frequency of the peak in the GLEAM band, $\nu_{\mathrm{p}}$, for the sources remaining after step 5 of the selection process, is presented in Figure \ref{fig:mwa_turnover}. The accumulation of black data points towards low $q$ and $\nu_{\mathrm{p}}$ is a function of noise, particularly in the highest sub-band in GLEAM. A concave spectrum is considered significant if $q \leqslant -0.2$ \citep{2012MNRAS.421..108D}.

Therefore, we select \ps sources based on the detection of a peak in the GLEAM band, and also separate into soft and hard samples depending on their value of \ah: 
\begin{enumerate}[resume]

\item[6d.] A source is added to the low frequency soft sample if it is found with $\alpha_{\mathrm{low}} < 0.1$, \ah~$\leqslant -0.5$, 72\,MHz\,$\leqslant$\,$\nu_{\mathrm{p}}$\,$\leqslant$\,231\,MHz, $q \leqslant -0.2$, and $\Delta q$ is less than that set by Equations \ref{eqn:curved_sources_lim1} and \ref{eqn:curved_sources_lim2}. Such a sample has a size of 394. 

\item[6e.] For sources to be added to the low frequency hard sample, it is required that $\alpha_{\mathrm{low}} < 0.1$, $-0.5 < \alpha_{\mathrm{high}} \leqslant 0$, 72\,MHz\,$\leqslant$\,$\nu_{\mathrm{p}}$\,$\leqslant$\,231\,MHz, $q \leqslant -0.2$, and $\Delta q$ is less than that set by Equations \ref{eqn:curved_sources_lim1} and \ref{eqn:curved_sources_lim2}. Similar to the high frequency hard sample selected in step 6b, this sample is more likely to contain variable flat-spectrum sources. There are a total of 115 sources in this sample.

\end{enumerate}

The hard and soft samples selected in this way are together referred to as the low frequency \ps sample to differentiate from the samples selected solely on radio color-color phase space, which are referred to as the high frequency \ps sample. The locations of more than 95\% of the low frequency \ps samples in radio color-color phase space are also displayed in Figure \ref{fig:color_color_guide}. 

The selection process also identified two known pulsars as \ps sources, PSR~J0630-2834 and PSR~J1645-0317 \citep{2005AJ....129.1993M}. These two sources were removed from the \ps samples, and we expect the contamination of pulsars in the total \ps sample to be less than 1\% based on the total number of pulsars detected by the GLEAM survey \citep{2016MNRAS.461..908B}.

Hence, a total of 1,483 extragalactic \ps candidates were selected from 96,628 sources. Note that all of the \ps candidates were classed as $\tt{isolated}$ by the cross-matching routine, as detailed in \S\,\ref{sec:crossmatch}. The spectra of a source from each of the samples discussed in step 6 of the selection process are presented in Figure \ref{fig:example_spectra}. In the appendix, the spectra for all \ps sources selected are plotted. The tables providing the characteristics for the high and low frequency \ps samples, and the GPS sample, in the style of the table presented in Appendix \ref{sec:appendix_tab}, are available online. 

\begin{figure*}
\begin{center}$
\begin{array}{cccc}
\includegraphics[scale=0.245]{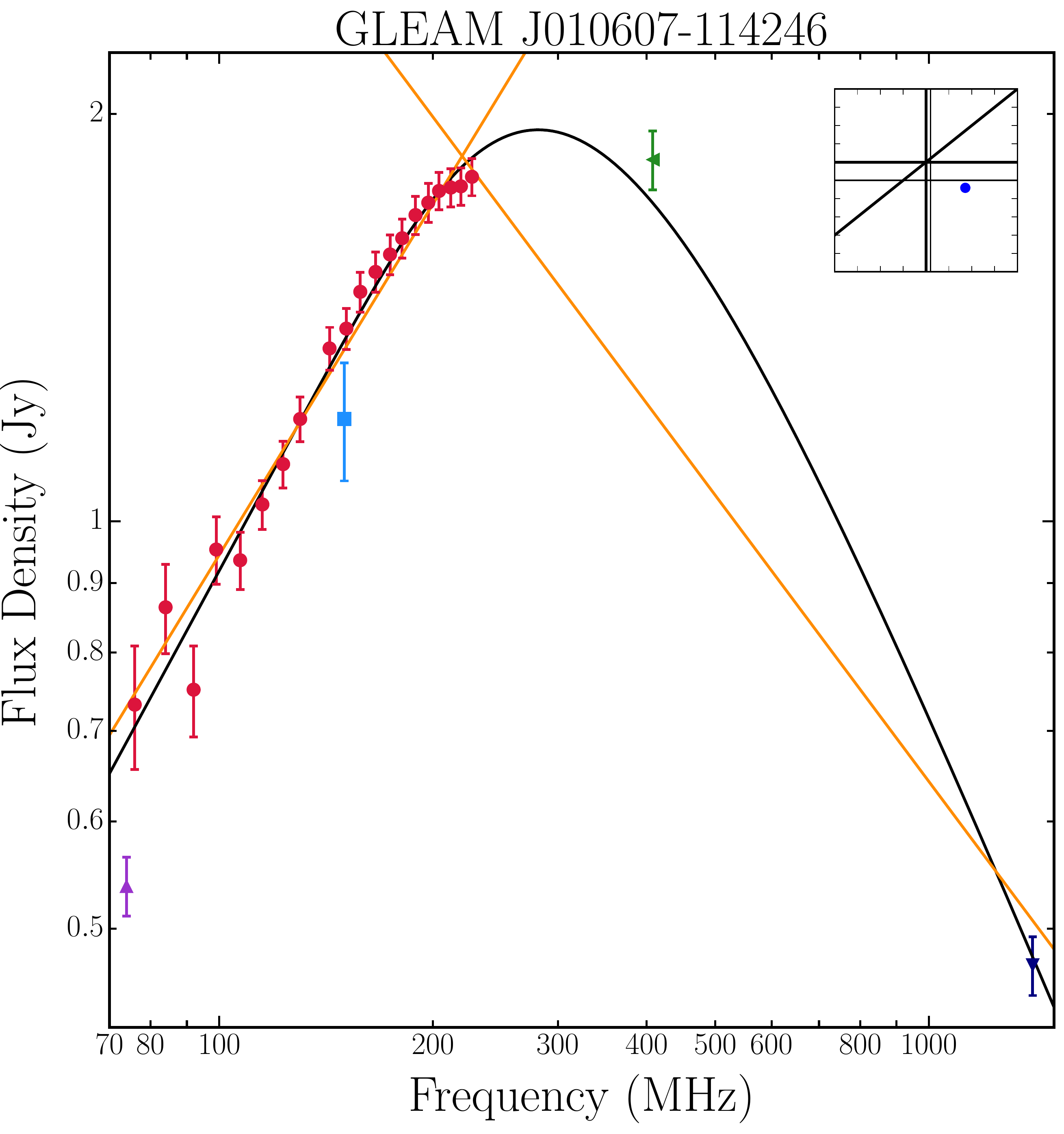} & 
\includegraphics[scale=0.245]{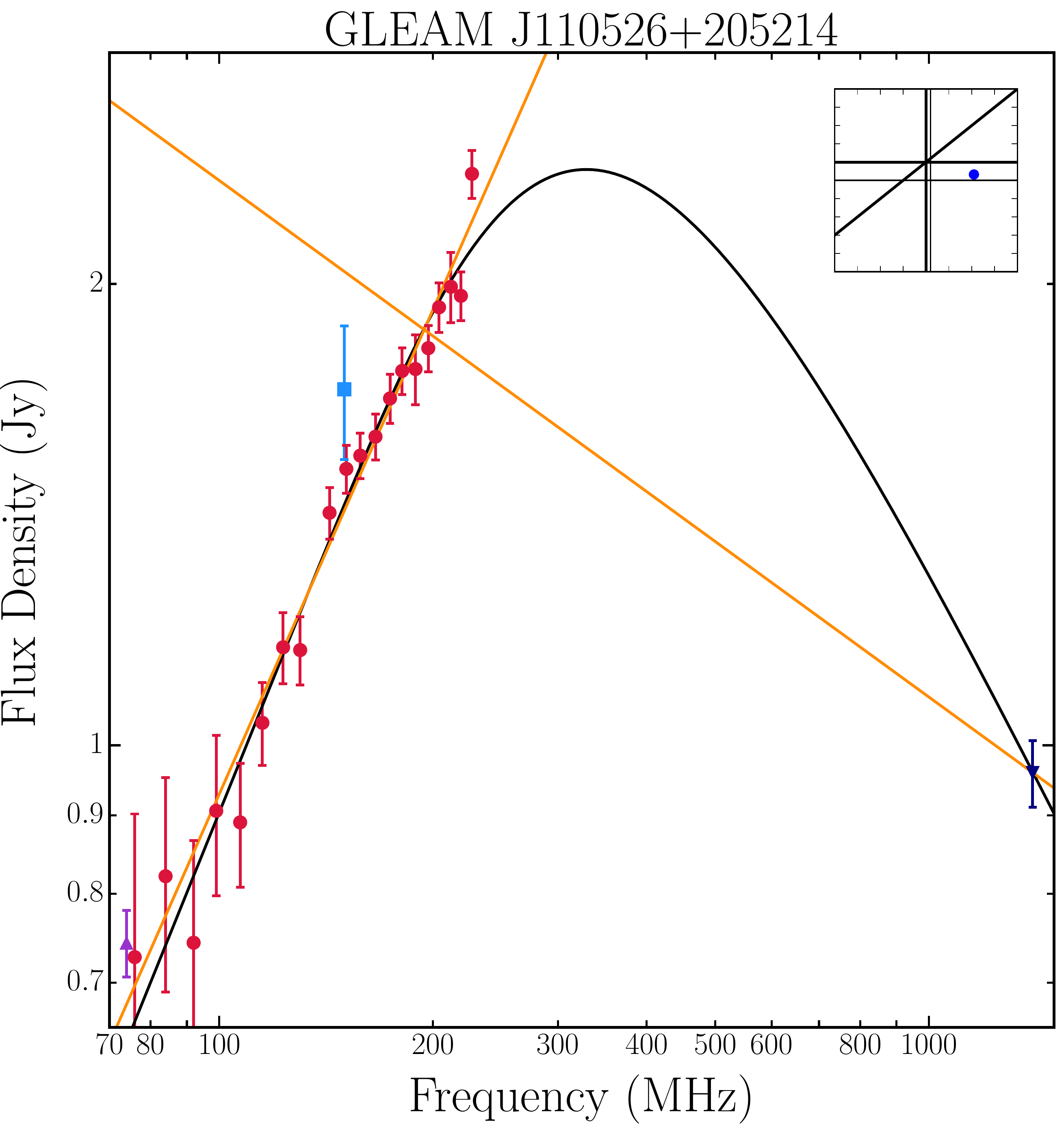}\\ 
\includegraphics[scale=0.245]{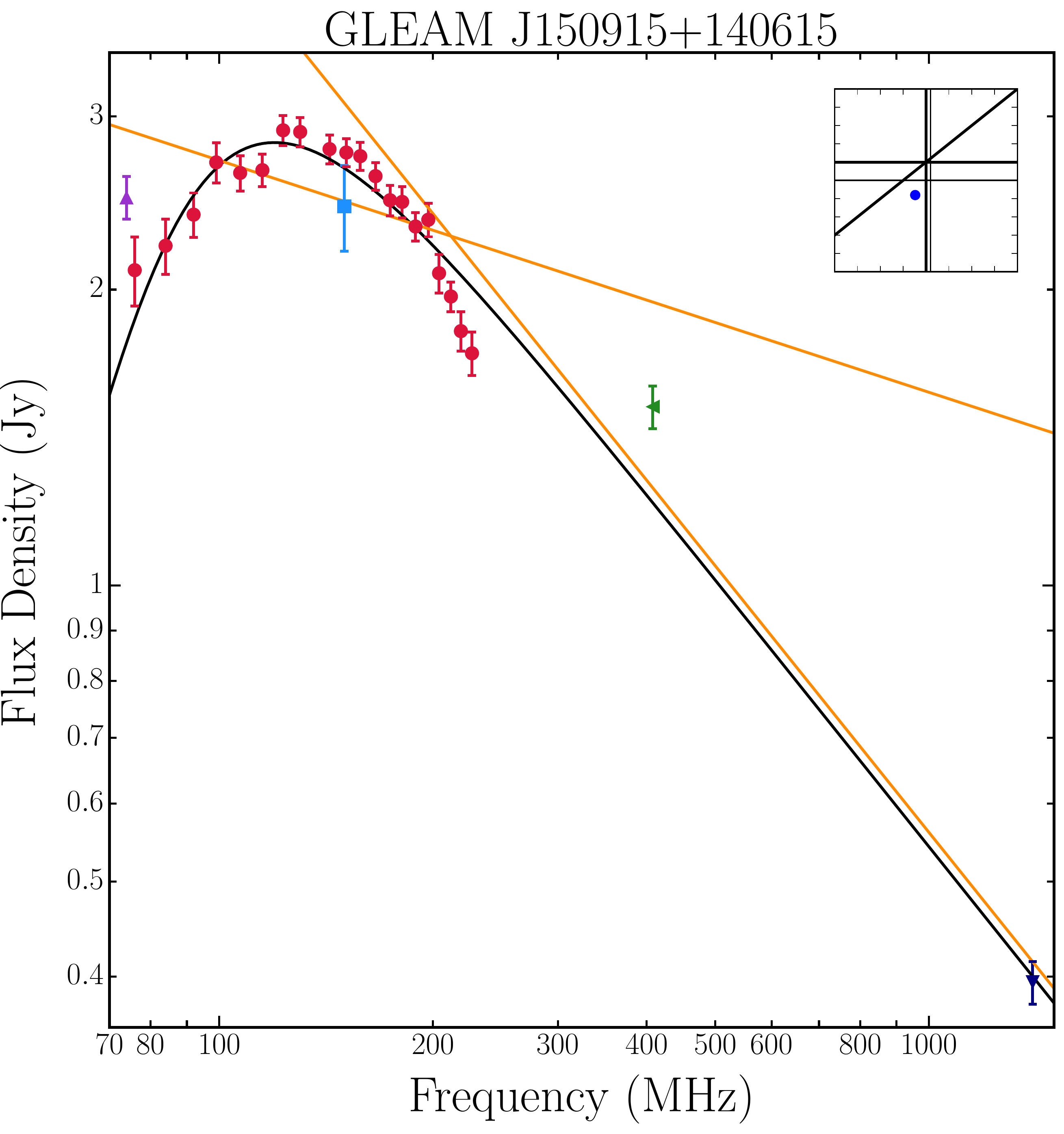} & 
\includegraphics[scale=0.245]{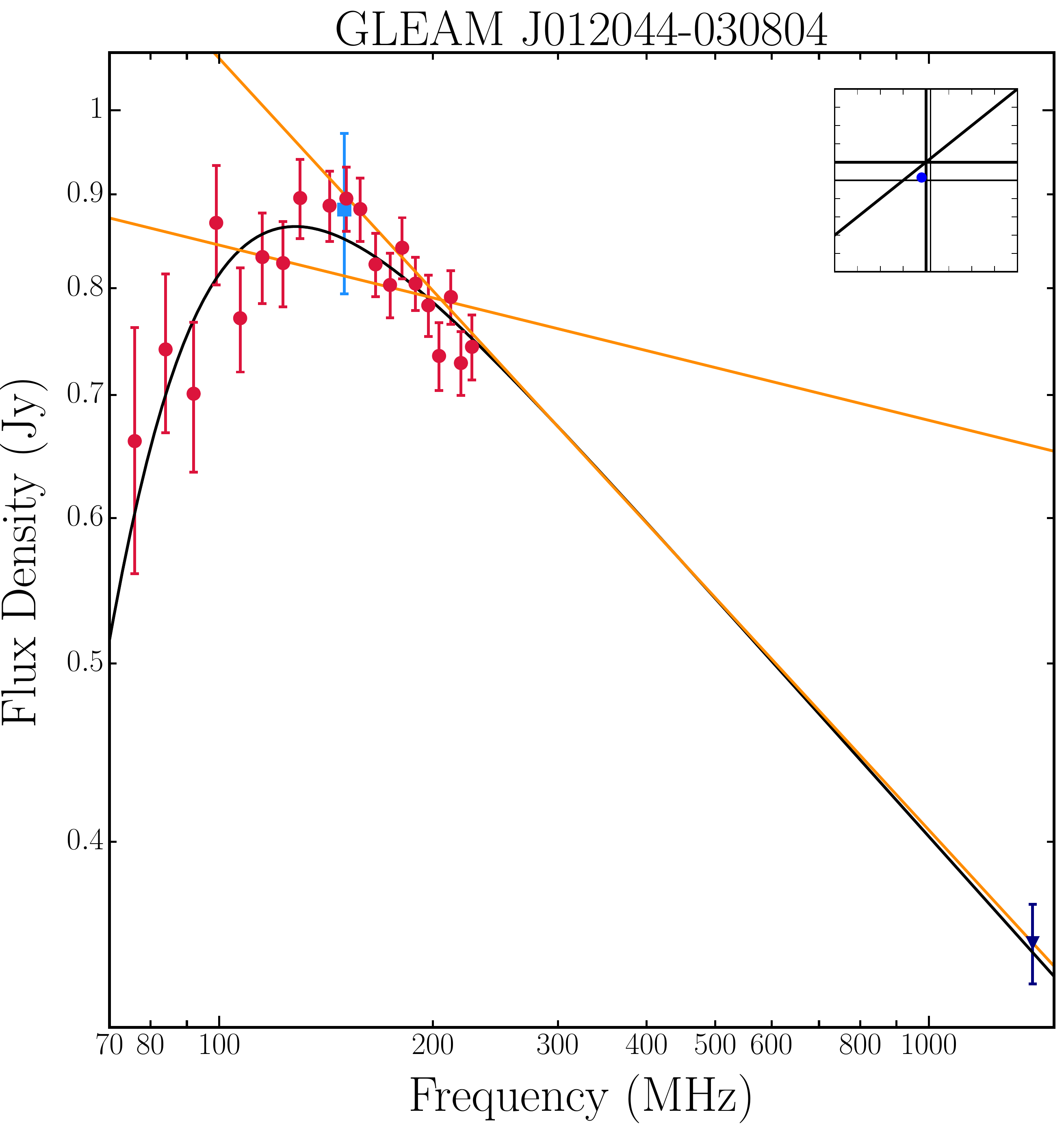}\\ 
\includegraphics[scale=0.245]{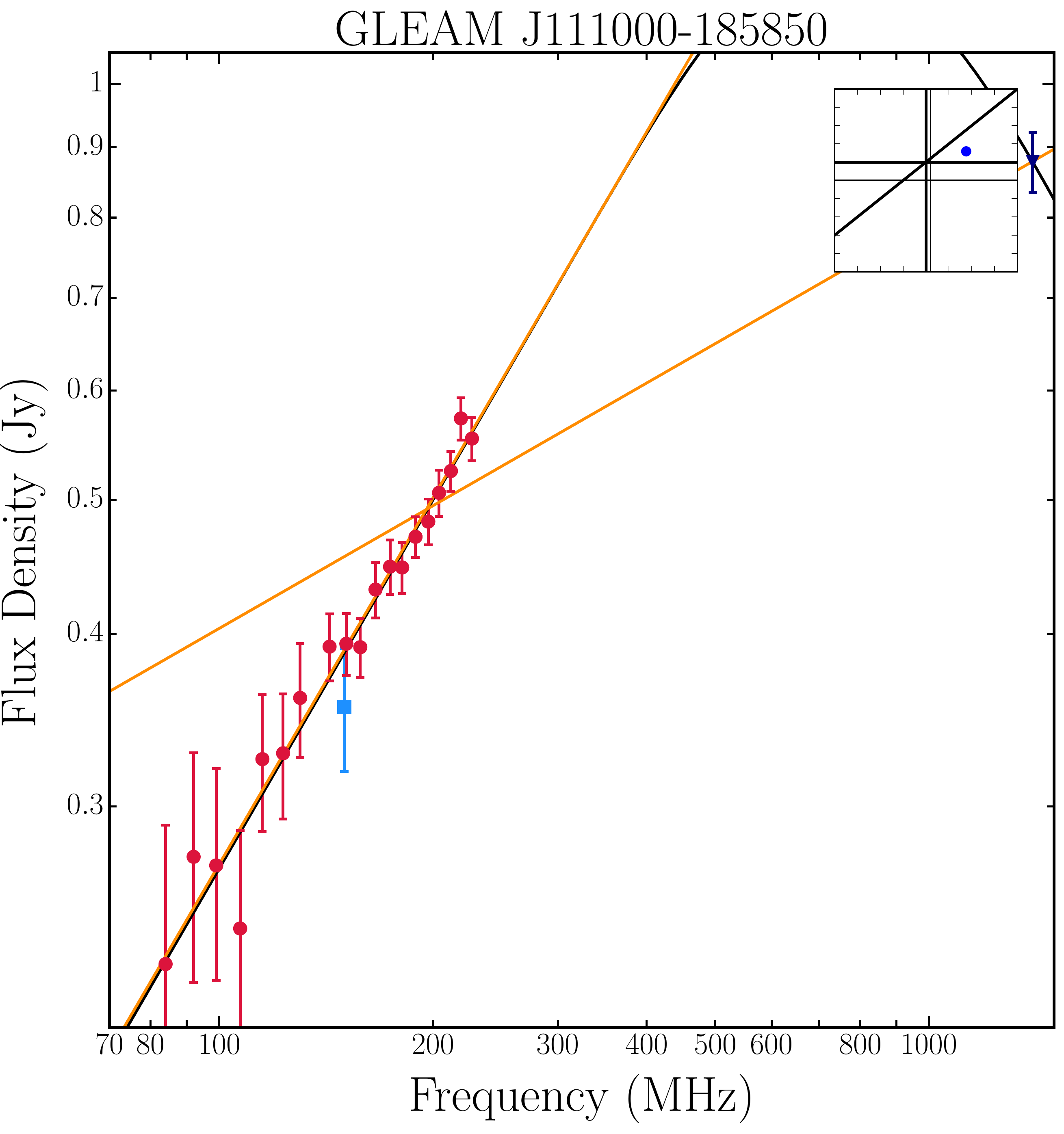} & 
\includegraphics[scale=0.245]{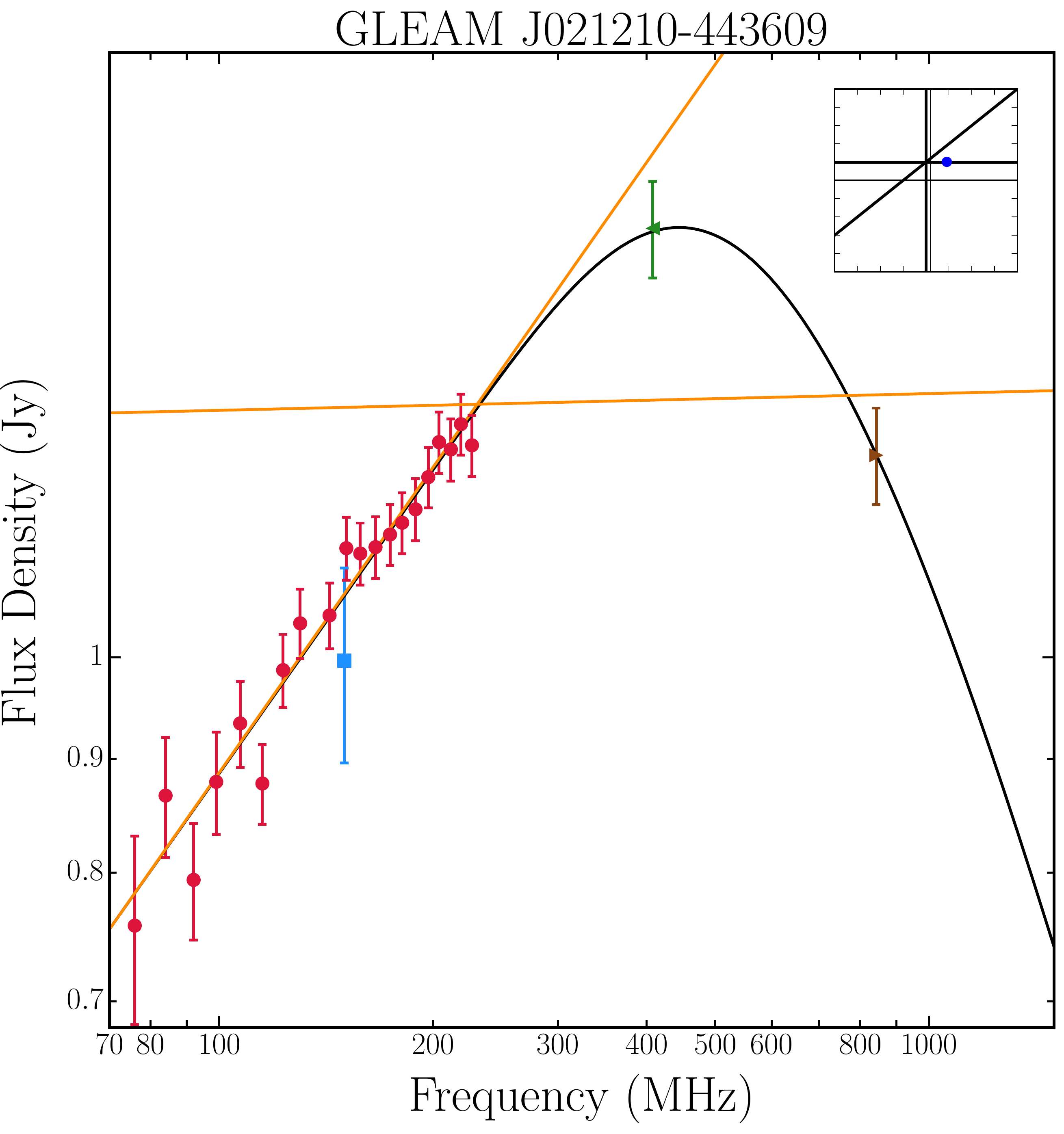} \\ 
\end{array}$
 \caption{Example spectra for sources drawn from the five different \ps samples identified via the selection process. The spectra at top-left, top-right, middle-left, middle-right, and lower-left are from the high frequency hard and soft, the low frequency hard and soft, and the GPS sample, respectively. The spectrum at bottom-right is to highlight a source that has \ah~$=0$. The purple upward-pointing triangle, red circles, blue square, green leftward-pointing triangle, brown rightward-pointing triangle, and navy downward-pointing triangle represent data points from VLSSr, GLEAM, TGSS-ADR1, MRC, SUMSS, and NVSS, respectively. The fit of the generic curved spectral model of Equation \ref{eqn:gen} to only the GLEAM and SUMSS and/or NVSS flux density points is shown by the black curve. The power-law fits from which \ah~and \al~have been derived are shown in orange. The plot inset in the top-right corner displays the position of the source in the color-color diagram of Figure \ref{fig:color_color} using a blue circle.}
\label{fig:example_spectra}
\end{center}
\end{figure*}

\subsection{Sources peaking below 72 MHz}

\begin{figure}
\begin{center}
\includegraphics[scale=0.3]{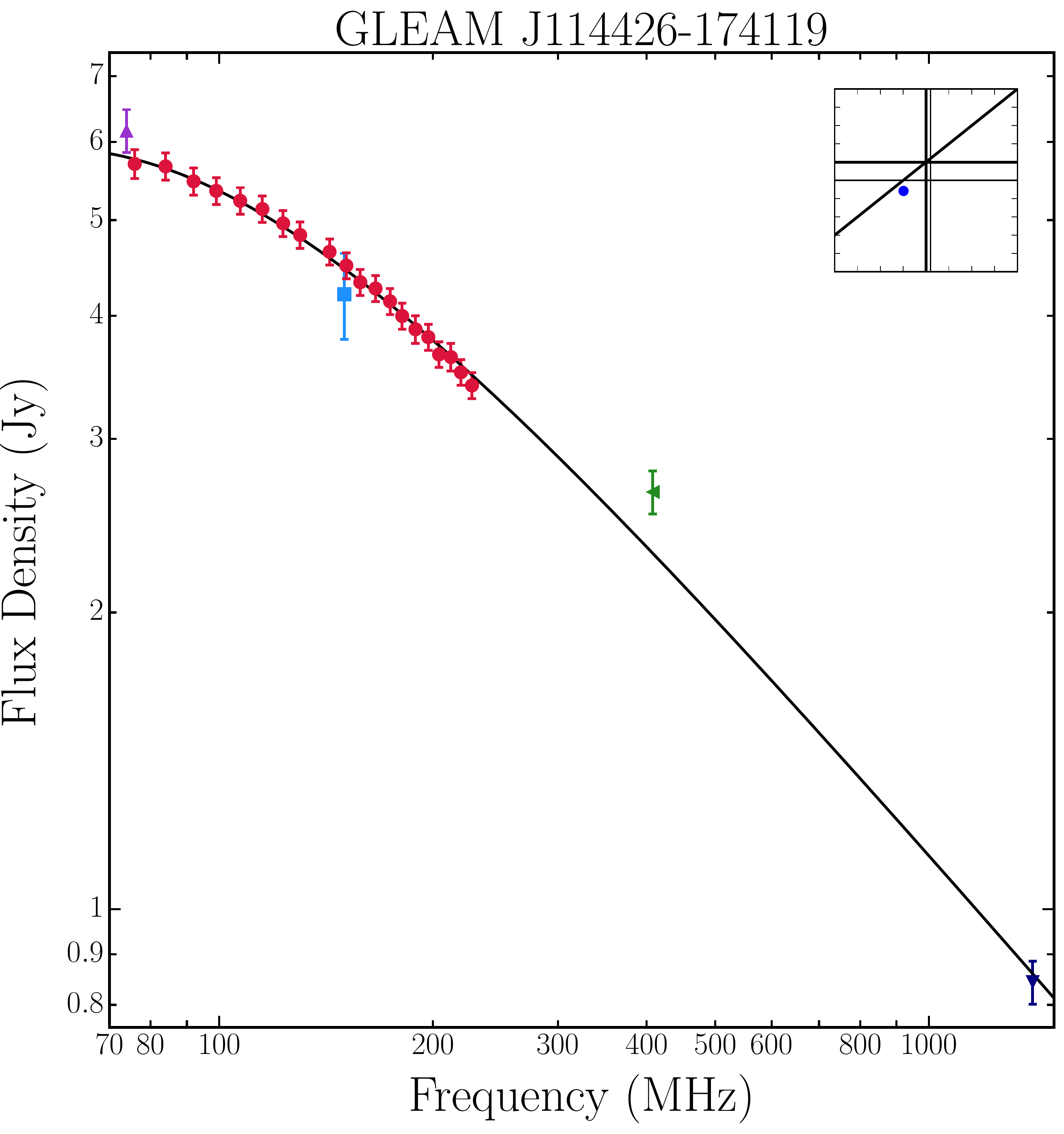}
 \caption{Spectral energy distribution of GLEAM~J114426-174119 from 72\,MHz to 1.4\,GHz. The symbols represent data from the same surveys as in Figure \ref{fig:example_spectra}. While no spectral turnover is detected between 72\,MHz and 1.4\,GHz, the high signal-to-noise spectrum of GLEAM~J114426-174119 allows us to derive $q = -0.31 \pm 0.03$, suggesting that a spectral turnover occurs below 72\,MHz. The plot inset in the top-right corner is the color-color diagram of Figure \ref{fig:color_color}, with the position of GLEAM~J114426-174119 marked by a blue circle.}
\label{fig:high_snr_no_turnover}
\end{center}
\end{figure}

It is also possible to identify sources that peak below 72\,MHz on the basis of significant curvature in high signal-to-noise spectra. An example of a source that is beginning to turn over, but peaks below the lowest GLEAM frequency, is shown in Figure \ref{fig:high_snr_no_turnover}. However, since we do not detect a peak in the spectrum, these sources are not used in any of the following analysis but are presented to encourage observations below 72\,MHz to confirm the spectral turnover. There are 36 sources identified with $q < -0.2$, a turnover below 72\,MHz, and a signal-to-noise greater than 100 in the GLEAM wideband image. We apply such a high signal-to-noise requirement to produce a reliable collection of sources that peak below 72\,MHz, rather than a complete sample. The properties of these sources are also provided online in the same format as the table presented in Appendix \ref{sec:appendix_tab}.

\subsection{Flux density variability and blazar contamination}

The reliability of proceeding studies in identifying \ps sources using multi-epoch survey data were significantly impacted by radio source variability \citep[e.g.][]{2000A&A...363..887D,2002MNRAS.337..981S}. Previously, the often single low frequency data point used to justify a spectral peak was provided by an observation of the source when it was in a less active phase compared to when high frequency data were taken. Hence, variable sources in previous multi-epoch studies could masquerade as \ps sources even though the intrinsic spectrum of the source was flat \citep{2005A&A...432...31T}. Since GLEAM surveyed the sky in four observing bands, with a two-minute cadence between observing bands, our estimate of the spectral peak and the slope below a turnover is not impacted by variability. While high-frequency studies have observed sources such as blazars varying on hour to month time-scales, the radio sky below 1\,GHz has been shown to be significantly less variable \citep{1990MNRAS.246..123M,2010AJ....140.1995L,2014MNRAS.438..352B,2016MNRAS.458.3506R}.

It is possible variability will impact our determination of the slope above the turnover since data from SUMSS and NVSS were utilized. However, one strength of having such a well sampled low-frequency survey is that the defining feature of a \ps source, whether the source has peak or a positive spectral slope, is characterized completely by the low frequency data. Hence, any high frequency variability will only move the source between the hard and soft samples, ensuring the source will still be identified as a \ps source. 

As mentioned above, blazars are known contaminants in \ps selections because of their variability. While our identification of a \ps source may not be significantly impacted by variability, the sample will contain some blazars. This is because a peak can occur in the spectrum of a blazar if it is observed during an AGN flare, such that a SSA component dominates the emission spectrum \citep{2007A&A...469..451T}, or due to a variation in the beaming angle. The two samples most likely to be contaminated with blazars are the hard and GPS samples, since the definition of these two samples include sources with \ah~$\sim\,0$. By comparing the reported nature of a source in the literature \citep[e.g.][]{2015Ap&SS.357...75M}, we estimate that blazars represent $\approx$\,10\% of  sources in the hard and GPS samples, and $<3$\% of sources in the soft samples. Future high resolution imaging, in particular with LOFAR, and low frequency multi-epoch observations from the MWA Transient Survey (MWATS; Bell et al., in prep.) survey, and the second year of GLEAM survey data, will isolate the sources in the samples presented that are jet-dominated.

\subsection{Obtaining redshifts}

We obtained redshift information for the \ps candidates selected in \S\,\ref{subsec:spec_class} by cross-matching our sample to previous targeted optical observations of GPS, CSS, and HFP sources \citep{Odea1991,1990A&A...231..333F,Labiano2007,2007A&A...464..879D,2008MNRAS.387..639H}, radio-optical studies that combined the large spectroscopic surveys of the Six-degree Field Galaxy Survey \citep[6dFGS;][]{2004MNRAS.355..747J,2009MNRAS.399..683J} and Sloan Digital Sky Survey \citep[SDSS;][]{2000AJ....120.1579Y} with NVSS and SUMSS \citep{2007MNRAS.375..931M,2012MNRAS.421.1569B}, targeted quasar surveys \citep{1987ApJS...63....1H,1994ApJ...436..678O,1997MNRAS.284...85D,2010MNRAS.405.2302H}, investigations that researched the optical properties of a radio source population that were not specifically GPS, CSS, and HFP sources \citep{1996ApJS..107...19M,2002A&A...386...97J,2006AJ....131..114B,2008ApJS..175...97H,2011MNRAS.417.2651M}, and the NASA/IPAC Extragalactic Database \citep[NED;][]{1991ASSL..171...89H}\footnote{\url{http://ned.ipac.caltech.edu/}}. 

\begin{figure}
\begin{center}
\includegraphics[scale=0.4]{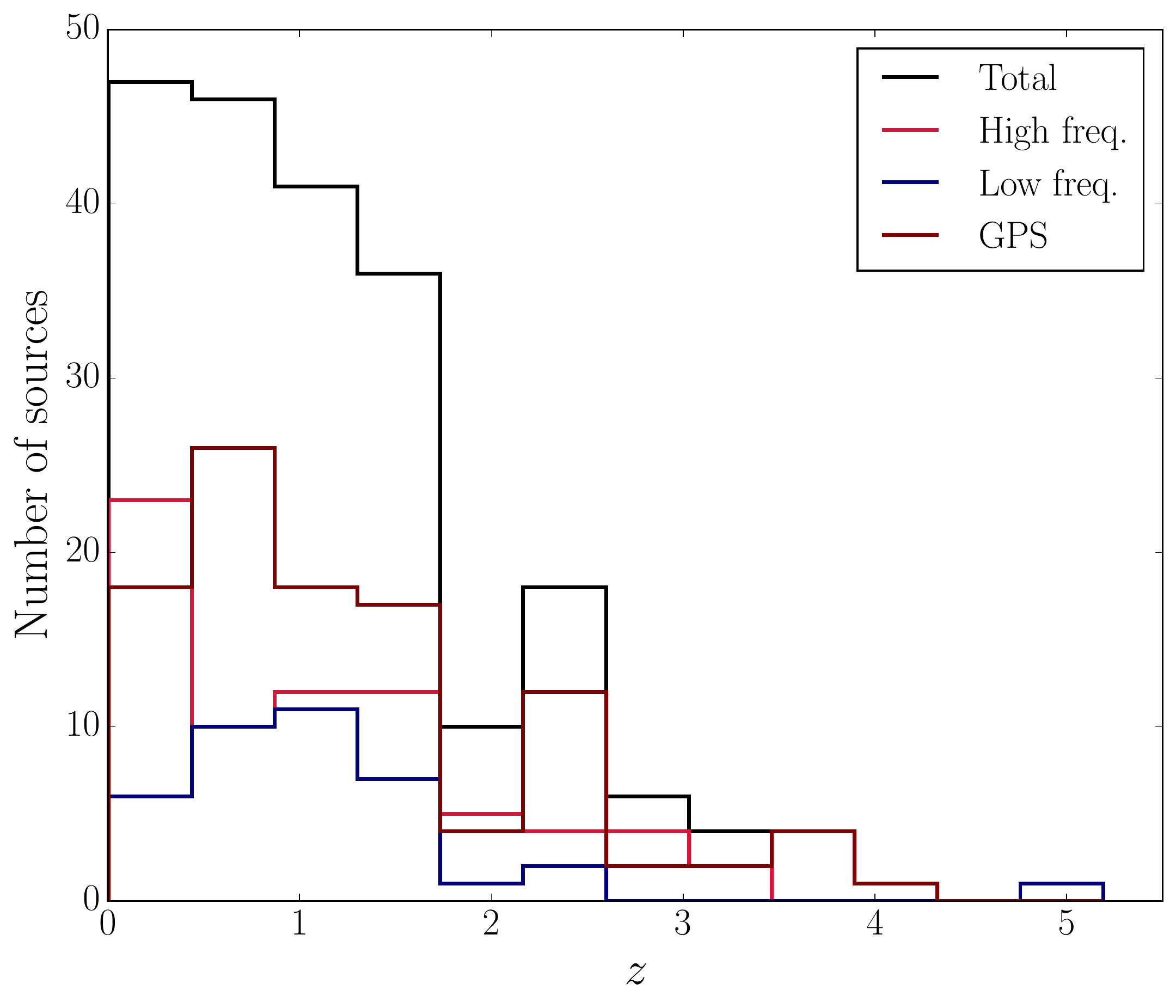}
 \caption{Redshift distribution of the 214 selected \ps sources that have reported spectroscopic redshifts. There are 72, 38, and 104 sources with spectroscopic redshifts in the low and high frequency \ps samples, and GPS sample, respectively. The median of the distribution is 0.98, with the most distant source at a redshift of 5.19. The total distribution of all identified \ps sources is shown in black, while the red, navy, and maroon distributions represent the distribution for the high frequency, low frequency, and GPS sample, respectively.}
\label{fig:redshift_dist}
\end{center}
\end{figure}

The positions of the NVSS or SUMSS counterparts were used to make the associations in the other catalogs. The NVSS positions were preferenced if the source had both a NVSS and SUMSS counterpart. An association was accepted if the literature source was within 3$\arcsec$ of the radio position. Such a simple cross-matching scheme was accurate for association since the proceeding studies have already made an association between the optical and radio sources. We obtain 214 spectroscopic redshifts for the total \ps sample, which breaks down into 61, 11, 15, 23, and 104 redshifts in the high frequency hard, high frequency soft, low frequency hard, low frequency soft, and GPS sample, respectively. The redshift distribution for the \ps samples is plotted in Figure \ref{fig:redshift_dist}, which has a median of 0.98 and a highest redshift of 5.19. The literature reference from which the redshift was obtained is detailed in Appendix \ref{sec:appendix_tab}.

Note that while NED contains the largest number of redshifts for extragalactic sources of any single database, it has known limitations \citep[see e.g.][]{2012arXiv1209.1438H}. Any redshift information derived from it for a population of sources will be inhomogeneous and incomplete. If multiple spectroscopic measurements of the redshift were reported in NED, the most reliable measurement was selected by inspecting the literature on the source. Additionally, we discarded any photometric redshifts and redshifts that had the ``$z$ Quality'' field flagged in NED.

\section{Comparison to known GPS, CSS, and HFP sources}
\label{sec:gpscsscomp}

To test the reliability of our \ps selection criteria, we compared our total \ps sample to the known GPS, CSS, and HFP sources from the literature. Only studies that identified GPS, CSS, and HFP sources below a declination of $+30\degree$ were considered. This included the GPS, CSS, and HFP samples identified by \citet{1990A&A...231..333F}, \citet{Odea1998}, \citet{1998A&AS..131..303S}, \citet{2000ApJ...534...90P}, \citet{2002MNRAS.337..981S}, \citet{2005A&A...432...31T}, \citet{Labiano2007}, \citet{2004A&A...424...91E}, and \citet{Randall2011}. After removing any duplicates, 157 of the 216 previously known GPS, CSS, and HFP sources that are below a declination of $+30\degree$ have a counterpart in the GLEAM sample from which the \ps sources were selected, and 73 of those 157 sources are in our \ps samples. The other 59 sources are too faint to be in the GLEAM extragalactic catalog.

The 84 known GPS, CSS, and HFP sources that are not in our total \ps sample, but are in the GLEAM sample used to isolate the \ps sources, can be sorted into three spectral categories: 36 sources show no deviation from a power-law with a negative slope, 34 sources have flat-spectra such that \al~$< 0.1$ or $q > -0.2$ in the GLEAM band, and 14 sources display a convex spectrum. We define a convex spectrum as having a negative spectral index in the GLEAM band and a positive spectral index between the end of the GLEAM band and the frequency of SUMSS/NVSS.

\begin{figure*}
\begin{center}$
\begin{array}{cc}
\includegraphics[scale=0.3]{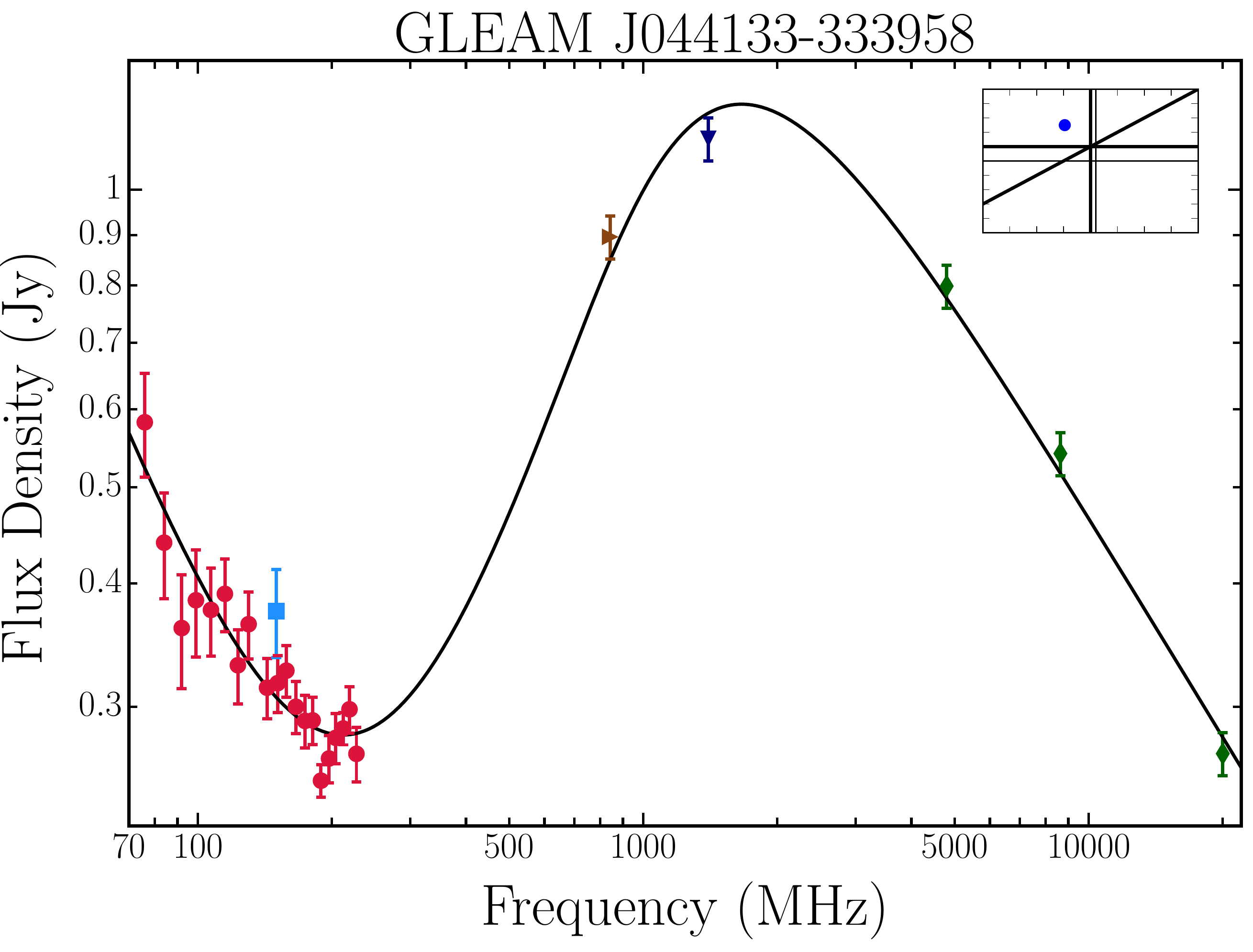} & 
\includegraphics[scale=0.3]{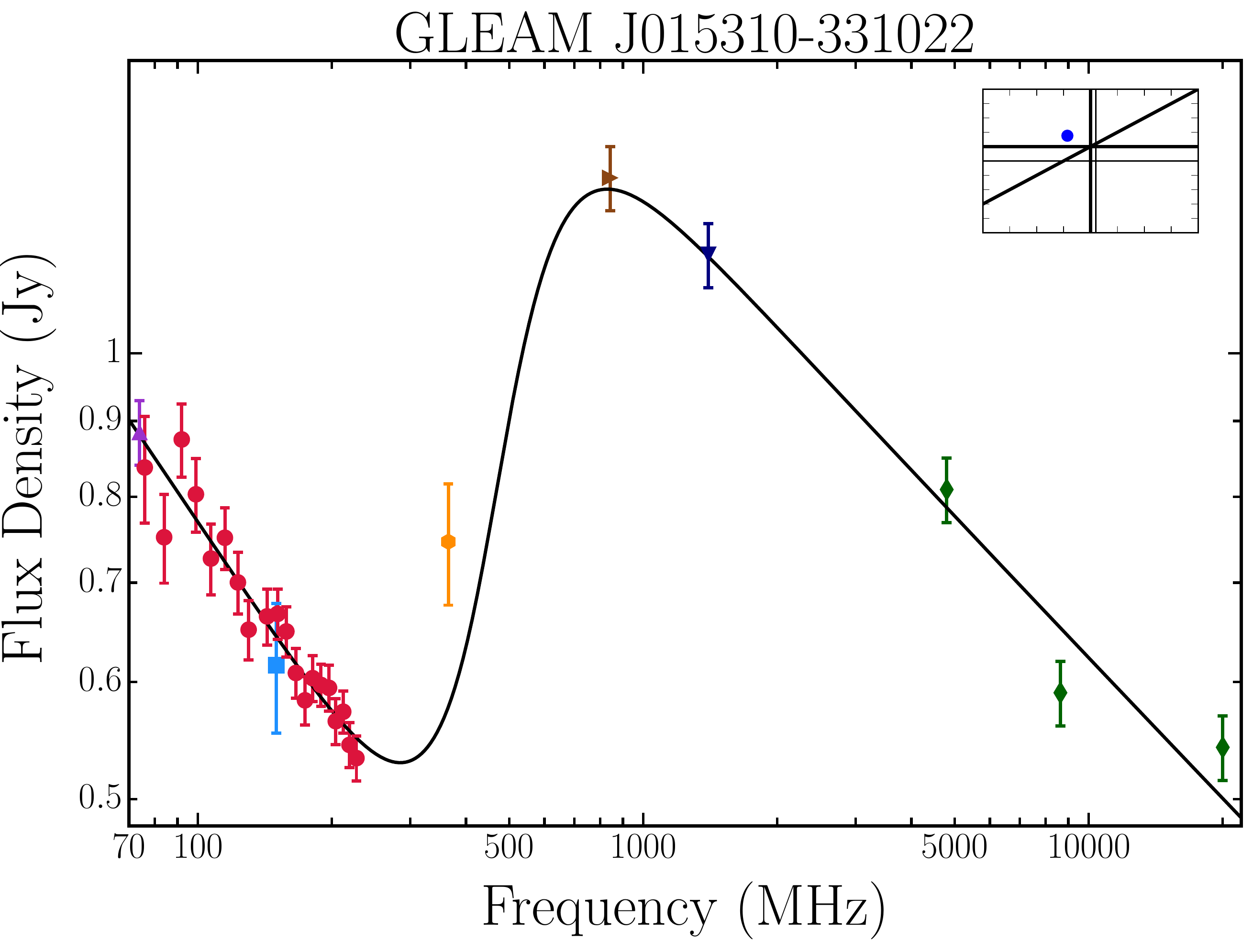} \\ 
\includegraphics[scale=0.3]{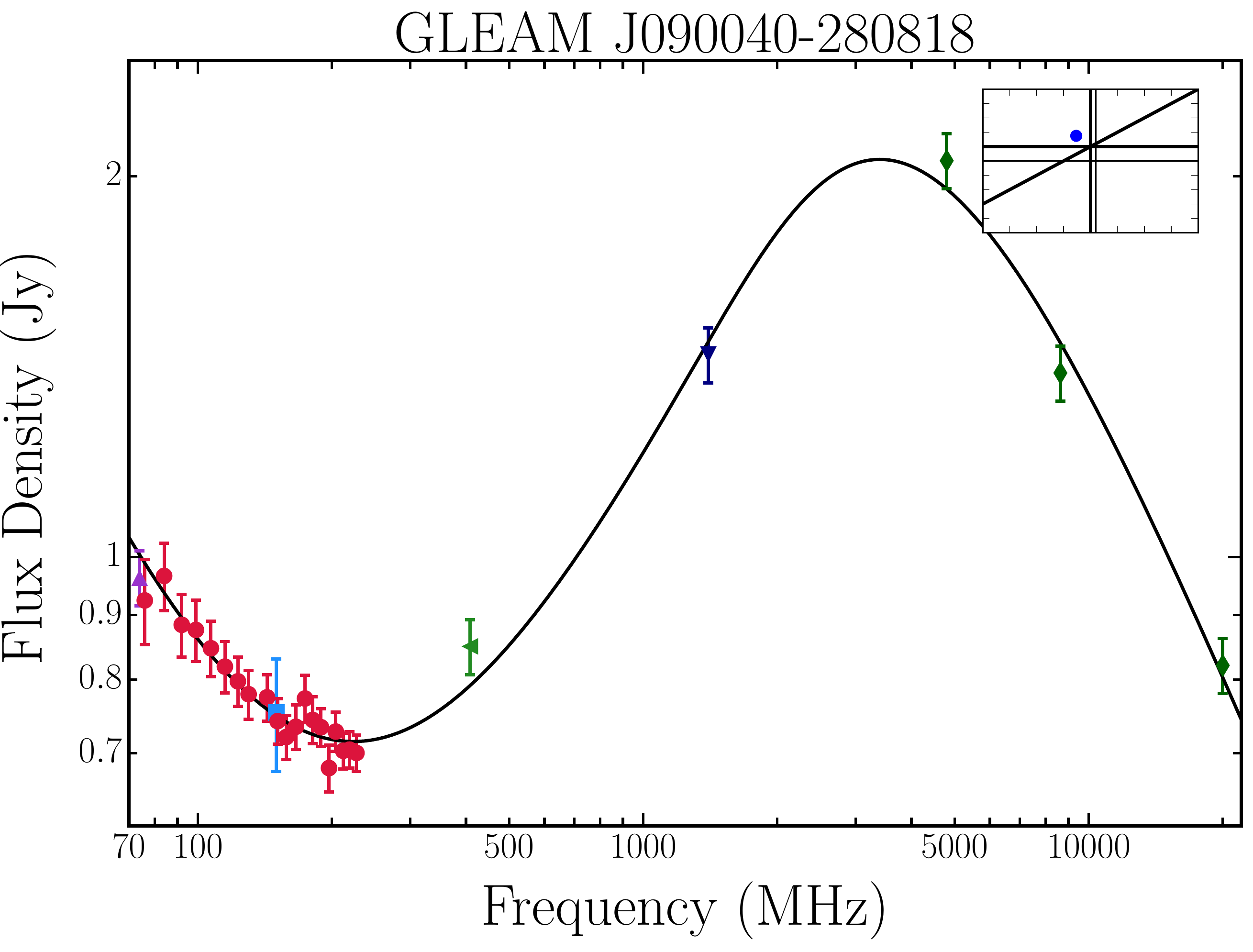}& 
\includegraphics[scale=0.3]{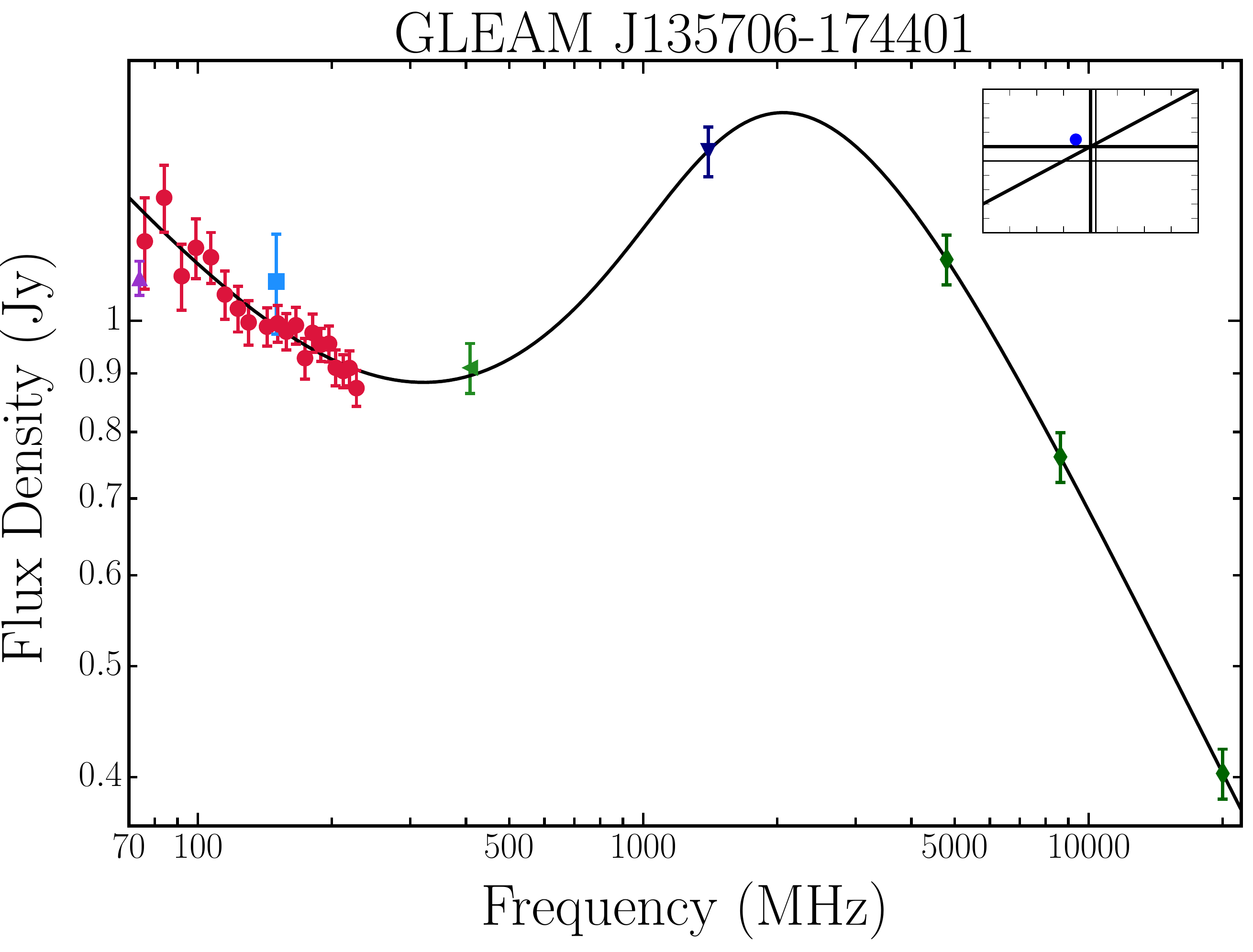} \\ 
\includegraphics[scale=0.3]{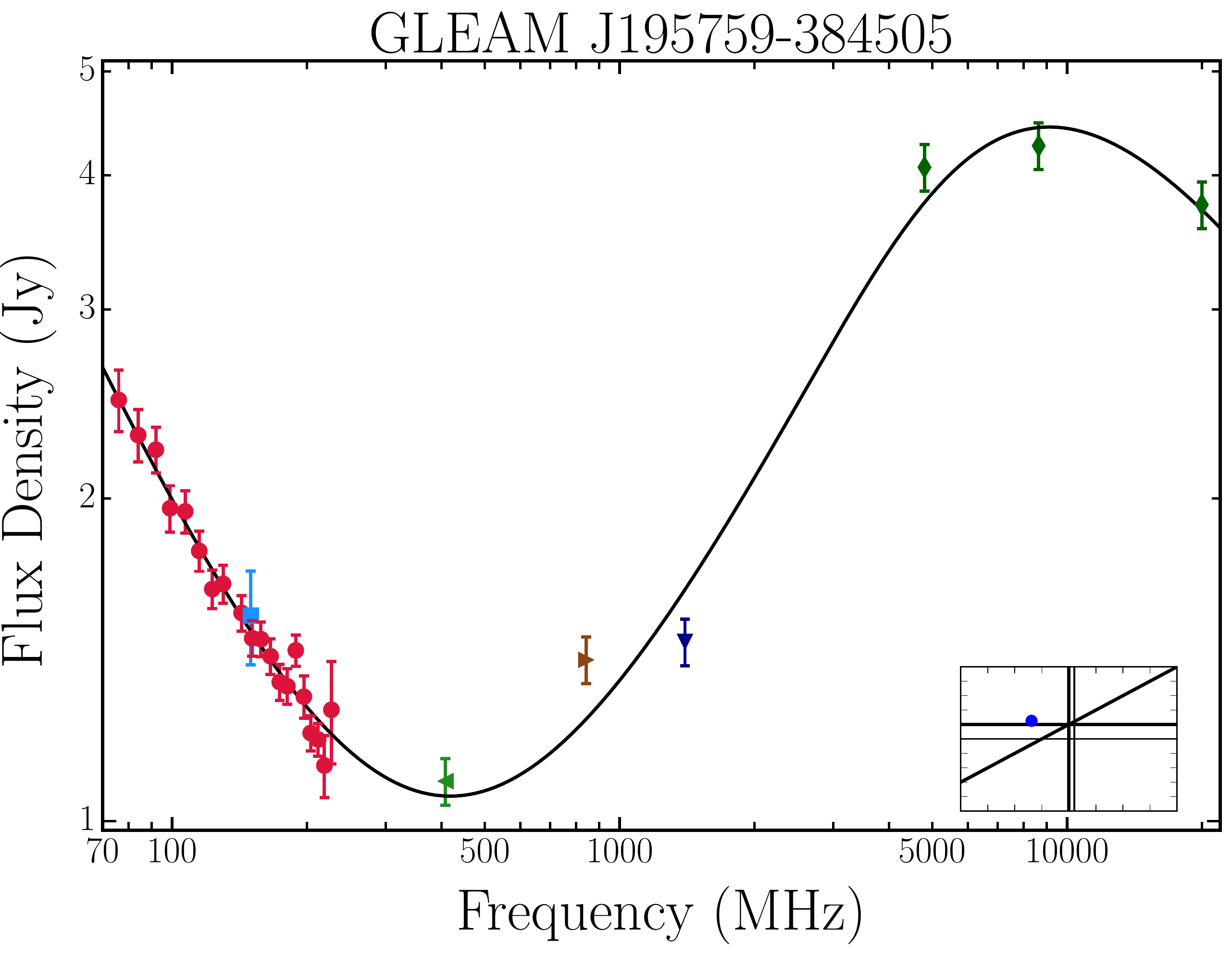}& 
\includegraphics[scale=0.3]{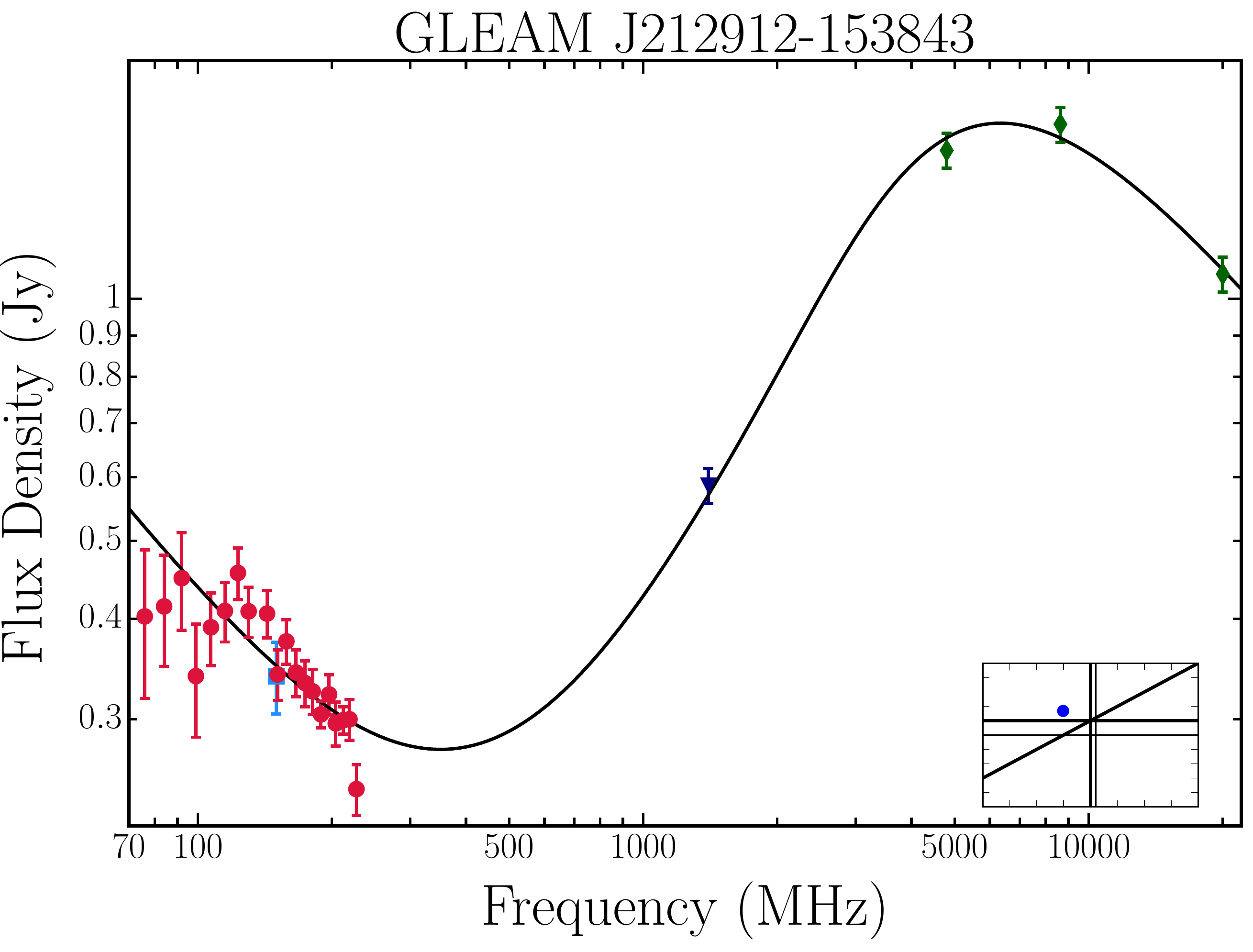}\\ 
\end{array}$
 \caption{Spectra of six known GPS sources, from 72\,MHz to 20\,GHz, that have a convex pattern between 72\,MHz and 1.4\,GHz. The symbols represent data from the same surveys as in Figure \ref{fig:example_spectra}, with the dark green diamonds and yellow hexagons depicting data from the AT20G and TXS surveys, respectively. The black curve represents the fit of the generic curved spectral model of Equation \ref{eqn:gen} with an addition of a low frequency power-law. The blue circle in the inset plot highlights the position of each source in the color-color diagram of Figure \ref{fig:color_color}.}
\label{fig:uptick}
\end{center}
\end{figure*}

Illustrated in Figure \ref{fig:uptick} are six of the sources that have convex spectra between 72\,MHz and 1.4\,GHz. The convex spectrum category is likely composed of sources that have had multiple epochs of AGN activity, with the peaked-spectrum component above 1\,GHz representing recent activity in the core, while the up-turn at frequencies below the turnover is suggestive of diffuse, older emission \citep{Baum1990,2004A&A...424...91E,2007A&A...469..451T,2010MNRAS.408.1187H}. Therefore, the presence of optically thin emission at low frequencies excludes these types of GPS sources from being truly young radio sources. In these types of sources, the peaked component of the spectrum should be interpreted as the radio source being re-started on short time scales, implying the sources have had a long life span but intermittent activity in the nuclear region. 

Sources with a peaked component and low frequency power-law, as shown in Figure \ref{fig:uptick}, have mostly been observed at the center of clusters \citep{2004rcfg.proc..335K,2015MNRAS.453.1223H}. While the multi-epoch activity for sources with this type of spectra is often explained by the high duty cycle expected for cool core cluster hosted AGN \citep{2015MNRAS.453.1223H}, only two of our sources are located in cluster environment. For the isolated sources, a varying acceleration rate, or the launching of unique knots in the radio jets, necessary to produce the observed convex spectra is likely related to interactions or mergers with nearby galaxies \citep[e.g.][]{2010MNRAS.408.1187H,2012MNRAS.427.1603T,2016A&A...585A..29B}.

Since sources with convex spectra could be helpful in understanding the duty-cycle of \ps sources, we identify 116 convex sources from the GLEAM sample using the criteria \al~$<-0.1$ and \ah~$>0.1$. These source are located in fourth quadrant of Figure \ref{fig:color_color}. Since surveys with frequencies above that of NVSS and SUMSS are required to confirm a turnover in these convex sources, where variability can be significant, we do not use the convex spectrum sources in any further analysis. A list of the convex spectrum sources, presented in a form similar to the table in Appendix \ref{sec:appendix_tab}, is available online. We note that high-resolution imaging at low frequencies will be needed to measure the size of the low frequency component and establish the activity time scales. 

On the basis that all of the previously known GPS, CSS, and HFP sources are either in our \ps sample, or have spectral characteristics that ensure they do not display a spectral turnover between 72\,MHz and 843\,MHz\,/\,1.4\,GHz, demonstrates that the selection criteria outlined in \S\,\ref{sec:selection} is reliable. This also implies we have identified 1,410 new \ps candidates, representing more than a factor of six increase in the number of known \ps sources below a declination of $+30\degree$, and doubling the number of known \ps in the entire sky. Additionally for $\approx$\,95\% of the \ps sources identified the peak is newly characterized.

\section{Comparison to ultra-steep-spectrum source samples}
\label{sec:uss}

Ultra-steep-spectrum (USS) sources, defined as compact radio sources with $\alpha < -1$, have been the focus of searches for high-redshift radio AGN \citep[e.g.][]{2001MNRAS.326.1585J,2006MNRAS.366...58D,Klamer2006,2009MNRAS.395.1099B}. Based on the study of the GPS source PKS~B0008-421, which was classed as a USS source until observations were conducted below 600\,MHz, \citet{Callingham2015} hypothesized that the GLEAM survey should find that a portion of the USS population is composed of GPS, CSS, and HFP sources that have ceased nuclear activity and entered a relic phase. They argued that PKS~B0008-421 had a steep optically thin slope because it had steepened by $-0.5$, as expected from the aging of electrons in the continuous injection model of \citet{Kardashev1962}, but that the discontinuous transition in the spectrum was not observed because the break frequency had moved to lower frequencies than the spectral turnover. Such a transition could help explain why \citet{Klamer2006} did not find any steepening in the spectra of USS sources above 1\,GHz. Additionally, if the injection of electrons has ceased in a \ps source, strong adiabatic cooling should quickly shift the peak frequency outside the gigahertz-regime \citep{Orienti2008}, ensuring a USS source identified above 1\,GHz would not be classed as \ps source without low frequency observations.

\begin{figure}
\begin{center}
\includegraphics[scale=0.275]{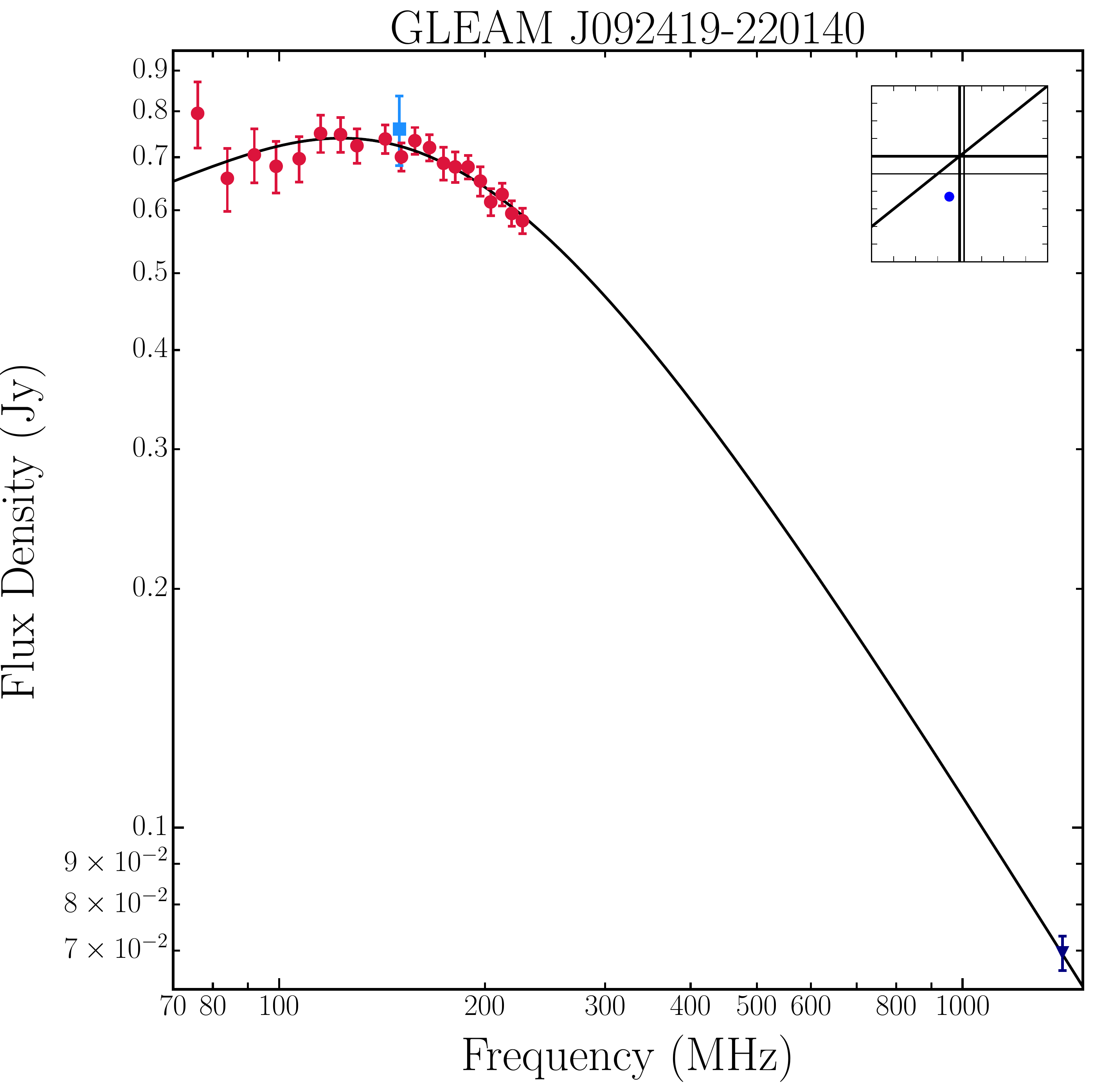}
 \caption{Spectrum of the USS source TN~J0924-2201, which is located at $z = 5.19$ \citep{1999ApJ...518L..61V}. The symbols represent the data from the same surveys detailed in Figure \ref{fig:example_spectra}. The spectral turnover occurs at $\nu_{\mathrm{p}} = 160\,\pm\,30$\,MHz.}
\label{fig:uss}
\end{center}
\end{figure}

Since the GLEAM survey has observed the sky with a large fractional bandwidth at low frequencies, we can test if known USS samples selected at high frequencies are significantly composed of sources that display spectral turnovers below 300\,MHz. We cross-matched our \ps samples with the southern USS catalogs of: \citet{2002A&A...394...59D}, which selected USS sources between 325\,MHz and 1.4\,GHz, and \citet{2007MNRAS.381..341B}, which selected sources between 408 and 843\,MHz. We find that eight and five sources from \citet{2002A&A...394...59D} and \citet{2007MNRAS.381..341B} show turnovers at or below 300\,MHz, respectively. In comparison, 113 of the 154 USS sources from \citet{2002A&A...394...59D} and 213 of the 239 USS sources from \citet{2007MNRAS.381..341B} are in the GLEAM sample used to select the \ps samples. Of the 13 USS sources in which we detect a spectral peak, two have reported spectroscopic redshifts at 2.204 and 5.19 \citep{2009MNRAS.395.1099B,1999ApJ...518L..61V}. The peaked-spectrum of the USS source TN~J0924-2201, located at a redshift of 5.19 \citep{1999ApJ...518L..61V}, is shown in Figure \ref{fig:uss}. 

Follow up observations of the USS sources with a spectral turnover, ideally with the wideband backends on the Australia Telescope Compact Array (ATCA) and the VLA, are required to confirm whether these sources display high frequency exponential breaks indicative of sources that have ceased nuclear activity \citep{Jaffe1973,Murgia2003}, as per PKS~B0008-421.

Alternatively, if these USS sources with low turnover frequencies are actively fueling their AGN and the `youth' scenario of \ps sources is correct, the turnover in their spectra could have shifted to low-frequencies because they are located at high redshift \citep{2004NewAR..48.1157F,2015MNRAS.450.1477C}. The idea that these sources are located at high redshift is supported if the steep optically-thin slope is a product of the sources being found in environments similar to the center of local, rich clusters. This is because it is possible that the high environmental densities in the early Universe both constrained the evolution of the radio jets \citep{Bicknell1997} and steepened the optically thin spectrum through first-order Fermi acceleration processes from slowed hot-spot advancement \citep{1998JApA...19...63A,Klamer2006}. Since extremely steep-spectrum radio galaxies and GPS/HFP sources are found to reside at the centres of rich clusters in the local universe, with greater than 8.5\% of all radio sources located at the center of cool-core clusters having a peaked-spectrum \citep{2003ApJ...592..755R,Klamer2006,2015MNRAS.453.1223H}, having both a low frequency turnover and steep spectral slope above the turnover could be excellent priors in searching for high redshift galaxies. This will be explored in greater detail in \S\,\ref{sec:spec_prop}.

While \citet{1999AJ....117..677B} posited that the correlation between redshift and steep optically thin spectral indices was a product of radio luminosity and Malmquist bias, adding the requirement of a turnover breaks the degeneracy of the evolution of radio luminosity and different survey flux density limits. Therefore, provided USS sources with low frequency turnovers are young radio AGN, they represent excellent high redshift candidates. 

\section{Spectral properties of the peaked-spectrum sample}
\label{sec:spec_prop}

Since we have derived a unique \ps sample by exploiting a wide band low frequency survey, we have an opportunity to use a new sample to extend on previous work in understanding the physical properties of \ps sources. It is also important to compare the spectral properties of this sample with archetypal \ps samples, such as those presented by \citet{Odea1998} and \citet{1998A&AS..131..435S}, to test whether the spectral properties of known GPS, CSS, and HFP sources are consistent with the newly characterized \ps sources, and to understand the biases in our selection method. A summary of the comparison of the total \ps sample with the samples studied by \citet{Odea1998} and \citet{1998A&AS..131..435S} is presented in Table \ref{table:comparison}.

\begin{table*}
	\small
	\caption{\label{table:comparison} A summary of the median spectral properties of the \ps sample presented in this study, and of the samples identified by \citet{Odea1998} and \citet{1998A&AS..131..435S}. The spread in the values quoted represents the difference between the median and the 16$^{\mathrm{th}}$ or 84$^{\mathrm{th}}$ percentiles in the distributions. The parameters listed for this study are for the total high and low frequency \ps samples, except for \athick, which is only for the high frequency \ps sample. The selection frequency for the \ps sample studied by \citet{Odea1998} is listed as inhomogeneous because the sample was assembled from different literature sources, such as \citet{1998A&AS..131..303S} and \citet{1990A&A...231..333F}.}
	\begin{center}
		\begin{tabular}{cccc}
		\hline
		\hline
Parameters & This study & O'Dea (1998) & Snellen et al. (1998) \\
		\hline			
Selection frequency (MHz) & 72\,$-$\,1400 & Inhomogeneous & 325\,$-$\,5000 \\
Faintest peak flux density (Jy) & 0.16 & 0.3 & 0.04 \\
Number of sources & 1483 & 69 & 47 \\
$\alpha_{\mathrm{thin}}$ & $-0.77^{+0.31}_{-0.38}$ & $-0.75 \pm 0.15$ & $-0.77 \pm 0.15$\\
$\alpha_{\mathrm{thick}}$ & $0.88^{+0.71}_{-0.49}$ & $0.56 \pm 0.20$ & $0.80 \pm 0.18$\\
Observed $\nu_{\mathrm{p}}$\,(MHz) & 190$^{+190}_{-90}$ & 750$^{+1200}_{-250}$ & 1500$^{+700}_{-500}$\\
$z$ & 0.98 & 0.84 & 1.01\\
Intrinsic $\nu_{\mathrm{p}}$\,(MHz) & 440$^{+560}_{-250}$ & 850$^{+2500}_{-100}$ & 2840$^{+1900}_{-860}$\\
$P_{5 \, \mathrm{GHz}}$\,($\log_{10}\,$W\,Hz$^{-1}$) & 26.5$^{+0.7}_{-1.2}$ & 27.6$^{+0.7}_{-1.0}$ & 26.2$^{+0.5}_{-0.4}$ \\

	\hline\end{tabular}                           
\end{center}                                                                               
\end{table*}

\begin{figure*}
\begin{center}
\includegraphics[scale=0.4]{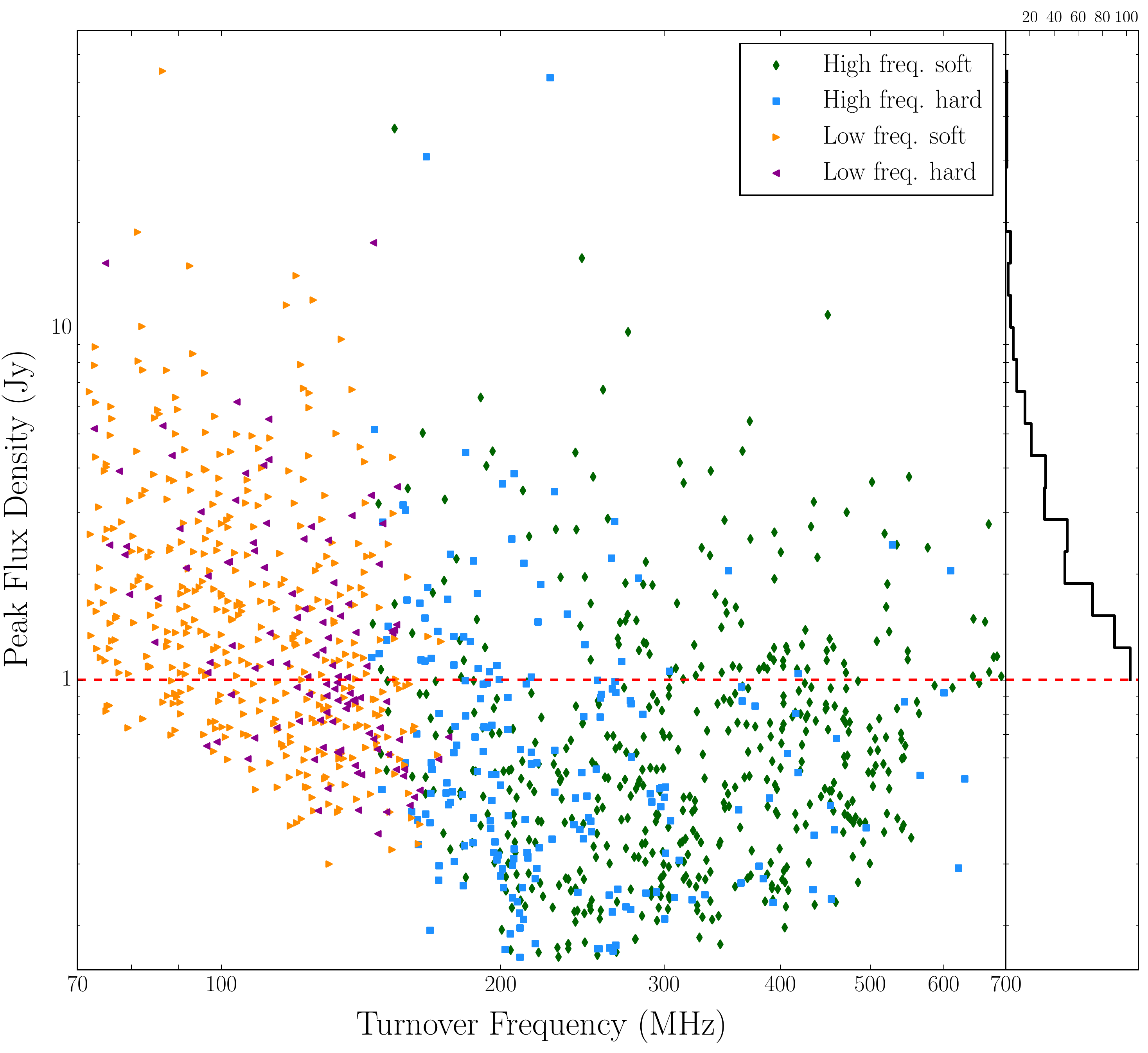}
 \caption{The distribution of the spectral turnover frequency and the peak flux density for the GLEAM sources in which a spectral turnover is detected. The colors represent the different \ps samples, as communicated by the legend. The red-dashed line represents the peak flux density in which the total \ps sample is complete, which corresponds to $\approx$\,1\,Jy. The histogram on the right represents the distribution of peak flux densities above 1\,Jy.}
\label{fig:peak_flux_v_peak_freq}
\end{center}
\end{figure*}

The observed distribution of the peak frequency and flux density for our \ps samples, with the different samples identified by their colors, is presented in Figure \ref{fig:peak_flux_v_peak_freq}. Clearly, the selection criteria employed are a function of both peak frequency and flux density, with no sources detected with a peak frequency below 72\,MHz or above 1.4\,GHz, or below the flux density cut made at 0.16\,Jy. Few sources are detected with peak flux densities below 0.3\,Jy beneath 100\,MHz or above 600\,MHz since the detection of a peak at the edge of the observed band is reliant on high signal-to-noise statistics. As expected from the definition of the different samples, the high frequency soft sample has the highest frequency peaks, with the transition between sources being selected based on radio color-color phase space and a peak in the GLEAM band occurring around $\approx$\,180\,MHz. Additionally, the continuity of the distribution of parameters in Figure \ref{fig:peak_flux_v_peak_freq} demonstrates that the cut using \ah~to separate the hard and soft population is arbitrary.

As is evident from the distribution of peaked-spectrum sources in \ref{fig:peak_flux_v_peak_freq}, the obvious biases introduced by the selection criteria in the flux density and frequency are eliminated above peak flux densities of 1\,Jy. Therefore, we infer the peaked-spectrum samples to be reasonably complete above 1\,Jy, allowing a comparison of the number of \ps sources that comprise the GLEAM sample. From the high and low frequency \ps samples, there are a total of 505 sources that have a peak flux density at or above 1\,Jy, compared to 11,400 sources from the sample from which these \ps were drawn. Therefore, approximately 4.5\% of the radio population in the GLEAM extragalactic catalog are identified as \ps sources. This is significantly less than the canonical 10\% of GPS sources occupying the total number of radio source population for surveys completed around or above 1\,GHz \citep{Odea1998}, implying that either the number of \ps sources declines with decreasing frequencies or the number of sources that follow a power-law increases.

The histogram on the right of Figure \ref{fig:peak_flux_v_peak_freq} demonstrates that the proportion of \ps sources to the total radio source population changes slightly with flux density. For example, a 4\,Jy peak flux density cut implies that 6.5\% of the radio source population is composed of \ps sources.  Additionally, the fraction of \ps sources will increase with survey frequency due to the different spectral index of two populations selected to be peaked at different frequencies. Such a change in frequency and luminosity is expected from the \citet{2000MNRAS.319..445S} evolutionary model of \ps sources. The frequency dependency of the source counts will be the focus of a future study.

\begin{figure}
\begin{center}
\includegraphics[scale=0.35]{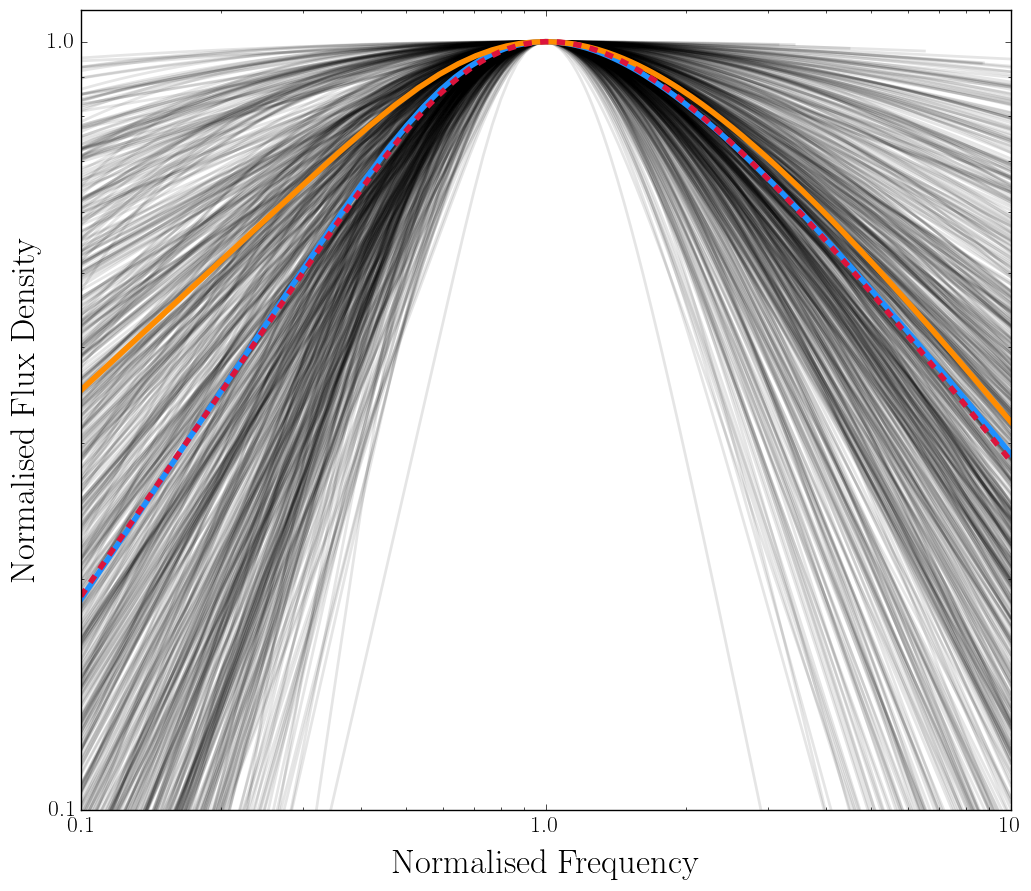}
 \caption{The spectra of the high frequency soft and hard \ps samples, as fit by Equation \ref{eqn:gen}, normalized by the peak flux density and frequency. The dashed red line represents the median spectra for the sample, with median thick and thin spectral indices of $0.88^{+0.71}_{-0.49}$ and $-0.77^{+0.31}_{-0.38}$, respectively. The solid blue and orange lines are the median values of the GPS samples presented by \citet{1998A&AS..131..435S} and \citet{Odea1998}, respectively.}
\label{fig:normalised_spectra}
\end{center}
\end{figure}

To understand the variation in the spectral shape of the selected \ps sources, the spectra normalized by the peak frequency and flux density, as described by the general curved spectral model of Equation \ref{eqn:gen}, for our high frequency \ps sample is shown in Figure \ref{fig:normalised_spectra}. The median spectrum of our high frequency sample, and the GPS and CSS samples presented by \citet{1998A&AS..131..435S} and \citet{Odea1998}, are over-plotted. The median spectrum of the high frequency \ps sample is almost identical to that derived by \citet{1998A&AS..131..435S} but shallower than the median spectrum identified by \citet{Odea1998}.

\begin{figure*}
\begin{center}$
\begin{array}{cc}
\includegraphics[scale=0.4]{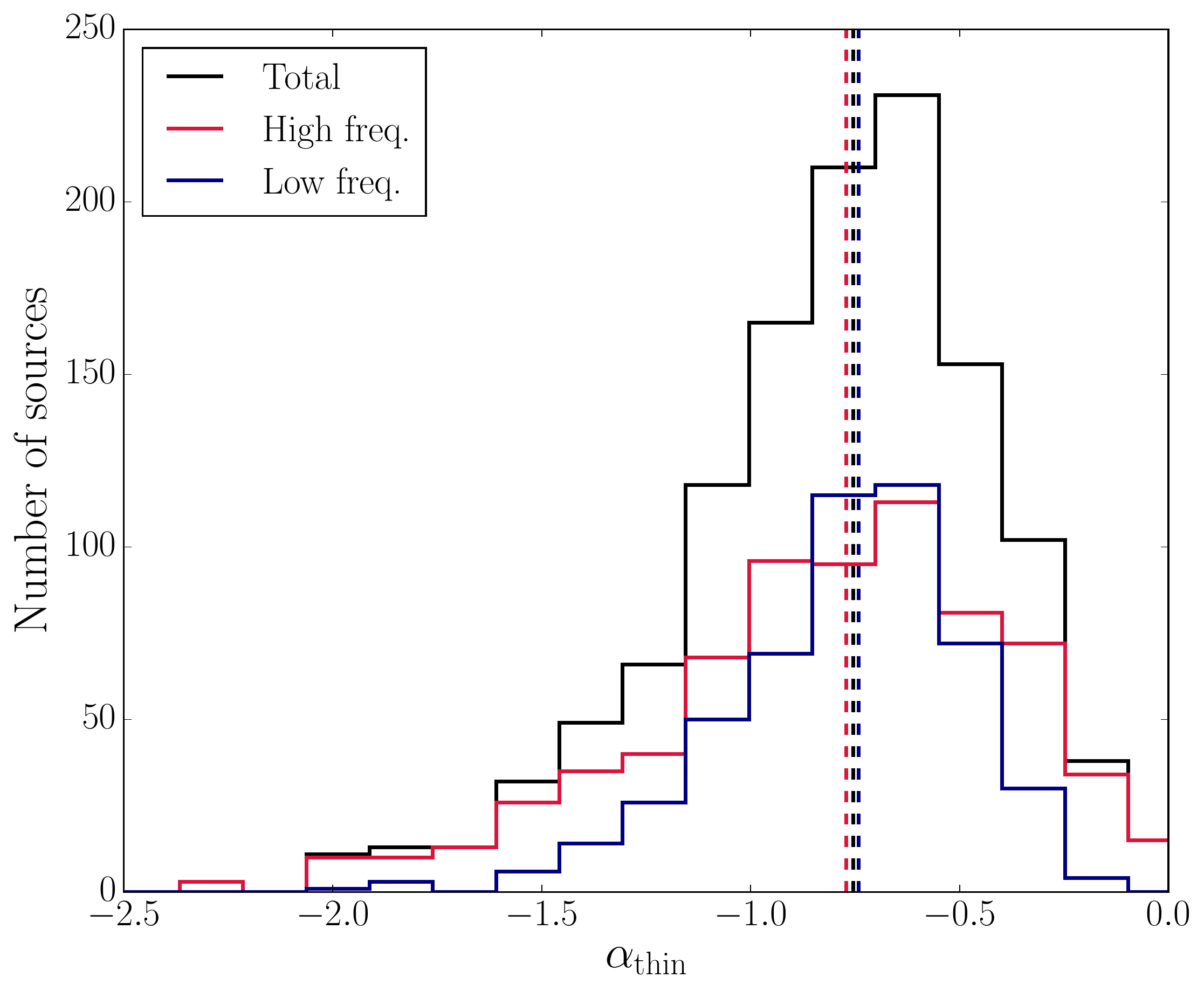} &
\includegraphics[scale=0.4]{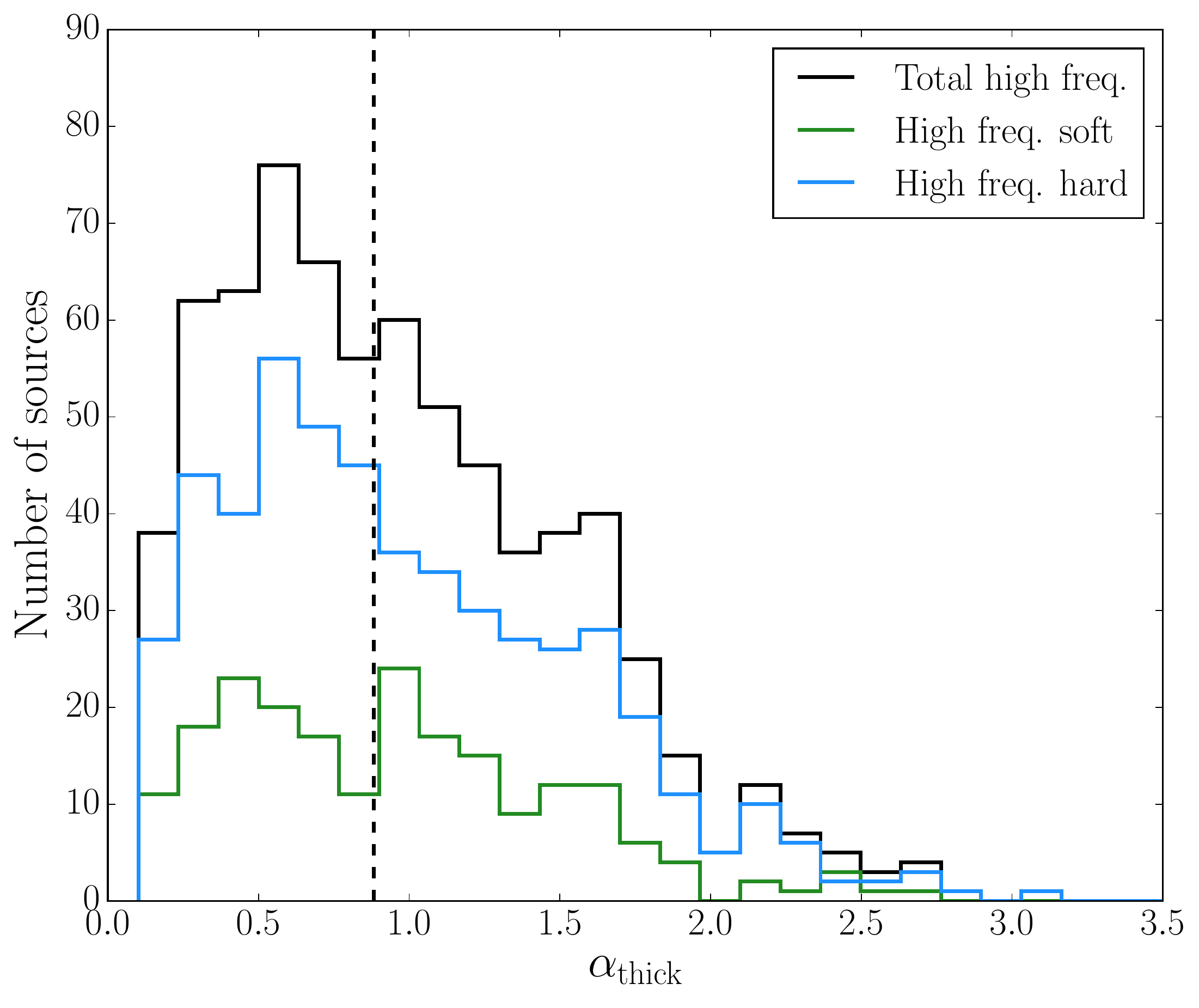} \\
\end{array}$
 \caption{Left panel: The spectral index distribution of $\alpha_{\mathrm{thin}}$. The distribution for the total, high frequency, and low frequency \ps samples are shown in black, red, and navy, respectively. The median for the total \ps sample is plotted as a black dashed line at $\alpha_{\mathrm{thin}} = -0.77$. The medians for the low and high frequency sample are presented as red and navy dashed lines, respectively, at approximately the same value as the median for the total sample. Right panel: The spectral index distribution of $\alpha_{\mathrm{thick}}$. The black, green, and blue histograms represent the distribution of $\alpha_{\mathrm{thick}}$ for the total high frequency, hard high frequency, and soft high frequency \ps sample, respectively. The dashed black line illustrates the median $\alpha_{\mathrm{thick}} = -0.88$ for the total high frequency sample.}
\label{fig:dist_alpha_thick_thin}
\end{center}
\end{figure*}

The distributions of \athin~and \athick, as derived from fitting Equation \ref{eqn:gen}, are presented in Figure \ref{fig:dist_alpha_thick_thin}. The distributions of \athin~and \athick~are more accurate and useful in understanding the properties of the \ps samples than that of \al~and \ah~since \athin~and \athick~are not artificially flattened or steepened by the presence of curvature in the spectrum, as evident in the spectra illustrated in Figure \ref{fig:example_spectra}. Note that only \ps sources that had a reduced $\chi^{2} < 3$ were used in the following analysis, discarding $\approx$\,3\% of the sample. Since the spectra in the low frequency \ps samples are not completely sampled below the turnover, only the distribution for the total high frequency sample is presented for \athick. The distribution for the total \ps sample is presented for \athin~because the high frequency spectral slope is completely sampled for most of the sources. The median value and the range to the 16$^{\mathrm{th}}$ and 84$^{\mathrm{th}}$ percentiles of the optically thick and thin spectral index distributions for the \ps sample are $0.88^{+0.71}_{-0.49}$ and $-0.77^{+0.31}_{-0.38}$, respectively. In comparison, \citet{1998A&AS..131..435S} and \citet{Odea1998} report median values of the optically thick spectral indices of $0.80 \pm 0.18$ and $0.56 \pm 0.20$, and optically thin spectral indices of $-0.75 \pm 0.15$ and $-0.77 \pm 0.15$, respectively. The comparison of these values for the different samples are presented succinctly in Table \ref{table:comparison}. 

The large dispersion in the optically thin and thick spectral indices is a product of our selection process being sensitive to the type of \ps sources identified previously and to both a spectrally flatter and steeper population than proceeding studies. NVSS and SUMSS are over two orders of magnitude more sensitive than the GLEAM survey, and the GLEAM survey observed sources with a wide fractional bandwidth, implying that we can select \ps sources that are extremely inverted or relatively flat when compared to the \ps sources identified by \citet{1998A&AS..131..435S} and \citet{Odea1998}. While \citet{1998A&AS..131..435S} estimated that their survey missed $\sim\,10\%$ of \ps sources that had steep optically thick spectra ($\alpha_\mathrm{thick} > 1)$, the large dispersion of the optically thick spectral index of our high frequency \ps samples suggests the number of sources missed could be be as high as 20\%.

In particular, the wide spread in optically thick spectral indices is strongly indicative that the turnover in the spectrum is caused by an inhomogeneous environment that differs from source to source \citep{deVries1997}, independent of whether the absorption is a product of ionized clouds of plasma acting as a screen or many synchrotron self-absorbed components. 

\subsection{Observed spectral turnover frequency relationships}

\begin{figure}
\begin{center}
\includegraphics[scale=0.35]{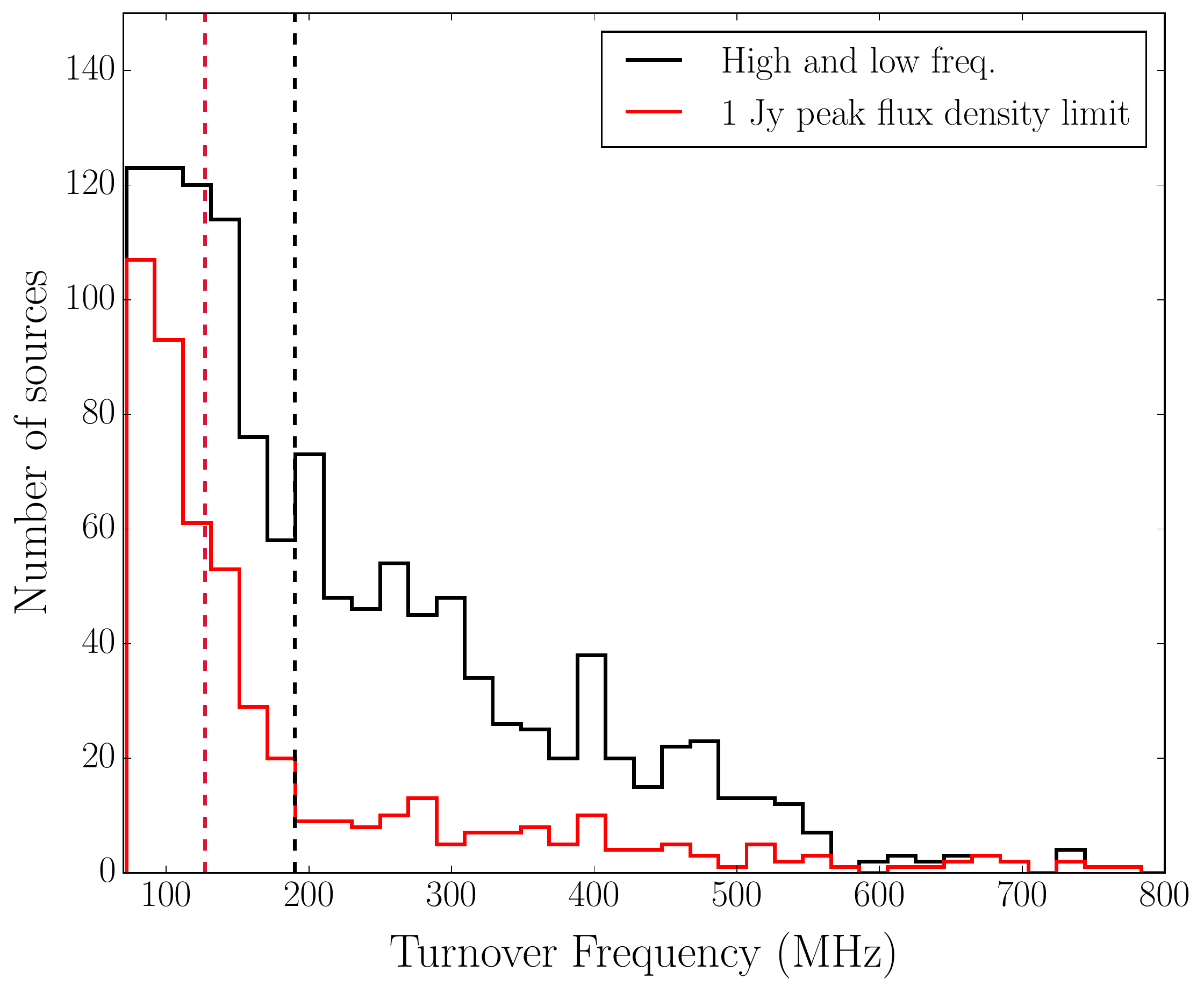}
 \caption{Distributions of the observed spectral turnover frequency for the low and high frequency \ps samples, and for the \ps sources in those samples with peak flux densities greater than 1\,Jy, shown in black and red, respectively. The median of the observed turnover frequency for the total low and high frequency \ps samples is plotted by a dashed black line at 190\,MHz. The median for the sources with peak flux densities greater than 1\,Jy is shown by the dashed red line at 130\,MHz. The linear bin sizes are 20\,MHz.}
\label{fig:turnover_dist}
\end{center}
\end{figure}

The distribution of the observed turnover frequencies for the low and high frequency \ps samples are presented in the left panel of Figure \ref{fig:turnover_dist}, with the median plotted at 190\,MHz. The distribution of the observed turnover frequencies is a function of the selection criteria, since a \ps source in this study is either identified by a spectral peak in the GLEAM band or by two power-law slopes that traced a concave curvature. The transition between the two selection methods is evident by the slight increase in the number of \ps sources identified above peak frequencies of $\approx$\,180\,MHz, corresponding to the small increase in sensitivity when using radio color-color phase space to assess whether a source is peaked. Somewhat surprisingly, there is a monotonic decline in turnover frequencies between 180 and 500\,MHz, where the sensitivity of the selection method is uniform. Above 500\,MHz, the decline in the number of \ps sources identified is a function of the decreasing frequency coverage above the turnover. Additionally, the flattening of the number of sources at $\approx$\,100\,MHz is also due to the selection criteria, implying the distribution will likely continue to monotonically increase at frequencies below 100\,MHz.

To minimize the impact biases introduced by the selection criteria, the distribution of observed turnover frequencies for the sources with peak flux densities greater than 1\,Jy is also plotted in red in Figure \ref{fig:turnover_dist}. The rise in the number of sources towards low turnover frequencies is still evident, suggesting a frequency evolution of the sources. It is also possible that the decline of sources towards higher turnover frequencies could be due to the interplay between the number of sources observed as a function of size and physical size being inversely proportional to turnover frequency in SSA sources. Such a explanation implies that there are many more unresolved radio sources with linear sizes greater than $\approx$\,1\,kpc than with linear sizes than less than $\approx$\,1\,kpc at low radio frequencies. The decline of the number of \ps sources with increasing turnover frequency will be discussed in more detail in \S\,\ref{sec:lumin}.

\begin{figure*}
\begin{center}$
\begin{array}{cc}
\includegraphics[scale=0.4]{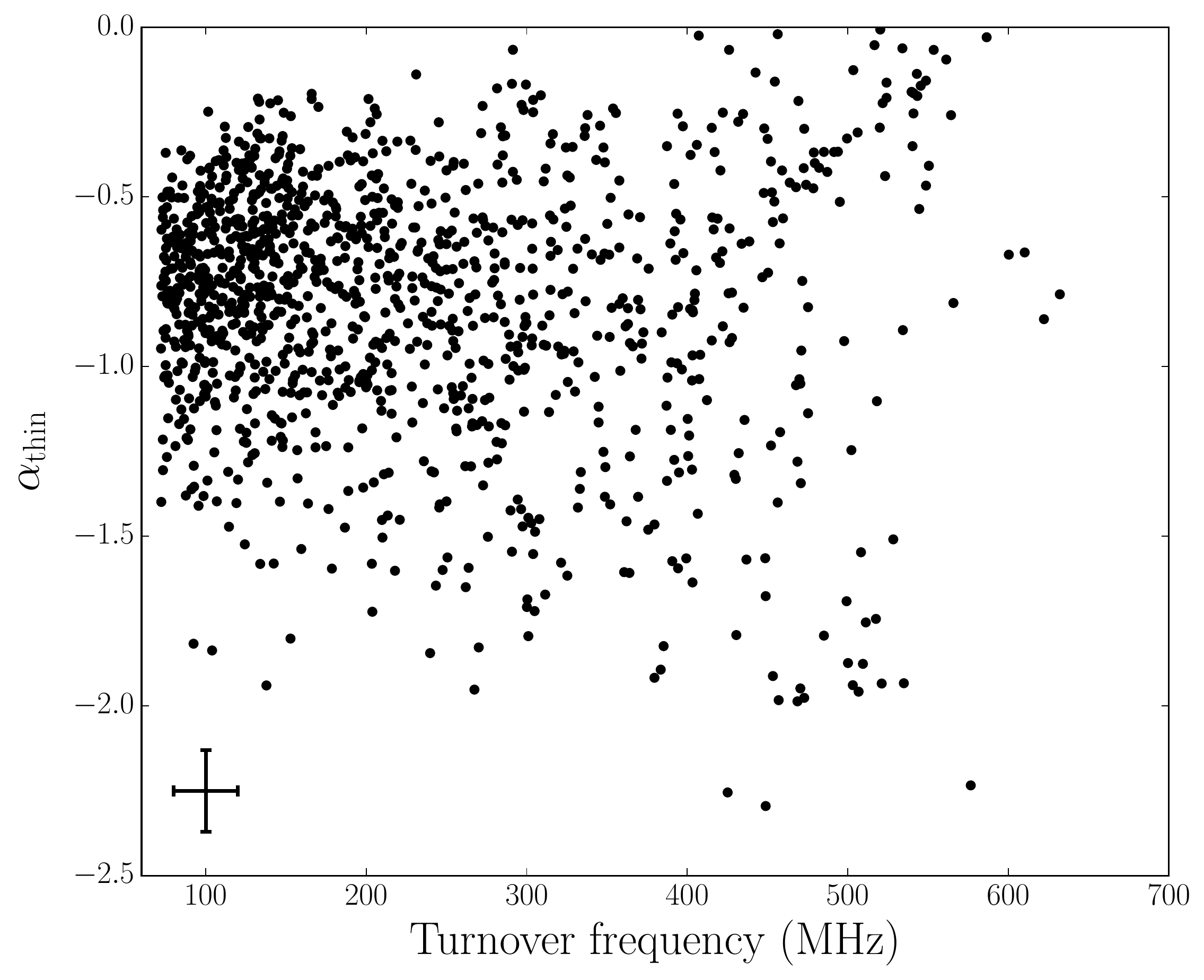} &
\includegraphics[scale=0.4]{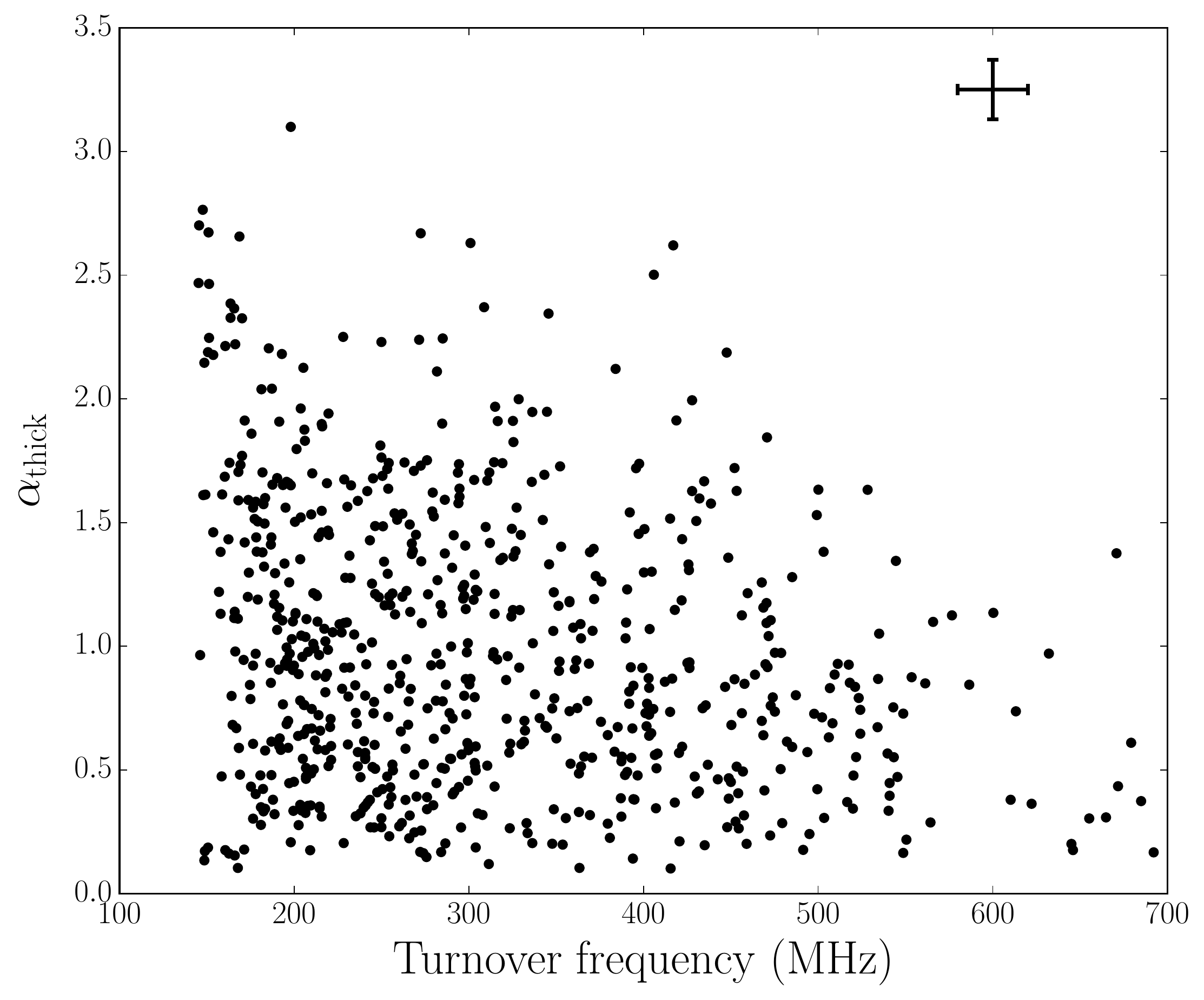} \\
\end{array}$
 \caption{The variation of the spectral turnover frequency with $\alpha_{\mathrm{thin}}$ (left panel), for the high and low frequency \ps source samples, and $\alpha_{\mathrm{thick}}$ (right panel), for the high frequency \ps source sample. While there is a trend of higher turnover frequency having steeper $\alpha_{\mathrm{thin}}$, this is a product of the selection method. $\alpha_{\mathrm{thick}}$ has no dependence on the turnover frequency, consistent with $\alpha_{\mathrm{thick}}$ being a product of the individual environments. The median uncertainties of  the values are located in the bottom left and top right corner of the plots.}
\label{fig:turnover_alpha_high_low}
\end{center}
\end{figure*}

How the distribution of the observed turnover frequencies varies with the optically thin and thick spectral indices is shown in Figure \ref{fig:turnover_alpha_high_low}. The concentration of sources at values of \athin\,$< -1$ and \athin\,$\approx$\,$-0.1$ at higher turnover frequencies is also a product of the selection method because either a turnover will be correctly detected at frequencies greater than 500\,MHz if the optically thin spectrum is steep (i.e. \athin\,$< -1$), or the NVSS/SUMSS point is sampling the spectrum near the turnover, causing the calculation of \athin~to be artificially flattened. Additionally, the optically thick spectral index shows no dependence with observed turnover frequency, as evident in the right panel of Figure \ref{fig:turnover_alpha_high_low}.

\subsection{Intrinsic spectral turnover frequency relationships}

\begin{figure}
\begin{center}
\includegraphics[scale=0.4]{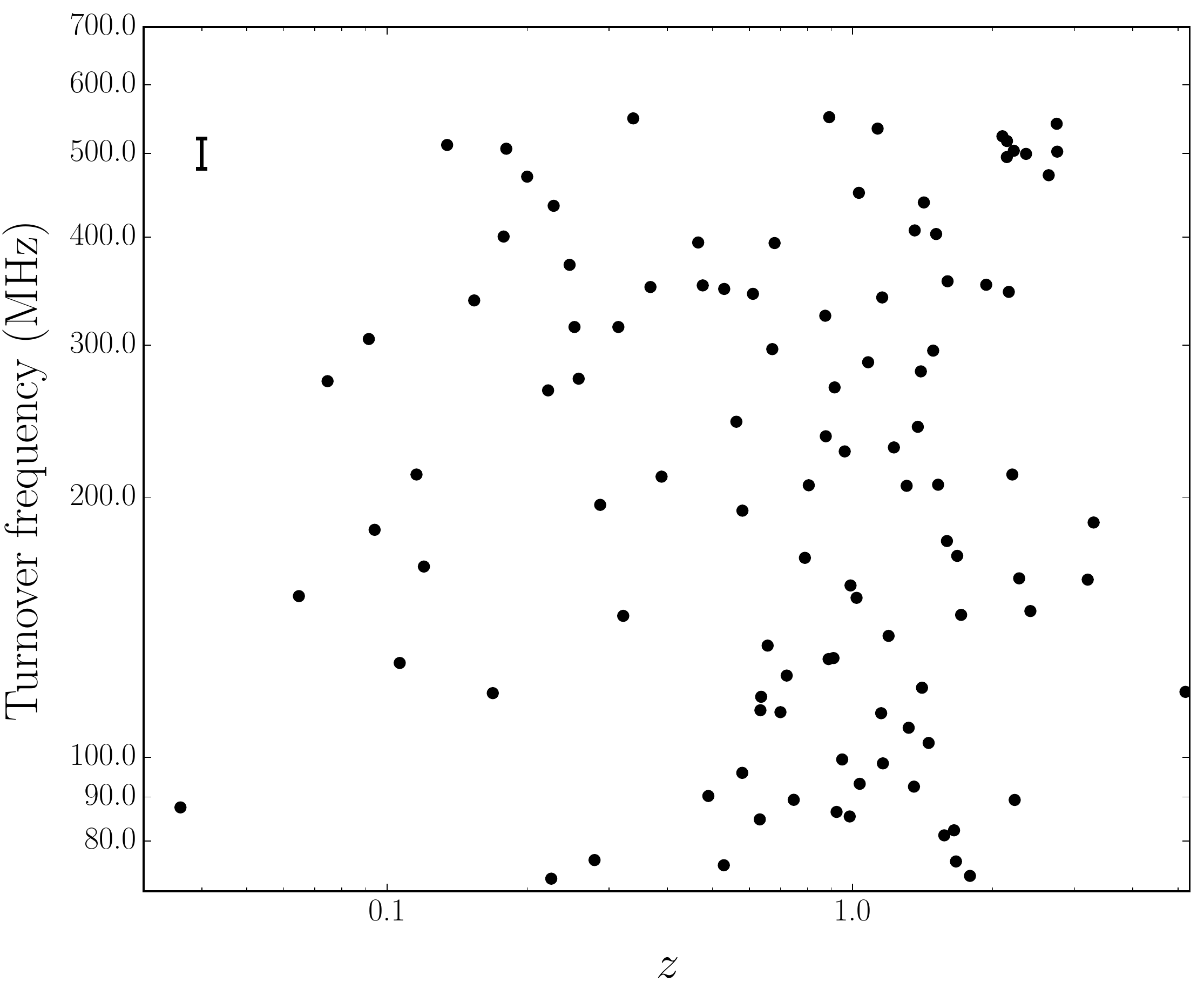}
 \caption{The variation of redshift with the observed frequency of the spectral turnover for the 110 \ps sources with reported spectroscopic redshifts and detected spectral turnovers. The random spread is consistent with the sample population being composed of both low redshift CSS sources and high redshift GPS sources. The median uncertainty of 20\,MHz in the observed turnover frequency is shown in the top left of the plot.}
\label{fig:z_turnover}
\end{center}
\end{figure}

Since the observed turnover frequency is convolved with source evolution and redshift, we repeat the same analysis presented in the previous section with the 110 sources in the high and low frequency \ps samples that have reported spectroscopic redshifts. We find no trend with the observed turnover frequency with redshift, as evident in Figure \ref{fig:z_turnover}. While \citet{deVries1997} noted that GPS sources with spectral peaks above 5\,GHz could be found preferentially at high redshift, our sample is not sensitive to these high frequency sources. The lack of a trend in turnover frequency with redshift, but with a wide spread in redshift, is consistent with the hypothesis that sources with observed turnover frequencies below 500\,MHz are composed of high redshift GPS-like sources and local CSS-like sources \citep{2016MNRAS.459.2455C}. Higher resolution low frequency imaging to measure the linear sizes of the sources is necessary to decouple the low redshift CSS-like population and high redshift GPS-like population. This will be the focus of a follow-up study. Additionally, the dearth of \ps sources at redshifts less than 0.1 is consistent with the luminosity function of GPS and CSS sources being flatter than that of large-size radio sources \citep{2000MNRAS.319..445S,2007MNRAS.375..931M,2012MNRAS.421.1569B}. 

\begin{figure}
\begin{center}
\includegraphics[scale=0.4]{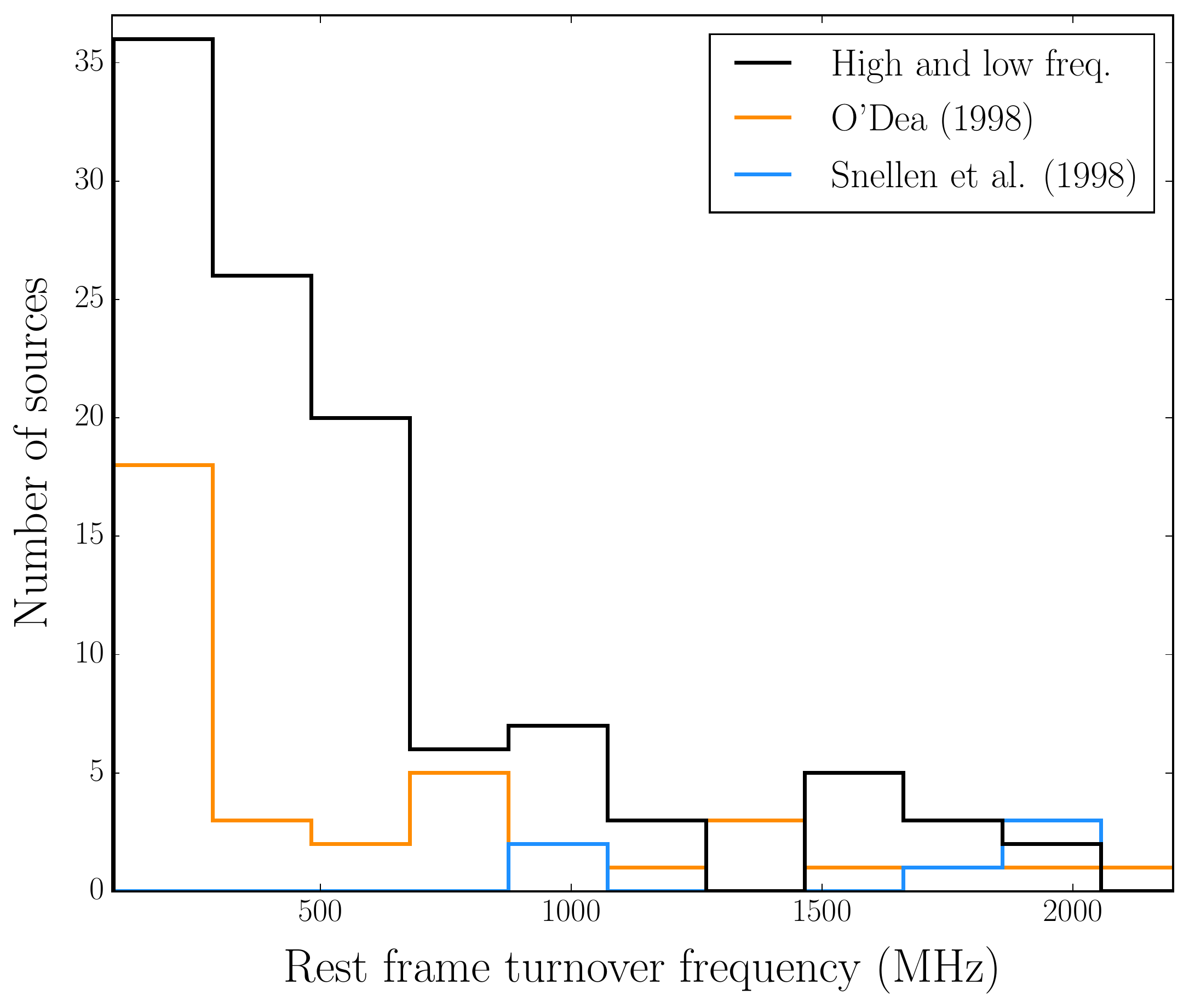}
 \caption{The distribution the rest frame turnover frequency for the 110 sources from the low and high frequency \ps samples that have reported spectroscopic redshifts, shown in black. The dark orange histogram represents the rest frame turnover frequency distribution of the GPS and CSS of the \citet{Odea1998} sample, which is mostly composed of the GPS and CSS sources identified by \citet{1998A&AS..131..303S} and \citet{1990A&A...231..333F}, respectively. In particular, the low frequency bin is dominated by the \citet{1990A&A...231..333F} CSS source sample. The blue histogram represents the distribution of the sample of \ps sources identified by \citet{1998A&AS..131..435S}. Note that the highest rest frame turnover frequency for a source in our  high and low frequency \ps samples is $\approx$\,2000\,MHz, while the rest frame turnover frequency distribution of the sources presented by \citet{Odea1998} and \citet{1998A&AS..131..435S} extend beyond 2000\,MHz.}
\label{fig:restframe_turnover_dist}
\end{center}
\end{figure}

The distribution of the rest frame turnover frequencies for the low and high frequency \ps samples is presented in Figure \ref{fig:restframe_turnover_dist}, and compared to the distributions presented by \citet{Odea1998} and \citet{1998A&AS..131..435S}. Note that the sample presented by \citet{Odea1998} is largely composed of the GPS sources studied by \citet{1998A&AS..131..303S} and of the CSS sources identified by \citet{1990A&A...231..333F}. The \ps sample has a median rest frame turnover frequency of 440$^{+560}_{-250}$\,MHz, compared to 850$^{+2500}_{-100}$ and 2840$^{+1900}_{-860}$\,MHz, for the samples identified by \citet{Odea1998} and \citet{1998A&AS..131..435S}, respectively.  Our sample has a surplus of sources with intrinsic turnover frequencies between $\approx$\,150 and 500\,MHz, compared to the other two \ps samples, consistent with the observing frequencies of the surveys used to compose our sample. 

\begin{figure*}{}
\begin{center}$
\begin{array}{cc}
\includegraphics[scale=0.37]{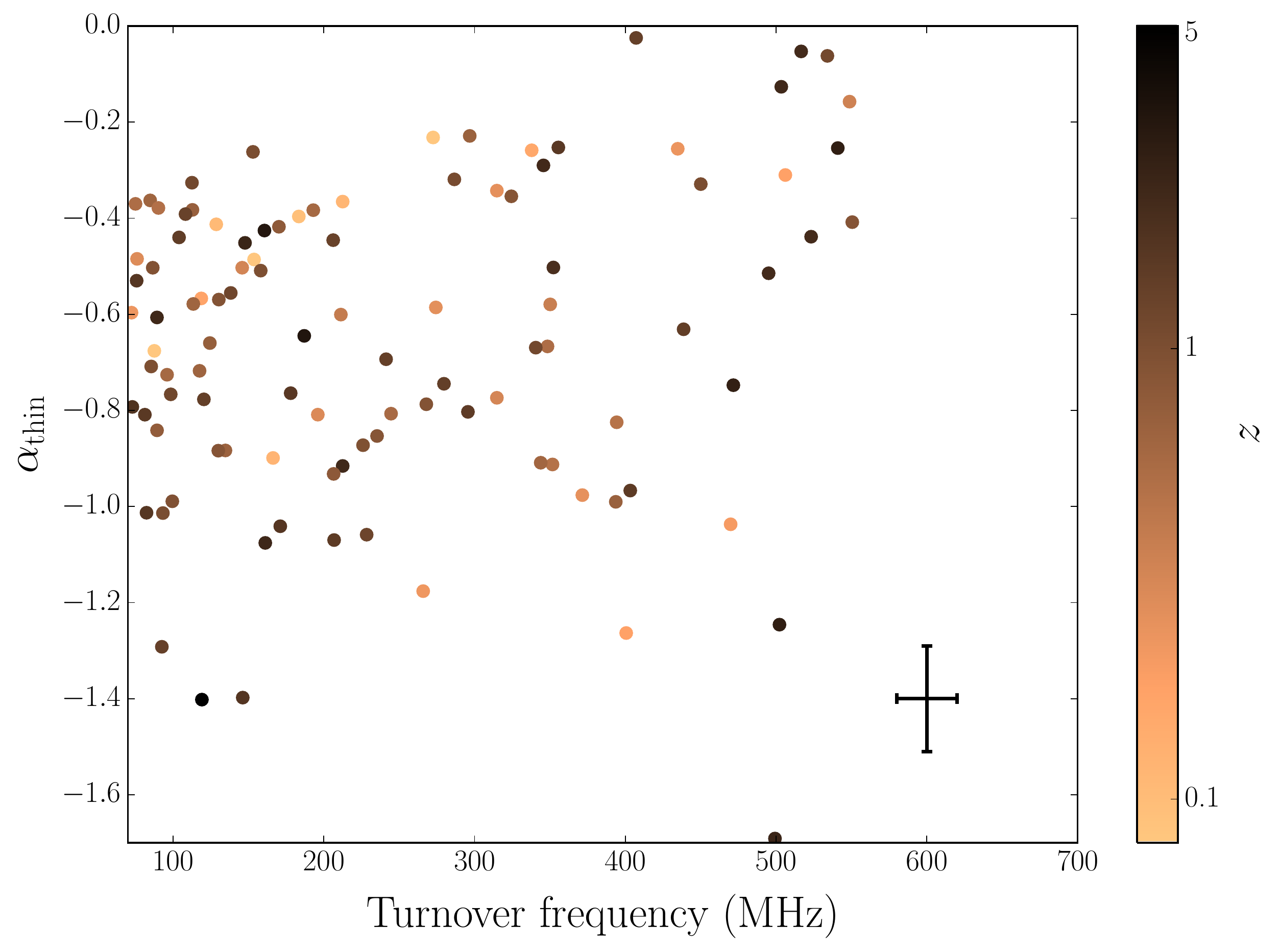} &
\includegraphics[scale=0.37]{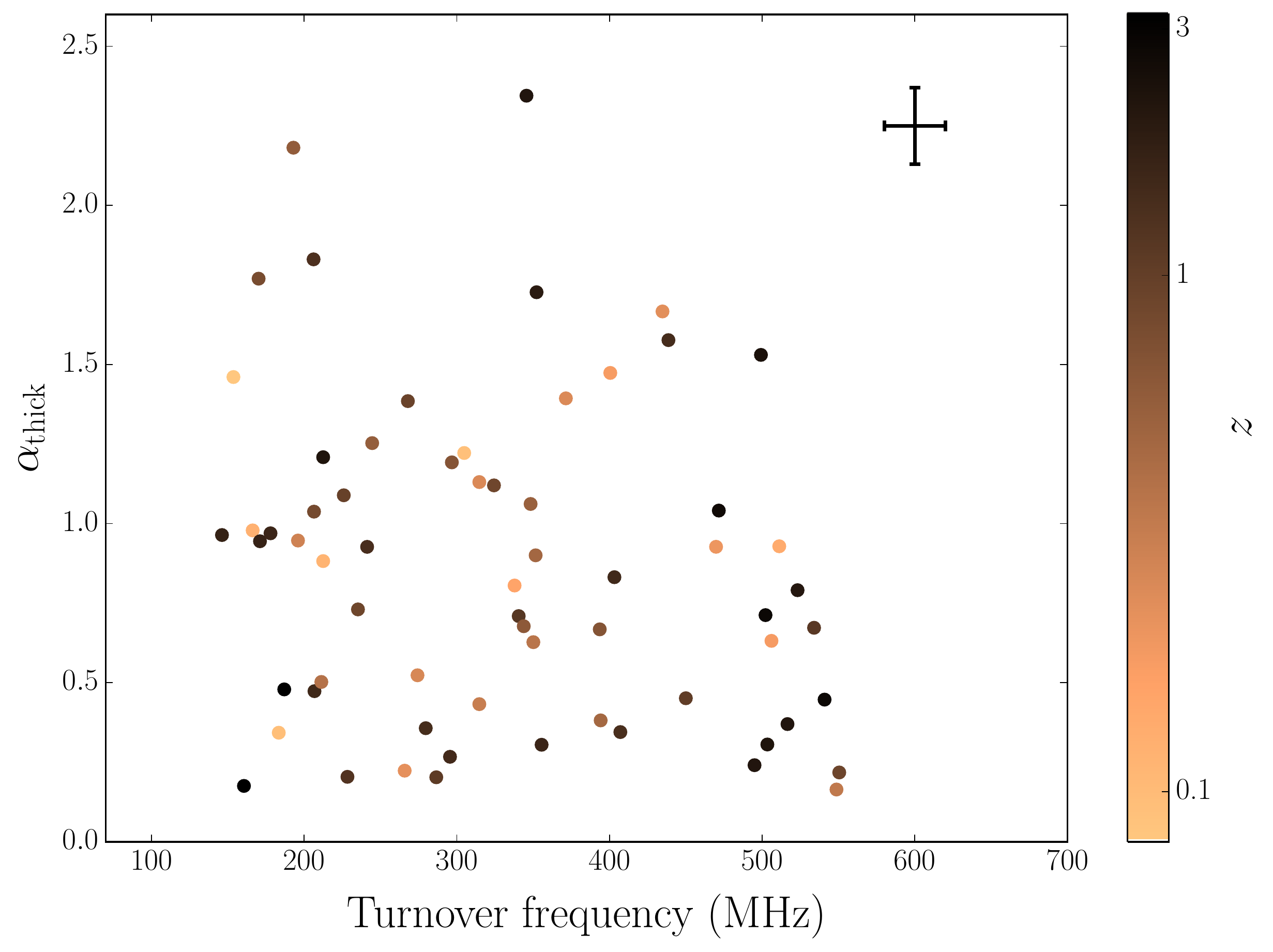} \\
\end{array}$
 \caption{Observed turnover frequency against \athin~(left panel) for both the high and low frequency \ps samples, and \athick~(right panel) for the high frequency \ps sample, that have reported spectroscopic redshifts. The color of each point corresponds to the redshift of the source. The median uncertainties  for the various parameters are plotted in the corner of the diagrams.}
\label{fig:turnover_alpha_high_low_z}
\end{center}
\end{figure*}

As suggested in \S\,\ref{sec:uss}, sources that have an optically thin spectral index $< -1$ and observed turnovers below 300\,MHz could be preferentially found at high redshifts. To test this hypothesis, the left panel of Figure \ref{fig:turnover_alpha_high_low_z} shows how \athin varies with the observed turnover frequency, colored by the redshift of the source. All sources with a turnover frequency less than 250\,MHz and \athin\,$<-1.2$ are located at $z>2$. However, there is significant amount of scatter in the dependence and a targeted optical follow-up campaign for the entire \ps sample is required to make a conclusive statement. Note that the right panel of Figure \ref{fig:turnover_alpha_high_low_z} shows no dependence of turnover frequency and the optically thick spectral index with redshift, consistent with the spectral turnover being a product of the individual environment of each source. 

\begin{figure*}
\begin{center}$
\begin{array}{cc}
\includegraphics[scale=0.4]{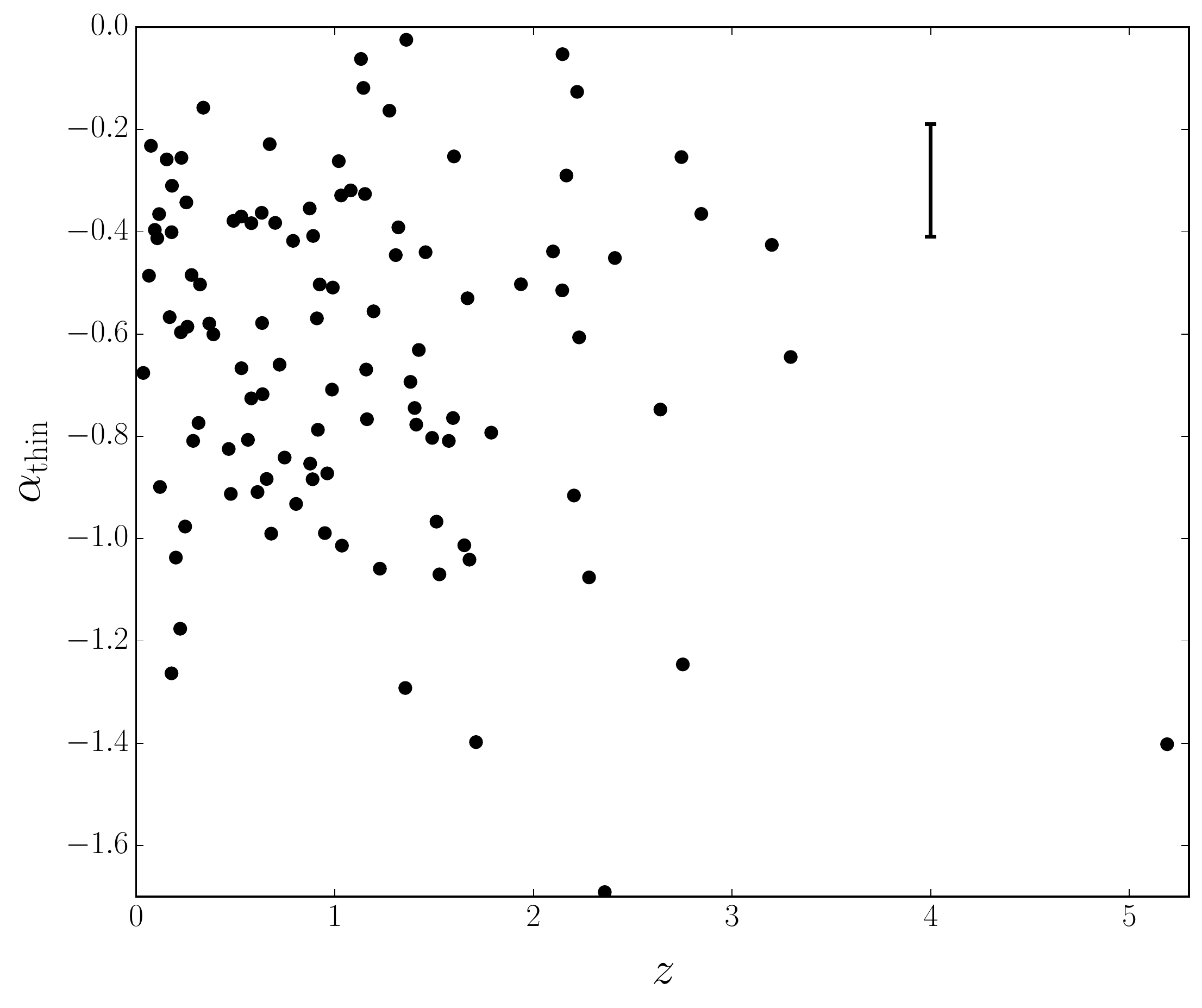} &
\includegraphics[scale=0.4]{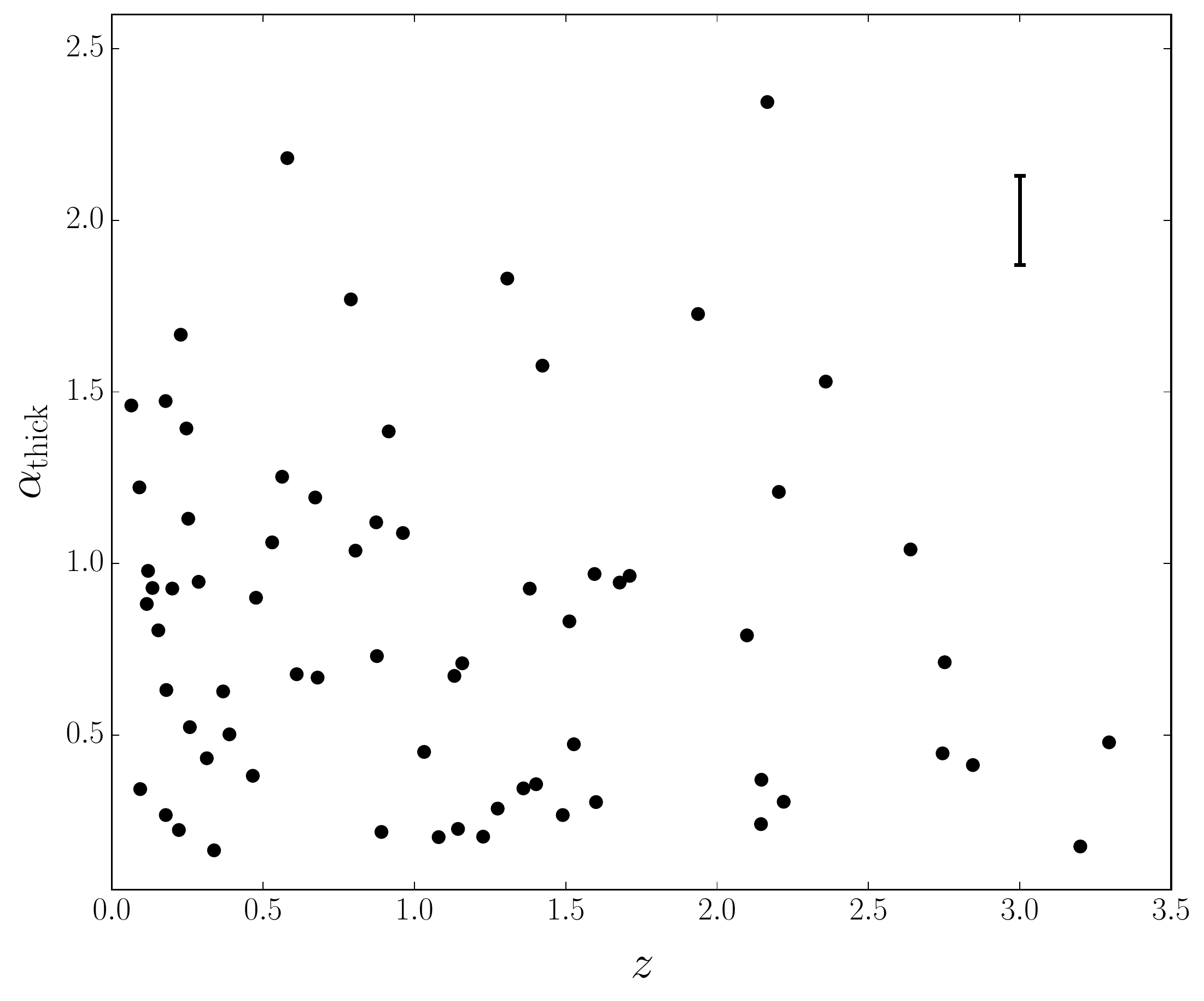} \\
\end{array}$
 \caption{The variation of redshift with \athin~(left panel), for the high and low frequency \ps samples, and \athick (right panel), for the high frequency \ps sample. The median uncertainty for the respective spectral index is plotted in the top right corner of the diagrams.}
\label{fig:z_alpha_high_low}
\end{center}
\end{figure*}

The relationships between the observed spectral indices with redshift are provided in Figure \ref{fig:z_alpha_high_low}. As shown by other authors \citep[e.g.][]{1997A&A...326..505R}, there is a slight trend of steeper optically thin slopes with higher redshift. The steepening of the optically thin spectral index has been variously explained to be due to increasing importance of inverse Compton losses off the cosmic microwave background photons in the early Universe or due to first order Fermi accelerations in the hotspots when encountering a dense medium \citep{Klamer2006}. It is also evident in the right panel of Figure \ref{fig:z_alpha_high_low} that there is no dependence of \athick~with redshift, further indicative that the slope of the optically thick spectral index is dependent on the individual physics of a source.

\subsection{5 GHz radio power relationships}
\label{sec:lumin}

We used the 110 sources from the high and low frequency \ps samples that have spectroscopic redshifts to investigate the distribution of the 5\,GHz luminosity and its dependence on redshift. The comparison frequency of 5\,GHz was chosen because past literature samples of GPS and CSS sources, such as those of \citet{Odea1998} and \citet{1998A&AS..131..435S}, had their radio luminosities evaluated at 5\,GHz. The 5\,GHz radio luminosities $P_{5 \, \mathrm{GHz}}$ was calculated as

\begin{equation}\label{eqn:lumin}
	P_{5 \, \mathrm{GHz}} = 4 \pi D_{L}^{2} S_{5\,\mathrm{GHz}} (1+z)^{-(1+\alpha_{\mathrm{thin}})},
\end{equation}

\noindent where $D_{L}$ is the luminosity distance in the adopted cosmological model, $S_{5\,\mathrm{GHz}}$ is the flux density of the source at 5\,GHz, and the term $(1+z)^{-(1+\alpha_{\mathrm{thin}})}$ is the $k$-correction used in radio astronomy. $S_{5\,\mathrm{GHz}}$ was evaluated by assuming the fit of Equation \ref{eqn:gen}, in particular \athin~derived from that fit, is an accurate description of a \ps source's spectrum to 5\,GHz. Note that it is possible that the spectrum of a source could deviate from the power-law description identified between the peak frequency and 843\,MHz\,/\,1.4\,GHz by 5\,GHz, either due to a high frequency spectral break or because the emission from the radio core begins to exceed the radio lobe emission. However, such deviations are not expected to be significant until frequencies greater than 5\,GHz \citep{2012MNRAS.422.2274C}.

\begin{figure}
\begin{center}
\includegraphics[scale=0.4]{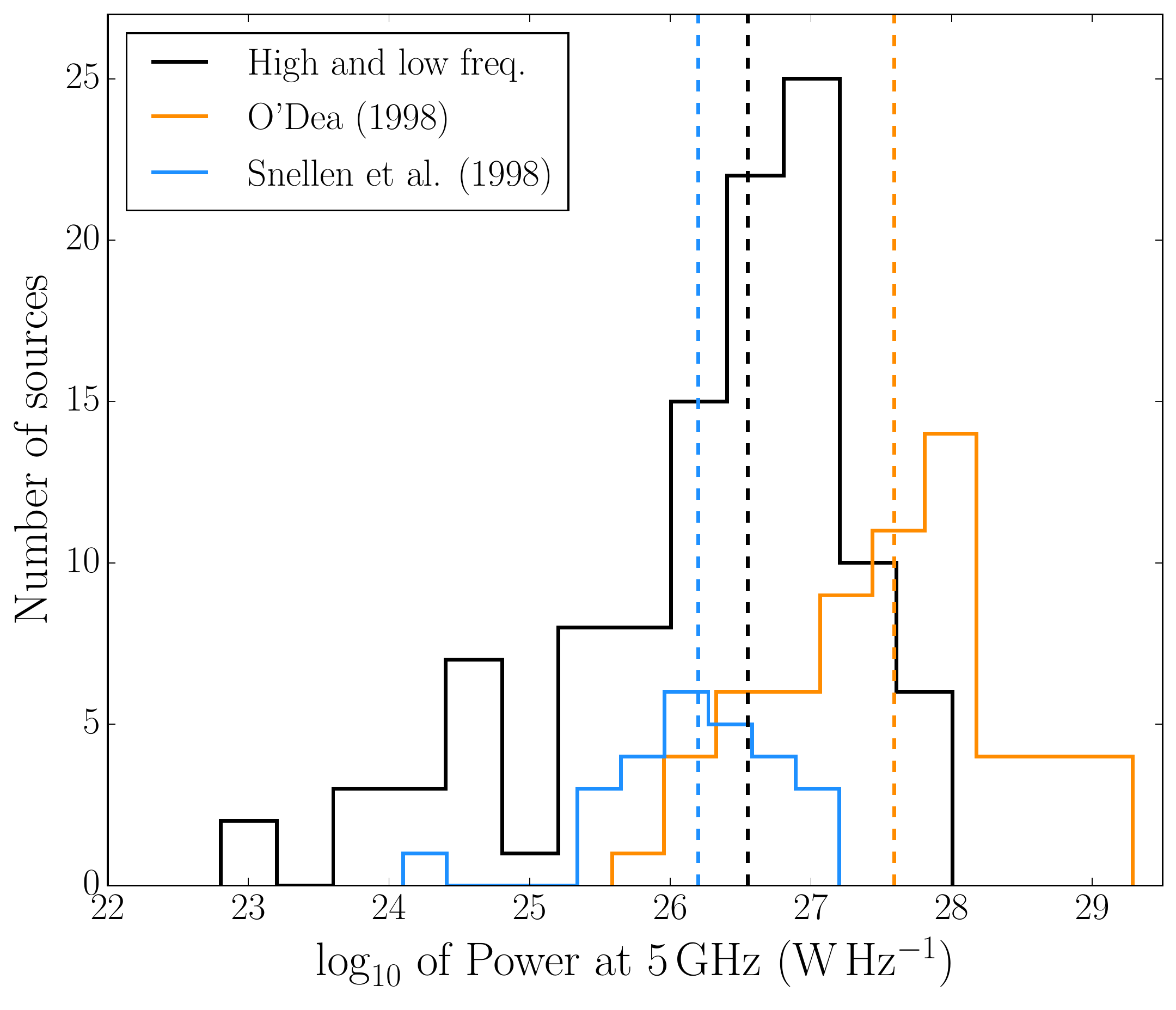}
 \caption{Distribution of the 5\,GHz radio power for the 110 sources from the low and high frequency \ps samples that have reported spectroscopic redshifts is shown in black. The orange and blue histograms represent the 5\,GHz power distribution of the \citet{Odea1998} sample and \citet{1998A&AS..131..435S} sample, respectively. The median $P_{5 \, \mathrm{GHz}}$ values, and the range to the 16$^{\mathrm{th}}$ and 84$^{\mathrm{th}}$ percentiles of the distribution, for samples identified by \citet{Odea1998}, \citet{1998A&AS..131..435S}, and this study, are 27.6$^{+0.7}_{-1.0}$, 26.2$^{+0.5}_{-0.4}$, and 26.5$^{+0.7}_{-1.2}$ ($\log_{10}$\,W\,Hz$^{-1}$), respectively. The median values are plotted as dashed lines in each respective sample's color.}
\label{fig:power_dist}
\end{center}
\end{figure}

\begin{figure}
\begin{center}
\includegraphics[scale=0.4]{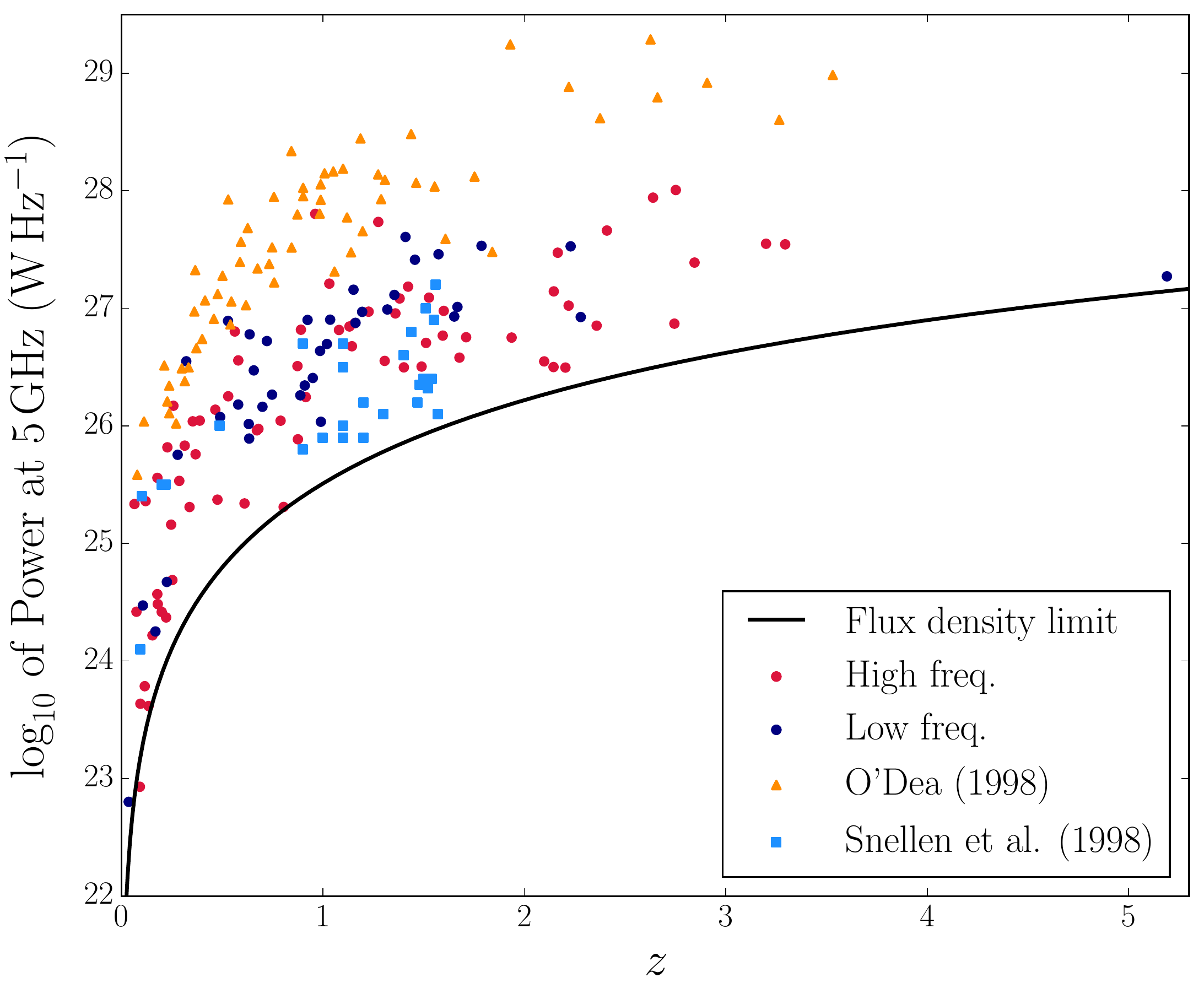}
 \caption{5\,GHz radio power against redshift for the 110 sources from the low and high frequency \ps samples that have reported spectroscopic redshifts. The sources from the high and low frequency samples are shown as red and navy circles, respectively. The orange triangles and light blue squares represent the GPS and CSS sources from the \citet{Odea1998} and \citet{1998A&AS..131..435S} samples, respectively. The black line corresponds to the 5\,GHz luminosity limit for a source that has a peak flux density of 0.16\,Jy at 230\,MHz, approximately the flux density cut employed in the selection process, assuming the median optically thin spectral index of $-0.77$.}
\label{fig:z_v_power_dist}
\end{center}
\end{figure}

The distribution of the 5\,GHz radio power for the high and low frequency \ps samples, and how it varies with redshift, is provided in Figure \ref{fig:power_dist} and Figure \ref{fig:z_v_power_dist}, respectively. The sample presented by \citet{1998A&AS..131..435S}, as shown by the blue histogram and squares, has sources with similar 5\,GHz radio luminosities as the \ps samples presented in this study. However, our low and high frequency \ps sample span a wider range in luminosity than the sample identified by \citet{1998A&AS..131..435S}. Additionally, our low frequency \ps sample contains the weakest \ps source ever identified with $P_{5 \, \mathrm{GHz}} = 6.3 \times 10^{22}$ W\,Hz$^{-1}$.

When compared to the GPS and CSS sample presented by \citet{Odea1998}, as shown by the orange histogram and triangles in Figures \ref{fig:power_dist} and \ref{fig:z_v_power_dist}, the \ps sources identified in this study are on average an order of magnitude fainter at 5\,GHz, and there are no \ps sources in the high or low frequency samples that exceed a 5\,GHz power of $1 \times 10^{28}$ W\,Hz$^{-1}$. The selection criteria employed in this study do not bias against selecting sources at $P_{5 \, \mathrm{GHz}} > 10^{28}$ W\,Hz$^{-1}$. Therefore, the low and high frequency \ps samples lack the similar highly luminous sources found in the \citet{Odea1998} sample because of evolutionary or redshift effects, or some combination of both. 

\begin{figure}
\begin{center}
\includegraphics[scale=0.35]{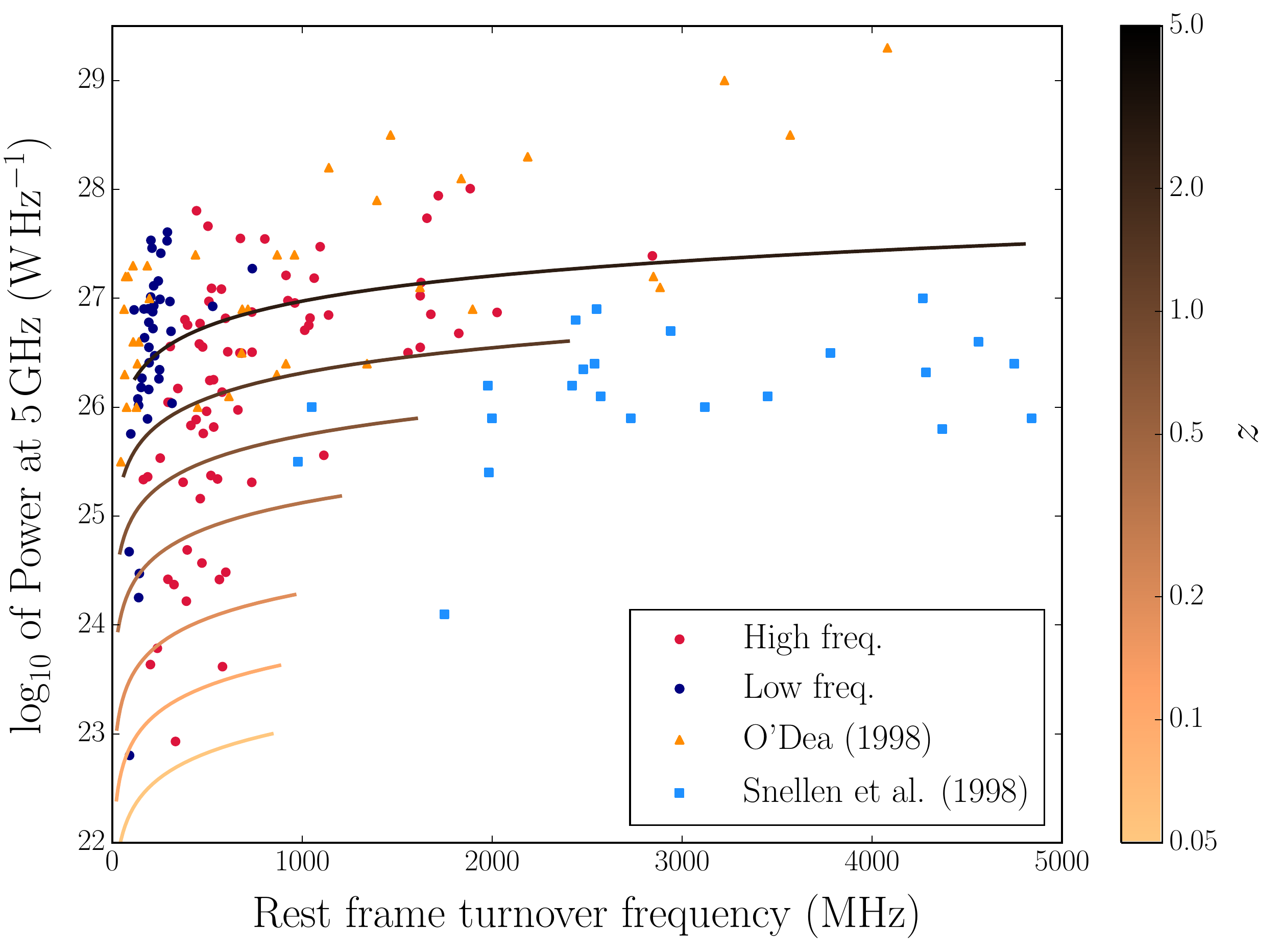}
 \caption{The variation of the 5\,GHz radio power with the rest frame turnover frequency for the high and low frequency \ps samples, shown by red and navy circles, respectively. The orange triangles and blue squares represent the GPS and CSS sources from the \citet{Odea1998} and \citet{1998A&AS..131..435S} samples, respectively. The colored curves are the limit in which \ps sources at various redshifts will be identified in this study, assuming the minimum peak flux density of 0.16\,Jy, since the selection process is only sensitive to turnovers between 72 and $\approx$\,800\,MHz. The color bar corresponds to the redshift at which the limit is evaluated, with each curve from bottom to top corresponding to redshifts 0.05, 0.1, 0.2, 0.5, 1, 2, and 5. Note that a number of sources from the \citet{Odea1998} and \citet{1998A&AS..131..435S} samples extend to rest frame frequencies above 5000\,MHz.}
\label{fig:intrinsic_turnover_v_radio_power}
\end{center}
\end{figure}

It is expected from various evolutionary models of CSS and GPS sources \citep[e.g.][]{2000MNRAS.319..445S,2010MNRAS.408.2261K} that the luminosity of a source declines with increasing linear size, which also corresponds to a shift of the intrinsic turnover frequency to lower frequencies. The dependence of the 5\,GHz luminosity for the low and high frequency \ps samples on rest frame turnover frequency, also including the samples presented by \citet{1998A&AS..131..435S} and \citet{Odea1998}, is presented in Figure \ref{fig:intrinsic_turnover_v_radio_power}. Note that the general shape of the distribution, with no sources with low power and high rest frame turnover in the high or low frequency \ps samples, is a product of redshift evolution of the flux density limit and also results from the selection process which is only sensitive to \ps sources that display an observed spectral turnover between 72 and $\approx$\,800\,MHz. Such redshift evolution of the flux density and turnover frequency limits is communicated by the colored curves in Figure \ref{fig:intrinsic_turnover_v_radio_power}.

It is promising that there are a number of sources from the low and high frequency \ps samples that have $P_{5 \, \mathrm{GHz}} < 10^{25}$ W\,Hz$^{-1}$ and rest frame turnover frequencies less than 1\,GHz, consistent with the evolutionary path of a self-similar evolving, ram pressure confined radio source \citep{1997MNRAS.286..215K,2000MNRAS.319..445S}. However, there is a large number of \ps sources that have $P_{5 \, \mathrm{GHz}} > 10^{26}$ W\,Hz$^{-1}$ and rest frame turnover frequencies less than 1\,GHz, which is inconsistent with this evolutionary model for \ps sources. It is possible the high power, low intrinsic turnover sources could be confined to small spatial scales due to a dense ambient medium, or they could be jet-dominated sources, as opposed to the sources at lower power being lobe-dominated. Future studies are required to test if the optical hosts of these high power, low intrinsic turnover sources are quasars or if their linear sizes are small for the respective turnover frequency, as expected from SSA theory.

It is also possible that the surplus of sources with high power and low intrinsic turnover frequency could be the product of the method of assembling the redshift information of the \ps sources from inhomogeneous literature sources, each with varying degrees of completeness. To confirm the trend of lower power with lower intrinsic turnover frequencies, it is vital to repeat this analysis using a spectroscopic redshift survey with known completeness, such as 6dFGS and SDSS, and to then produce a local luminosity function.

\section{Extreme spectra}
\label{sec:extreme_spec}

\begin{figure*}
\begin{center}$
\begin{array}{cccc}{}
\includegraphics[scale=0.32]{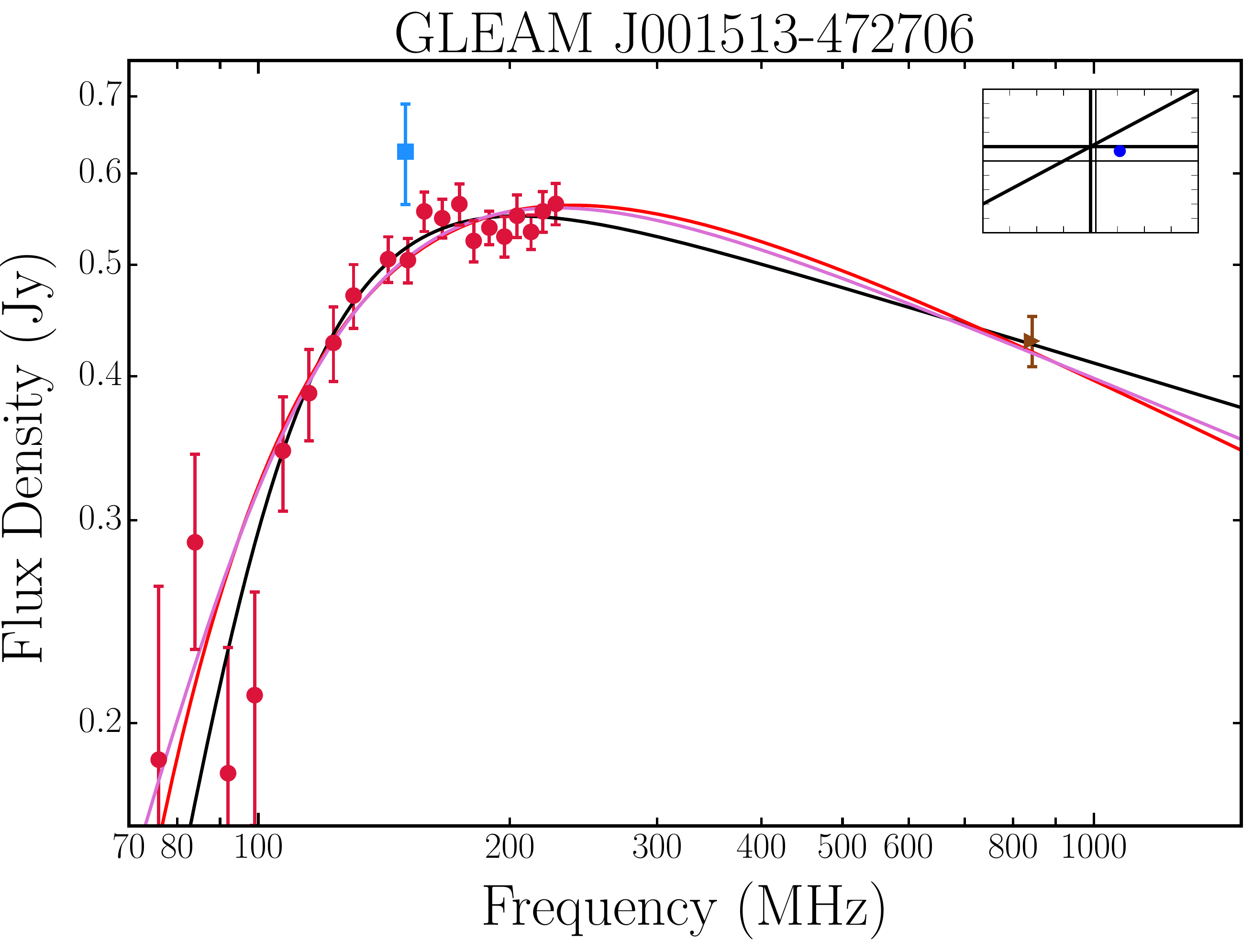} & 
\includegraphics[scale=0.32]{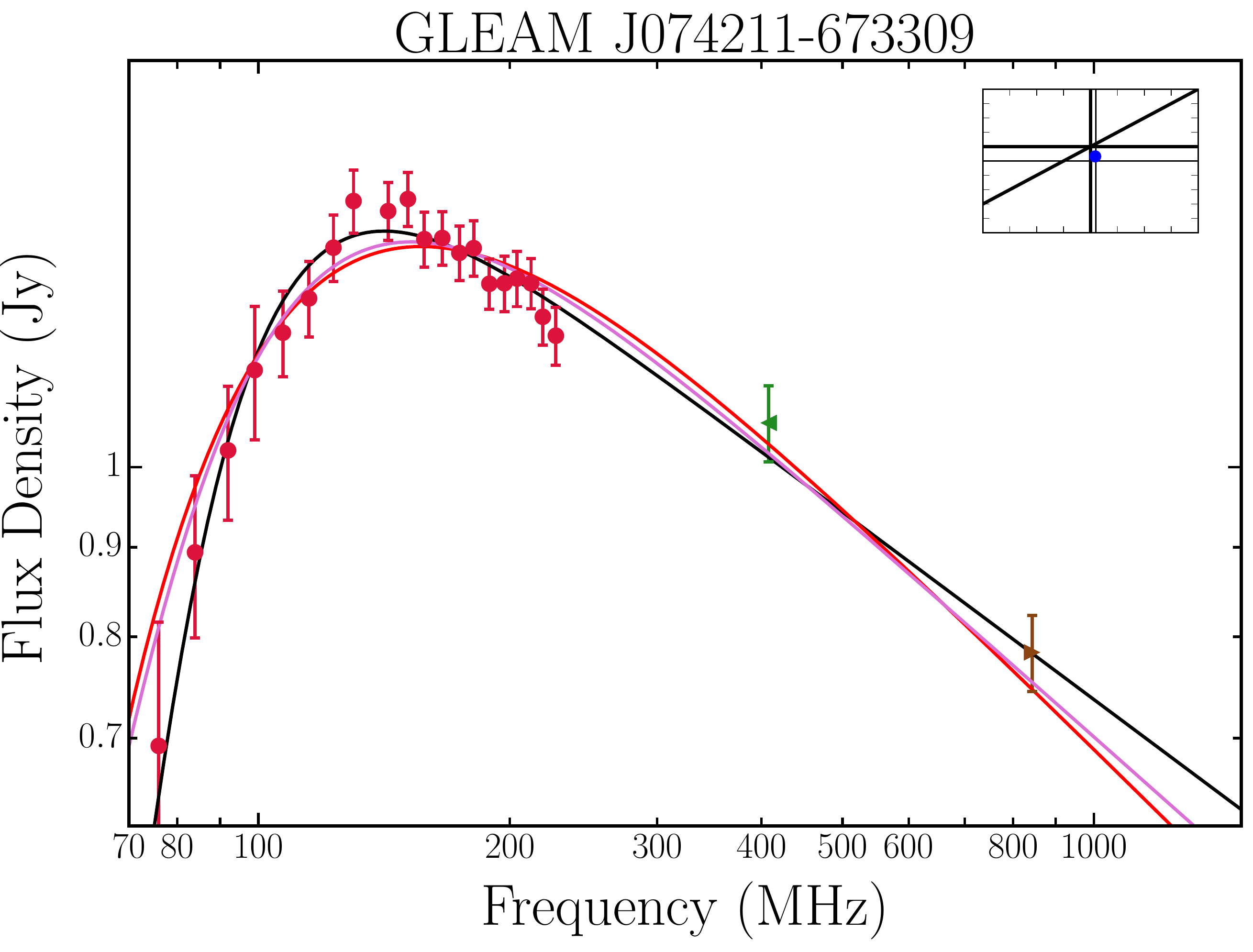} \\ 
\includegraphics[scale=0.32]{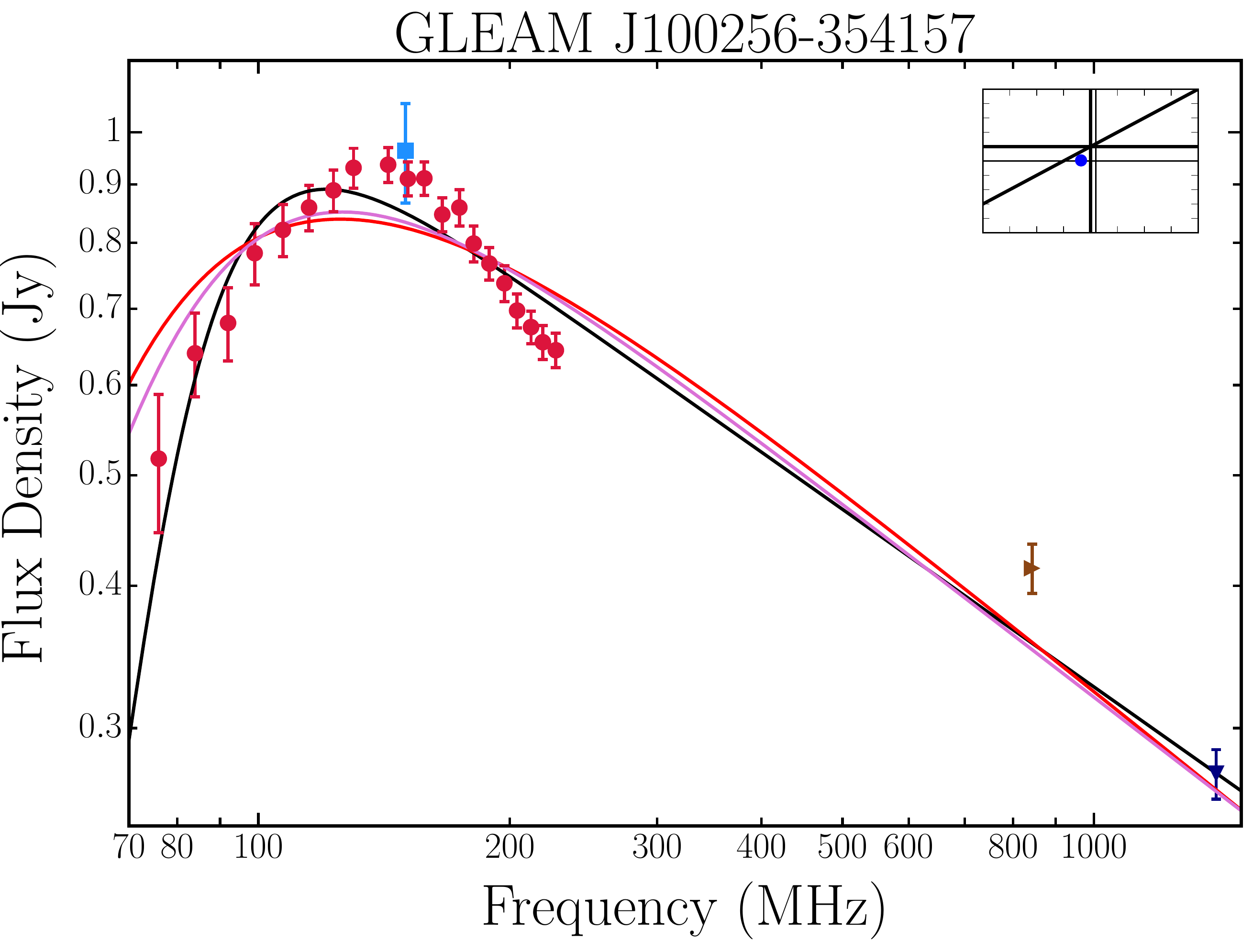} & 
\includegraphics[scale=0.32]{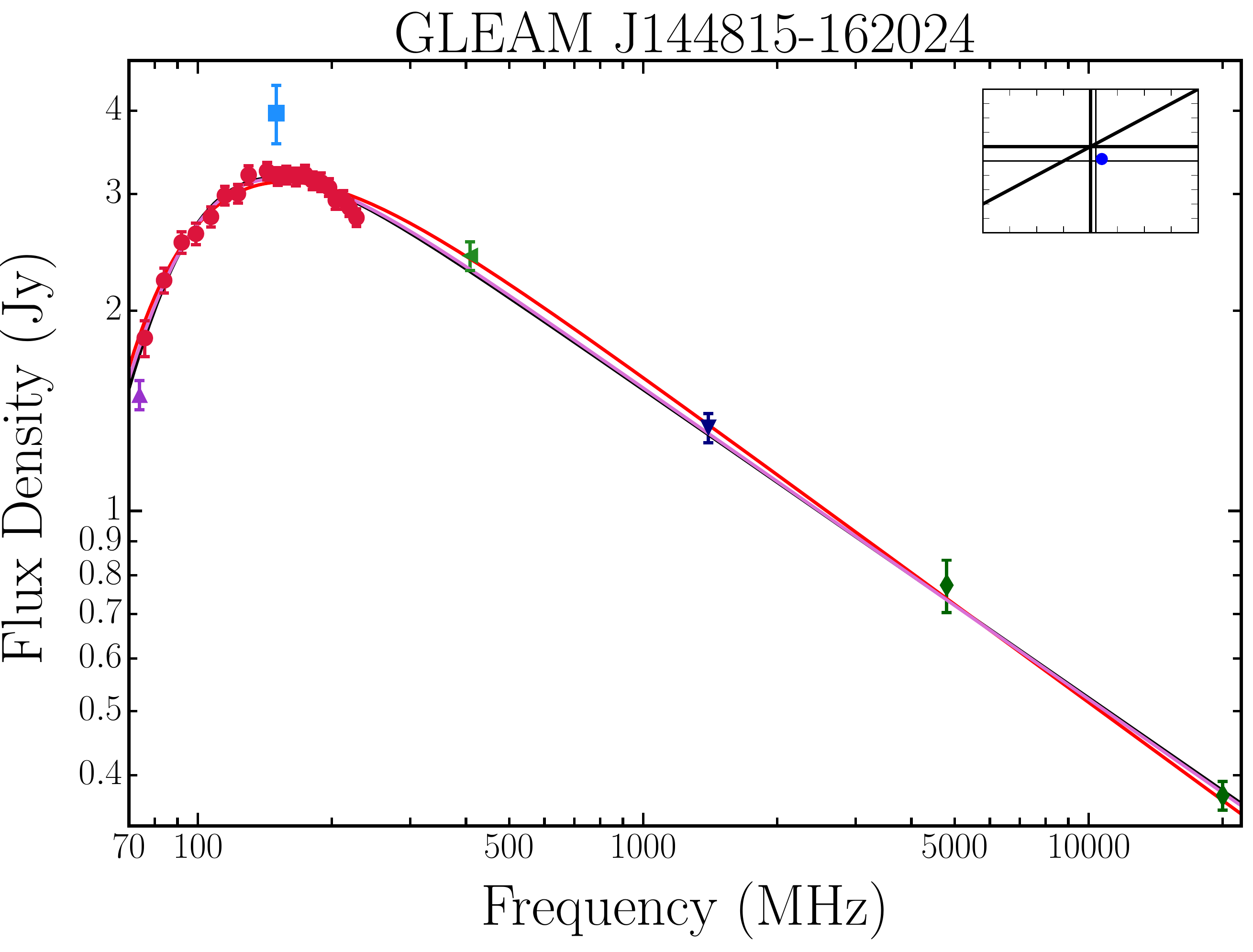} \\
\includegraphics[scale=0.32]{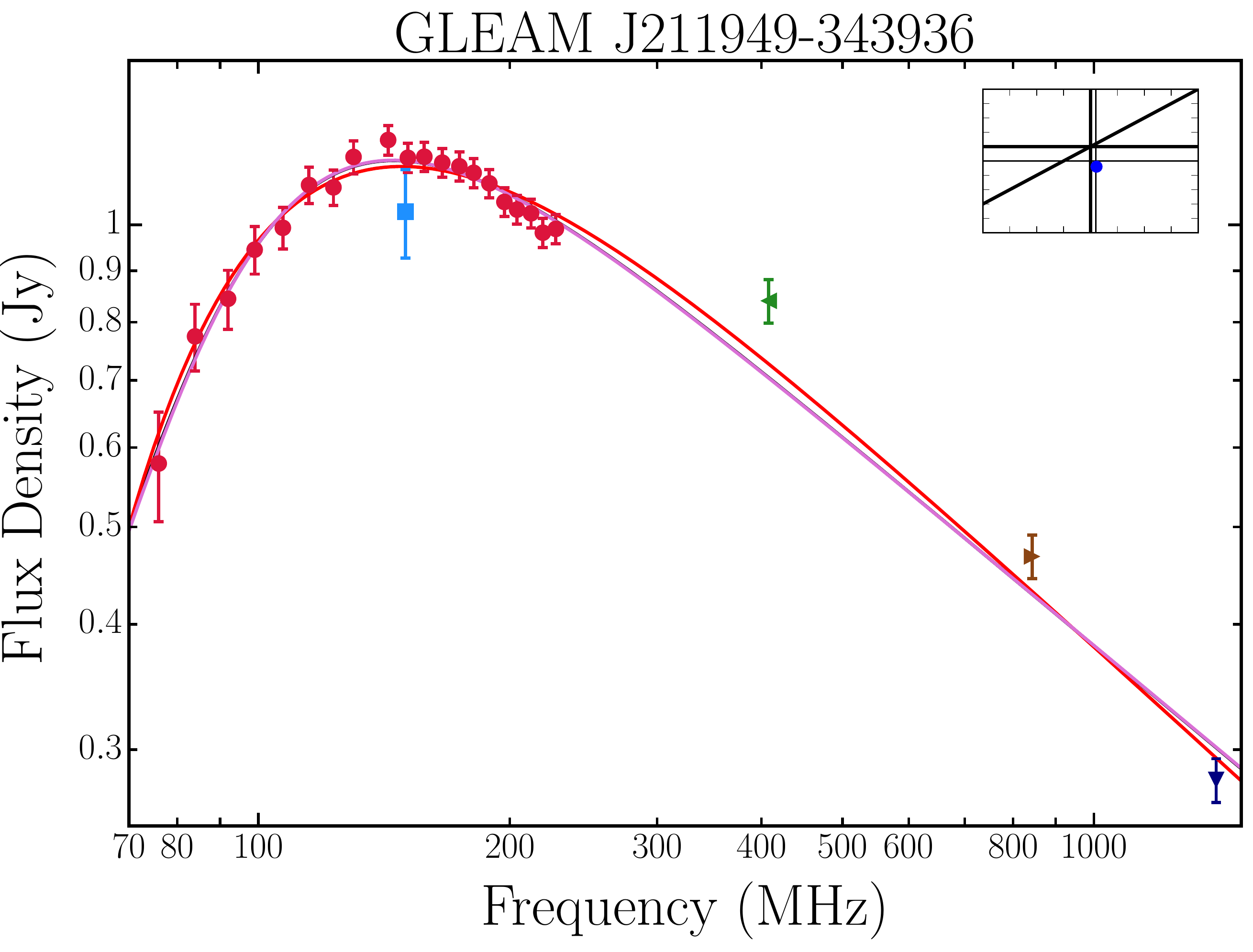} & 
\includegraphics[scale=0.32]{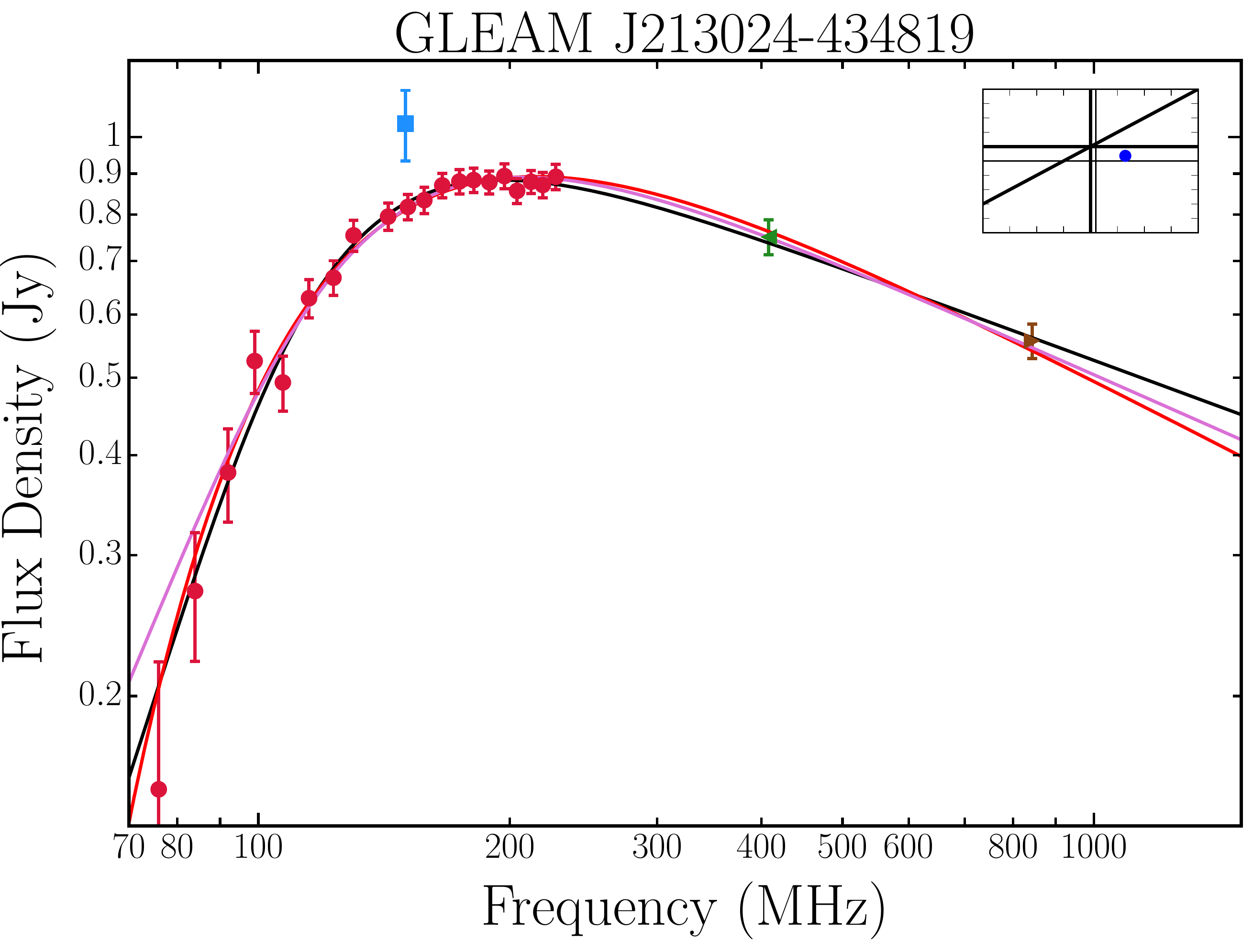} \\ %
\end{array}$
 \caption{The spectra of six extreme \ps sources that have $\alpha_{\mathrm{thick}} \gtrsim 2.5$. The symbols represent the same survey data as in Figure \ref{fig:example_spectra}. The black, purple, and red curves represent the fit of the general curved model described by Equation \ref{eqn:gen}, homogeneous SSA, and the single homogeneous FFA model of Equation \ref{eqn:ffa}, respectively. In all cases, excluding GLEAM~J100256-354157, the FFA model is the statistically favored model. Note that the plot for GLEAM~J144815-162024 extends to 20\,GHz since it has AT20G measurements represented by the green diamonds. The inset plot located in the top-right corner of each plot displays the position of the source in the color-color diagram of Figure \ref{fig:color_color} by a blue circle.}
\label{fig:extreme_spectra}
\end{center}
\end{figure*}

As is evident from the right panel of Figure \ref{fig:dist_alpha_thick_thin}, there are approximately fifteen sources that have a reliable optically thick spectral index near or above the theoretical SSA limit of 2.5 \citep{Pacholczyk1970}. The spectra for six of the brightest sources that have an optically thick spectral index near or exceeding 2.5 are shown in Figure \ref{fig:extreme_spectra}, with the fit of the general curved model, SSA, and homogeneous FFA model shown in black, purple, and red, respectively.

The values of $\alpha_{\mathrm{thick}}$ and the peak frequency $\nu_{\mathrm{p}}$ from Equation \ref{eqn:gen} for the six sources are provided in Table \ref{table:extreme_spectra}. Due to the steepness of the spectral index below the turnover, we also fit the homogeneous FFA model of Equation \ref{eqn:ffa} since it provides exponential attenuation at low frequencies, and a SSA model by forcing $\alpha_{\mathrm{thick}} = 2.5$ in Equation \ref{eqn:gen}. The reduced $\chi^{2}$-value for the model fits are also provided in Table \ref{table:extreme_spectra}. While the SSA model provides similar fitting statistics, the FFA model is statistically favoured over the SSA model for all the sources, excluding GLEAM~J100256-354157. None of these sources have been previously identified as \ps sources and most lack optical or higher radio frequency counterparts.

If these \ps sources are confirmed to have similar physical morphologies as the HFP, GPS and CSS sources which are characterized as young, such as core-lobe structures on milliarcsecond scales, the spectra presented in Figure \ref{fig:extreme_spectra} could be the first conclusive evidence of FFA acting in previously identified young sources. Such a detection of FFA absorption would unambiguously imply that at least a portion of the HFP, GPS and CSS population is confined to small spatial scales due to a dense ambient nuclear medium.

Alternatively, these sources could be jet-dominated, with the FFA absorption occurring as the radio emission from the jet passes through a dense circumnuclear medium. Such a dense circumnuclear medium, possibly associated with the broad line region \citep{1998A&A...330...79L}, has been suggested to exist in Centaurus A \citep{1996ApJ...466L..63J}, Cygnus A \citep{1998A&A...329..873K,2001ApJ...546..210T}, and 3C\,84 \citep{1994ApJ...430L..45W}, because the counter-jets in these sources are significantly fainter than the forward jet or not observed at all. The only source that has a AT20G and optical counterpart, GLEAM~J144815-162024, is associated with a quasar \citep{2011MNRAS.417.2651M} and has a linear size less than 0.5\,kpc based on its 6\,km 20\,GHz visibilities \citep{2013MNRAS.434..956C}. Jet-dominated sources have been shown to display low-frequency variability on time-scales of several months \citep{2014MNRAS.438..352B}, so it is vital to assess if these sources are variable before fitting multi-epoch data since the trends induced by variability may dominate physical relationships. 

Finally, it is possible that such spectra are the product of an extreme scattering event (ESE), since ESEs haven been shown to produce irregular spectra and steep spectral slopes \citep[e.g.][]{2016Sci...351..354B}. An ESE is likely due to the propagation of a compact source's radio emission through dense plasma lenses in the Galaxy \citep{1994ApJ...430..581F}. The densities required to produce such extreme lensing are inconsistent with our current understanding of gas conditions in the Milky Way. In particular, GLEAM~J100256-354157 is more likely a potential ESE, over the other sources presented in Figure \ref{fig:extreme_spectra}, as the spectrum is even inconsistent with FFA absorption. However, the spectral properties of ESEs at low frequencies are unknown, with limited variability seen below 1\,GHz \citep{2016Sci...351..354B}. The low frequencies provided by the GLEAM survey could place the most stringent constraints on the density of the inter-stellar medium if a source is shown to be experiencing an ESE.

During the first year of the GLEAM survey, many sources were observed on multiple occasions due to the wide field of view of the MWA and the highly degenerate observing strategy. This allows us to put constraints on the variability of some of these sources below 231\,MHz. GLEAM~J001513-472706, GLEAM~J074211-673309, and GLEAM~J144815-162024 were observed by the MWA twice three-months apart, and GLEAM~J211949-343936, GLEAM~J213024-434819 were observed twice nine-months apart. There is no evidence of variability in the low frequency spectra for any of the sources. Unfortunately, the best ESE candidate GLEAM~J100256-354157, was only observed once during the GLEAM survey.

Independent of the nature of these sources, they represent excellent candidates for targeted follow-up to observe intrinsic H\,\textsc{i} absorption due to the inference of a dense ambient medium from the spectrum. To assess the nature of these sources, low frequency monitoring to constrain any variability, and high resolution observations with LOFAR or the GMRT, are necessary. We note that the optically thick slope is completely determined from the contemporaneous GLEAM data, so the extreme spectral slope can not be due to variability. In particular, observations with the ATCA or VLA above the turnover will be important to ensure the spectral width is properly sampled, providing an additional parameter for model differentiation, as shown by \citet{Callingham2015}, and confirmation of the extreme spectral slope.

\begin{table}
	\small
	\caption{\label{table:extreme_spectra} A list of the parameters derived from the spectral fits of the general curved model, SSA, and FFA models to the specra presented in Figure \ref{fig:extreme_spectra}. The parameters listed in the table are: the optically thick spectral index $\alpha_{\mathrm{thick}}$ from the fit of Equation \ref{eqn:gen}, spectral turnover frequency $\nu_{\mathrm{p}}$ from the fit of Equation \ref{eqn:gen}, and the reduced $\chi^{2}$-value of the model fit for FFA and SSA $\chi^{2}_{\mathrm{red,FFA}}$ and $\chi^{2}_{\mathrm{red,SSA}}$, respectively.}
	\begin{center}
		\begin{tabular}{ccccc}
		\hline
		\hline
GLEAM name & $\alpha_{\mathrm{thick}}$ & $\nu_{\mathrm{p}}$\,(MHz) & $\chi^{2}_{\mathrm{red,SSA}}$ & $\chi^{2}_{\mathrm{red,FFA}}$ \\
		\hline			
J001513-472706 & $3.7 \pm 1.1$ & $240 \pm 20$ & 1.31 & 1.24 \\
J074211-673309 & $4.1 \pm 0.9$ & $160 \pm 30$ & 1.06 & 0.81\\
J100256-354157 & $5.0 \pm 1.7$ & $110 \pm 40$ & 5.40 & 4.52\\
J144815-162024 & $2.6 \pm 0.2$ & $160 \pm 10$ & 1.10 & 0.80 \\
J211949-343936 & $2.5 \pm 0.3$ & $150 \pm 20$ & 1.29 & 1.25\\
J213024-434819 & $3.2 \pm 0.6$ & $220 \pm 30$ & 0.80 & 0.70\\
		\hline\end{tabular}                           
\end{center}                                                                               
\end{table}

\section{Conclusions}\label{conclusions}

In this study, we identify 1,483 \ps sources that display a spectral turnover between 72\,MHz and 1.4\,GHz, representing $\approx$\,4.5\% of the GLEAM radio source population. A source was selected as \ps if it either displayed a spectral peak within the GLEAM band, a concave spectrum in radio color-color phase space when a SUMSS or NVSS flux density measurement was added to the spectrum, or a positive power-law that extended from the top of the GLEAM band to a SUMSS or NVSS counterpart. The \ps candidates were also required to be unresolved in the GLEAM extragalactic catalog, located between the declinations of $-80\degree$ and $+30\degree$, and brighter than 0.16\,Jy in the GLEAM wideband image, which is centered at 200\,MHz. Utilizing the wide fractional bandwidth of the GLEAM survey as a basis of identifying \ps sources implies the observations at and below the spectral peak are minimally impacted by variability.

All of the 73 known GPS, CSS, and HFP sources that have a spectral turnover between 72\,MHz and 1.4\,GHz, and are bright enough to be detected in the GLEAM survey, are contained within our \ps sample, demonstrating the high reliability of the selection criteria. This study provides 1,410 new \ps sources, representing over a factor of six increase in the number of \ps sources known below a declination of $+30\degree$. We also highlight six the GPS sources that peak at 5\,GHz or above display a convex spectrum between the GLEAM survey and NVSS/SUMSS, implying that at least a portion of the GPS and HFP population is likely to go through multiple epochs of activity.

The spectra of a number of USS sources are found to be turning over at GLEAM frequencies. We suggest that a steep optically thin spectral index and a spectral peak below 230\,MHz are potential indicators of high redshift candidates since both spectral properties could be produced by radio lobes and hot spots being confined by dense ambient nuclear region. All of the sources with a reported spectroscopic redshift that have an optically thin spectral index less than $-1.2$ and a turnover frequency in the GLEAM band are located at redshifts greater than two. However, such a trend needs to be confirmed with optical or H\,\textsc{i} absorption follow-up campaigns. 

In terms of the spectral properties of the \ps sample, the distributions of the optically thin and thick spectral indices are similar to previous studies but with a much wider spread, mostly due to the high sensitivity of SUMSS/NVSS and the wide fractional bandwidth of the GLEAM survey. In particular, the large spread in the optically thick spectral indices is suggestive that the turnover in the spectrum is caused by an inhomogeneous environment that differs from source to source. There is also no dependence between the observed turnover frequency with redshift, as expected if the sources selected with observed low frequency spectral turnovers are composed of both local CSS sources and high redshift GPS sources. To measure linear sizes and decouple these two source populations, high resolution follow-up at low frequencies, for example with the GMRT or LOFAR, to will be required. 

The 5\,GHz luminosity distribution of the \ps sample lacks the brightest GPS and CSS sources of previous samples, suggesting a convolution of evolution and redshift effects for sources identified at low frequencies relative to \ps sources selected above 1\,GHz. While the trends evident in intrinsic turnover frequency and radio luminosity are consistent with redshift evolution, there are a number of sources at high power and low observed turnover frequencies that are inconsistent with the stereotypical evolutionary model of a \ps source. The small scale structure and optical properties of these sources need to be investigated to confirm if these sources have the same properties as young precursors to large radio galaxies.

Finally, we highlight six sources that have optically thick spectral indices near or above the SSA limit. If the sources are shown to be minimally variable and have milliarcsecond scale double morphologies, such a detection will be the first clear violation of SSA theory in GPS, CSS, and HFP sources. Independently, such sources represent excellent candidates for targeted follow-up to observe intrinsic H\,\textsc{i} absorption due to the inference of a dense ambient medium.

\section*{Acknowledgments}

This scientific work makes use of the Murchison Radio-astronomy Observatory, operated by CSIRO. We acknowledge the Wajarri Yamatji people as the traditional owners of the Observatory site. Support for the operation of the MWA is provided by the Australian Government (NCRIS), under a contract to Curtin University administered by Astronomy Australia Limited. We acknowledge the Pawsey Supercomputing Centre which is supported by the Western Australian and Australian Governments. This research was conducted by the Australian Research Council Centre of Excellence for All-sky Astrophysics (CAASTRO), through project number CE110001020. J.~R.~C. acknowledges the support of the Australian Postgraduate Award. The Dunlap Institute is funded through an endowment established by the David Dunlap family and the University of Toronto. B.~M.~G. acknowledges the support of the Natural Sciences and Engineering Research Council of Canada (NSERC) through grant RGPIN-2015-05948, and of the Canada Research Chairs program. This research has made use of the NASA/IPAC Extragalactic Database (NED), which is operated by the Jet Propulsion Laboratory, California Institute of Technology, under contract with the National Aeronautics and Space Administration, and the VizieR catalog access tool, CDS, Strasbourg, France. Topcat \citep{2005ASPC..347...29T}, SAOImage DS9, NASA's Astrophysics Data System bibliographic services, and Astropy, a community-developed core Python package for Astronomy \citep{2013A&A...558A..33A}, were also used in this study.

\bibliographystyle{apj}
\bibliography{Callingham_mwa_peakedspectrumsrcs_arxiv.bbl}

\begin{appendix}

\section{List of column headings for tables of the peaked-spectrum and convex source samples}\label{sec:appendix_tab}

The column numbers, names, and units for the tables presenting the different \ps source samples are outlined below. The tables are available online. The optically thick spectral index is not presented for sources that have a spectral peak below 100\,MHz due to the large uncertainty in the values. The parameters derived from the generic curvature model of Equation 3, such as $S_{\mathrm{p}}$, $\nu_{\mathrm{p}}$, $\alpha_{\mathrm{thin}}$, and $\alpha_{\mathrm{thick}}$, are not reported for the GPS sample, convex source sample, and for the sources that peak below 72\,MHz. Redshift values are not listed for sources that peak below 72\,MHz or the convex source sample. Note that all of the sub-band GLEAM flux densities are presented but only flux densities that had a SNR\,$>$\,3 are used in the spectral fitting and plotted in the corresponding source spectra. 
\newpage

\begin{longtable}{cccc}
\hline
\hline
Number & Name & Unit & Description \tabularnewline
\hline
1  & Gleam\_name  & -- & Name of the source in the GLEAM extragalactic catalog  \tabularnewline 
2  & RA\_gleam & degrees & Right ascension of the source in the the GLEAM extragalactic catalog  \tabularnewline 
3  & Dec\_gleam & degrees & Declination of the source in the GLEAM extragalactic catalog \tabularnewline 
4  & S\_p & Jy & Flux density at the spectral peak  \tabularnewline 
5  & S\_p\_error & Jy & Uncertainty in the spectral peak flux density  \tabularnewline 
6  & nu\_p & MHz & Frequency of the spectral peak  \tabularnewline 
7  & nu\_p\_error & MHz & Uncertainty in the spectral peak frequency  \tabularnewline 
8  & alpha\_thin & -- & Optically thin spectral index from fitting Equation 3 to the entire spectrum  \tabularnewline 
9  & alpha\_thin\_error & -- & Uncertainty in the optically thin spectral index  \tabularnewline 
10 & alpha\_thick & -- & Optically thick spectral index from fitting Equation 3 to the entire spectrum  \tabularnewline 
11 & alpha\_thick\_error & -- & Uncertainty in the optically thick spectral index  \tabularnewline 
12 & alpha\_low & -- & Low frequency spectral index derived from fitting a power-law to data between 72 and 231\,MHz  \tabularnewline 
13 & alpha\_low\_error & -- & Uncertainty in the low frequency spectral index  \tabularnewline 
14 & alpha\_high & -- & High frequency spectral index derived from fitting a power-law \\ & & & to SUMSS and/or NVSS flux density point(s) and to GLEAM data at 189 and 212\,MHz \tabularnewline 
15 & alpha\_high\_error & -- & Uncertainty in the high frequency spectral index   \tabularnewline 
16 & q & -- & Curvature parameter derived from fitting Equation 2 to only the GLEAM data  \tabularnewline
17 & q\_error & -- & Uncertainty in the curvature parameter  \tabularnewline 
18 & z & -- & Spectroscopic redshift  \tabularnewline 
19 & Ref. z & -- & Reference for the reported spectroscopic redshift  \tabularnewline 
20 & S\_200 & Jy & Integrated flux density measurement of the source in the GLEAM wideband image  \tabularnewline 
21 & S\_200\_error & Jy & Uncertainty in the GLEAM wideband flux density measurement  \tabularnewline 
22 & RA\_vlssr & degrees & Right ascension of the source in VLSSr \tabularnewline 
23 & Dec\_vlssr  & degrees & Declination of the source in VLSSr \tabularnewline 
24 & S\_vlssr & Jy & VLSSr flux density at 74\,MHz  \tabularnewline 
25 & S\_vlssr\_error & Jy & Uncertainty in the VLSSr flux density measurement, convolved with calibration uncertainty  \tabularnewline 
26 & TGSS-ADR1\_name & -- & Name of the source in TGSS-ADR1 \tabularnewline 
27 & RA\_tgss & degrees & Right ascension of the source in TGSS-ADR1 \tabularnewline 
28 & Dec\_tgss & degrees & Declination of the source in TGSS-ADR1 \tabularnewline 
29 & S\_tgss & Jy & TGSS-ADR1 flux density at 150\,MHz   \tabularnewline 
30 & S\_tgss\_error & Jy & Uncertainty in the TGSS-ADR1 flux density measurement  \tabularnewline
31 & MRC\_name & -- & Name of the source in MRC \tabularnewline  
32 & RA\_mrc  & degrees & Right ascension of the source in MRC \tabularnewline 
33 & Dec\_mrc  & degrees & Declination of the source in MRC \tabularnewline 
34 & S\_mrc  & Jy & MRC flux density at 408\,MHz \tabularnewline 
35 & S\_mrc\_error & Jy & Uncertainty in the MRC flux density measurement, convolved with calibration uncertainty  \tabularnewline 
36 & RA\_sumss & degrees & Right ascension of the source in SUMSS \tabularnewline 
37 & Dec\_sumss  & degrees & Declination of the source in SUMSS \tabularnewline 
38 & S\_sumss & Jy & SUMSS flux density at 843\,MHz \tabularnewline 
39 & S\_sumss\_error & Jy & Uncertainty in the SUMSS flux density measurement, convolved with calibration uncertainty  \tabularnewline 
40 & RA\_nvss  & degrees & Right ascension of the source in NVSS  \tabularnewline 
41 & Dec\_nvss  & degrees & Declination of the source in NVSS  \tabularnewline 
42 & S\_nvss & Jy & NVSS flux density at 1.4\,GHz  \tabularnewline 
43 & S\_nvss\_error & Jy & Uncertainty in the NVSS flux density measurement, convolved with calibration uncertainty  \tabularnewline 
44 & S\_76  & Jy & Integrated flux density at 76\,MHz from the GLEAM extragalactic catalog \tabularnewline 
45 & S\_76\_error  & Jy & Uncertainty in the integrated flux density at 76\,MHz, which is the convolution of fitting and systematic errors  \tabularnewline 
46 & S\_84 & Jy & Integrated flux density at 84\,MHz from the GLEAM extragalactic catalog \tabularnewline 
47 & S\_84\_error & Jy & Uncertainty in the integrated flux density at 84\,MHz, which is the convolution of fitting and systematic errors  \tabularnewline 
48 & S\_92 & Jy & Integrated flux density at 92\,MHz from the GLEAM extragalactic catalog \tabularnewline 
49 & S\_92\_err& Jy & Uncertainty in the integrated flux density at 92\,MHz, which is the convolution of fitting and systematic errors  \tabularnewline 
50 & S\_99 & Jy & Integrated flux density at 99\,MHz from the GLEAM extragalactic catalog \tabularnewline 
51 & S\_99\_error& Jy & Uncertainty in the integrated flux density at 99\,MHz, which is the convolution of fitting and systematic errors  \tabularnewline 
52 & S\_107  & Jy & Integrated flux density at 107\,MHz from the GLEAM extragalactic catalog \tabularnewline 
53 & S\_107\_error & Jy & Uncertainty in the integrated flux density at 107\,MHz, which is the convolution of fitting and systematic errors  \tabularnewline 
54 & S\_115  & Jy & Integrated flux density at 115\,MHz from the GLEAM extragalactic catalog \tabularnewline 
55 & S\_115\_error & Jy & Uncertainty in the integrated flux density at 115\,MHz, which is the convolution of fitting and systematic errors  \tabularnewline 
56 & S\_122  & Jy & Integrated flux density at 122\,MHz from the GLEAM extragalactic catalog \tabularnewline 
57 & S\_122\_error & Jy & Uncertainty in the integrated flux density at 122\,MHz, which is the convolution of fitting and systematic errors  \tabularnewline 
58 & S\_130  & Jy & Integrated flux density at 130\,MHz from the GLEAM extragalactic catalog \tabularnewline 
59 & S\_130\_err & Jy & Uncertainty in the integrated flux density at 130\,MHz, which is the convolution of fitting and systematic errors  \tabularnewline 
60 & S\_143  & Jy & Integrated flux density at 143\,MHz from the GLEAM extragalactic catalog \tabularnewline 
61 & S\_143\_error & Jy & Uncertainty in the integrated flux density at 143\,MHz, which is the convolution of fitting and systematic errors  \tabularnewline 
62 & S\_151  & Jy & Integrated flux density at 151\,MHz from the GLEAM extragalactic catalog \tabularnewline 
63 & S\_151\_error & Jy & Uncertainty in the integrated flux density at 151\,MHz, which is the convolution of fitting and systematic errors  \tabularnewline 
64 & S\_158  & Jy & Integrated flux density at 158\,MHz from the GLEAM extragalactic catalog \tabularnewline 
65 & S\_158\_error & Jy & Uncertainty in the integrated flux density at 158\,MHz, which is the convolution of fitting and systematic errors  \tabularnewline 
66 & S\_166  & Jy & Integrated flux density at 166\,MHz from the GLEAM extragalactic catalog \tabularnewline 
67 & S\_166\_error & Jy & Uncertainty in the integrated flux density at 166\,MHz, which is the convolution of fitting and systematic errors  \tabularnewline 
68 & S\_174  & Jy & Integrated flux density at 174\,MHz from the GLEAM extragalactic catalog \tabularnewline 
69 & S\_174\_error & Jy & Uncertainty in the integrated flux density at 174\,MHz, which is the convolution of fitting and systematic errors  \tabularnewline 
70 & S\_181  & Jy & Integrated flux density at 181\,MHz from the GLEAM extragalactic catalog \tabularnewline 
71 & S\_181\_error & Jy & Uncertainty in the integrated flux density at 181\,MHz, which is the convolution of fitting and systematic errors  \tabularnewline 
72 & S\_189  & Jy & Integrated flux density at 189\,MHz from the GLEAM extragalactic catalog \tabularnewline 
73 & S\_189\_error & Jy & Uncertainty in the integrated flux density at 189\,MHz, which is the convolution of fitting and systematic errors  \tabularnewline 
74 & S\_197  & Jy & Integrated flux density at 197\,MHz from the GLEAM extragalactic catalog \tabularnewline 
75 & S\_197\_error & Jy & Uncertainty in the integrated flux density at 197\,MHz, which is the convolution of fitting and systematic errors  \tabularnewline 
76 & S\_204  & Jy & Integrated flux density at 204\,MHz from the GLEAM extragalactic catalog \tabularnewline 
77 & S\_204\_error & Jy & Uncertainty in the integrated flux density at 204\,MHz, which is the convolution of fitting and systematic errors  \tabularnewline 
78 & S\_212  & Jy & Integrated flux density at 212\,MHz from the GLEAM extragalactic catalog \tabularnewline 
79 & S\_212\_error & Jy & Uncertainty in the integrated flux density at 212\,MHz, which is the convolution of fitting and systematic errors  \tabularnewline 
80 & S\_220  & Jy & Integrated flux density at 220\,MHz from the GLEAM extragalactic catalog \tabularnewline 
81 & S\_220\_error & Jy & Uncertainty in the integrated flux density at 220\,MHz, which is the convolution of fitting and systematic errors  \tabularnewline 
82 & S\_227  & Jy & Integrated flux density at 227\,MHz from the GLEAM extragalactic catalog \tabularnewline 
83 & S\_277\_error & Jy & Uncertainty in the integrated flux density at 227\,MHz, which is the convolution of fitting and systematic errors  \tabularnewline 
\hline
\end{longtable}
The number references in the table correspond to the following studies:
(0) \citet{1975MNRAS.173p..87B}; (1) \citet{1978ApJ...219L...1J}; (2) \citet{1978MNRAS.185..149H}; (3) \citet{1979ApJ...229...73W}; (4) \citet{1983A&A...117...60F}; (5) \citet{1983MNRAS.205..793W}; (6) \citet{1984ApJ...286..498J}; (7) \citet{1986ChA&A..10..196C}; (8) \citet{1989A&AS...80..103S}; (9) \citet{1989ApJS...69....1H}; (10) \citet{1990PASP..102.1235T}; (11) \citet{1990PKS...C......0W}; (12) \citet{1991ApJS...75..297H}; (13) NED; (14) \citet{1993MNRAS.263..139S}; (15) \citet{1993MNRAS.263..999T}; (16) \citet{1994A&AS..105..211S}; (17) \citet{1994ApJ...428...65H}; (18) \citet{1994ApJ...436..678O}; (19) \citet{1994ApJS...95..157C}; (20) \citet{1994MNRAS.269..998d}; (21) \citet{1995A&AS..114..259d}; (22) \citet{1995AJ....110.2570L}; (23) \citet{1995ApJS..100...69F}; (24) \citet{1995RMxAA..31..119M}; (25) \citet{1995RMxAA..31..159M}; (26) \citet{1996ApJ...468..556S}; (27) \citet{1996ApJS..107...19M}; (28) \citet{1996MNRAS.279.1294S}; (29) \citet{1996MNRAS.281..425M}; (30) \citet{1997A&A...323L..21Z}; (31) \citet{1997A&A...326..505R}; (32) \citet{1997A&A...328...48W}; (33) \citet{Odea1997}; (34) \citet{1997IAUC.6639....2H}; (35) \citet{1997MNRAS.284...85D}; (36) \citet{1998ApJ...494..175C}; (37) \citet{1998ApJ...503..138D}; (38) \citet{1998ApJS..118..275K}; (39) \citet{1999AJ....117.1122S}; (40) \citet{1999AJ....117.2034T}; (41) \citet{1999ApJ...518L..61V}; (42) \citet{1999MNRAS.310..223B}; (43) \citet{2000A&AS..143..181D}; (44) \citet{2000MNRAS.313..237W}; (45) \citet{2001A&A...379..393E}; (46) \citet{2001AJ....121.1241D}; (47) \citet{2002A&A...386...97J}; (48) \citet{2002MNRAS.337..981S}; (49) \citet{2001MNRAS.328.1039C}; (50) \citet{2003AJ....125..572H}; (51) \citet{2003AJ....126.2237D}; (52) \citet{2003MNRAS.340..632L}; (53) \citet{2003MNRAS.346.1021B}; (54) \citet{2004A&A...418..813H}; (55) \citet{2004ApJ...609..564S}; (56) \citet{2004MNRAS.349.1397C}; (57) \citet{2004AJ....128..502A}; (58) \citet{2005ApJ...626...95S}; (59) \citet{2005ApJ...632..751R}; (60) \citet{2006ApJS..162...38A}; (61) \citet{2006A&A...445..889T}; (62) \citet{2006AJ....131..114B}; (63) \citet{2001MNRAS.322..486B}; (64) \citet{2006MNRAS.372..425C}; (65) \citet{Labiano2007}; (66) \citet{2007A&A...464..879D}; (67) \citet{2008ApJS..175..297A}; (68) \citet{2008ApJS..175...97H}; (69) \citet{2008MNRAS.385.1297B}; (70) \citet{2008MNRAS.387..639H}; (71) \citet{2008MNRAS.390..819G}; (72) \citet{2007MNRAS.375..931M}; (73) \citet{2009ApJS..180...67R}; (74) \citet{2009MNRAS.395.1099B}; (75) \citet{2010AJ....139.2360S}; (76) \citet{2010ApJ...712...14M}; (77) \citet{2010ApJS..187..272W}; (78) \citet{2010MNRAS.401..633D}; (79) \citet{2010MNRAS.401.1429D}; (80) \citet{2010MNRAS.403..906N}; (81) \citet{2010MNRAS.405.2302H}; (82) \citet{2010MNRAS.406.1435E}; (83) \citet{2011A&A...525A..51B}; (84) \citet{2011AJ....142..165T}; (85) \citet{2011MNRAS.417.2651M}; (86) \citet{2012ApJ...754...38C}; (87) \citet{2013A&A...556A.140Z}; (88) \citet{2014ApJ...795...63F}; (89) \citet{2012MNRAS.421.1569B}

\end{appendix}
	
\end{document}